\documentclass[final]{aa}
\usepackage[varg]{txfonts}
\usepackage{graphicx}
\usepackage{natbib}
\usepackage{subfigure}
\usepackage{rotating}
\usepackage{lscape}
\bibpunct{(}{)}{;}{a}{}{,}

\begin{document}

\title{The Fornax Deep Survey (FDS) with the VST.} 

\subtitle{IV. A size and magnitude limited catalog of dwarf galaxies in the area of the Fornax cluster} 

\author{Aku Venhola\inst{1,2}
        \and Reynier Peletier\inst{2}
        \and Eija Laurikainen\inst{1}
        \and Heikki Salo\inst{1}
        \and Enrichetta Iodice\inst{3}
        \and Steffen Mieske \inst{4}
        \and Michael Hilker \inst{5}
        \and Carolin Wittmann \inst{6}
        \and Thorsten Lisker\inst{6}
        \and Maurizio Paolillo \inst{3,7}
        \and Michele Cantiello \inst{8}
        \and Joachim Janz \inst{1,15}
        \and Marilena Spavone \inst{3}
        \and Raffaele D'Abrusco \inst{9}
        \and Glenn van de Ven \inst{5,10}
        \and Nicola Napolitano \inst{3}
        \and Gijs Verdoes Kleijn \inst{2}
        \and Natasha Maddox \inst{12}
        \and Massimo Capaccioli \inst{11}
        \and Aniello Grado \inst{3}     
        \and Edwin Valentijn \inst{2}
        \and Jes\'us Falc\'on-Barroso \inst{13,14}
        \and Luca Limatola \inst{3}
        }

\offprints{A. Venhola, \email{avenhola@univ.yo.oulu.fi}}

\institute{Astronomy Research Unit, University of Oulu, Finland 
  \and Kapteyn Institute, University of Groningen, Groningen, the Netherlands
  \and INAF - Astronomical Observatory of Capodimonte, Salita Moiariello 16, I80131, Naples, Italy 
  \and European Southern Observatory, Alonso de Cordova 3107, Vitacura,  Santiago, Chile
  \and European Southern Observatory, Karl-Schwarzschild-Strasse 2, D-85748 Garching bei München, Germany 
  \and Astronomisches Rechen-Institut, Zentrum f\"ur Astronomie der Universit\"at Heidelberg, M\"onchhofstra\ss e 12-14, 69120 Heidelberg, Germany
  \and University of Naples Federico II, C.U. Monte Sant'Angelo, Via Cinthia, 80126 Naples, Italy
  \and INAF Osservatorio Astronomico di Teramo,  Via Maggini, 64100, Teramo, Italy
  \and Smithsonian Astrophysical Observatory, 60 Garden Street, 02138 Cambridge, MA, USA
  \and Max-Planck-Institut für Astronomie, Königstuhl 17, D-69117 Heidelberg, Germany
  \and  University of Naples Federico II, C.U. Monte Sant'Angelo, Via Cinthia, 80126 Naples, Italy
  \and ASTRON, the Netherlands Institute for Radio Astronomy, Postbus 2, 7990 AA, Dwingeloo, the Netherlands
  \and Instituto de Astrofisica de Canarias, C/ Via L'actea s/n, 38200 La Laguna, Spain 
  \and Depto. Astrofisica, Universidad de La Laguna,  C/ Via L'actea s/n, 38200 La Laguna, Spain
  \and Finnish Centre of Astronomy with ESO (FINCA)
, University of Turku, Väisäläntie 20, FI-21500 Piikkiö, Finland  }

\date{Received \today / Accepted ---}

\abstract { The Fornax Deep Survey (FDS), an imaging survey in the u', g', r', and i'-bands, has a supreme resolution and image depth compared to the previous spatially complete Fornax Cluster Catalog (FCC). Our new data allows us to study the galaxies down to r'-band magnitude m$_{r'}$ $\approx$ 21 mag (M$_{r'}$ $\approx$ -10.5 mag), which opens a new parameter regime to investigate the evolution of dwarf galaxies in the cluster environment. After the Virgo cluster, Fornax is the second nearest galaxy cluster to us, and with its different mass and evolutionary state, it provides a valuable comparison that makes it possible to understand the various evolutionary effects on galaxies and galaxy clusters. These data provide an important legacy dataset to study the Fornax cluster.
} 
{ We aim to present the Fornax Deep Survey (FDS)        dwarf galaxy catalog, focusing on explaining the data reduction and calibrations, assessing the quality of the data, and describing the methods used for defining the cluster memberships and first order morphological classifications for the catalog objects. We also describe the main scientific questions that will be addressed based on the catalog. This catalog will also be invaluable for future follow-up studies of the Fornax cluster dwarf galaxies.
}
{ As a first step we used the SExtractor fine-tuned for dwarf galaxy detection, to find galaxies from the FDS data, covering a 26 deg$^2$ area of the main cluster up to its virial radius, and the area around the Fornax A substructure. We made 2D-decompositions of the identified galaxies using GALFIT, measure the aperture colors, and the basic morphological parameters like concentration and residual flux fraction. We used color-magnitude, luminosity-radius and luminosity-concentration relations to separate the cluster galaxies from the background galaxies. We then divided the cluster galaxies into early- and late-type galaxies according to their morphology and gave first order morphological classifications using a combination of visual and parametric classifications. 
}
{ Our final catalog includes 14,095 galaxies. We classify 590 galaxies as being likely Fornax cluster galaxies, of which 564 are dwarfs (M$_{r'}$ > -18.5 mag) consisting our Fornax dwarf catalog.  Of the cluster dwarfs we classify 470 as early-types, and 94 as late-type galaxies. Our final catalog reaches its 50\% completeness limit at magnitude M$_{r'}$ = -10.5 mag and surface brightness $\bar{\mu}_{e,r'}$ = 26 mag arcsec$^{-2}$, which is approximately three magnitudes deeper than the FCC. Based on previous works and comparison with a spectroscopically confirmed subsample, we estimate that our final Fornax dwarf galaxy catalog has  $\lesssim$  10\% contamination from the background objects.
}{}

\keywords{galaxies : dwarf galaxies : Fornax cluster : galaxy catalog : photometry} 
\maketitle   

\section{Introduction}

Understanding galaxy evolution is one of the major problems of astronomy. During recent  decades, our understanding of the basic processes involved in the evolution of galaxies in the context of the $\Lambda$CDM cosmology has taken great steps, but many details are not yet well understood. For example, the environmental dependence of the frequency of different galaxy morphologies was discovered by \citet{Dressler1980}. Despite this, the importance of the different mechanisms transforming star-forming late-type galaxies into quiescent and red early-type galaxies in the group and cluster environments (see e.g., \citealp{Peng2012},\citealp{Peng2014},\citealp{Jaffe2018}), is still unclear. 

\indent An important resource for studying galaxy evolution is the availability of homogeneous and complete samples of galaxy observations that can be used statistically to investigate how the properties of the galaxies change in different environments. The new deep surveys, such as the Next Generation Virgo Survey (NGVS, \citealp{Ferrarese2012}), the Next Generation Fornax Survey (NGFS, \citealp{Munoz2015}), VST Early-type GAlaxy Survey (VEGAS, \citealp{Capaccioli2015}) and  the Fornax Deep Survey (FDS, Peletier et al, in prep.) reveal a large number of previously unknown faint galaxies that are powerful probes to environmental processes. At the same time, large scale cosmological simulations such as  IllustrisTNG \citep{Pillepich2018}, have reached such a high resolution that direct comparisons down to dwarf sized galaxies with stellar mass of M$_*$ = 10$^{8-9}$ M$_\odot$ can be made.

\indent The faint galaxies found in the new imaging surveys typically lack distance information, and many of these galaxies have such a low surface brightness that obtaining their spectroscopic redshifts for a complete sample is not realistic with the currently available instruments. Thus, to be able to exploit these galaxies in a statistical way, one needs to assess cluster memberships using their photometric properties. Photometric redshifts (see e.g., \citealp{Bilicki2018} and references therein) are often used to obtain distances for a large samples of galaxies. Another way to obtain distances of the galaxies is to use the known scaling relations for galaxies. In clusters, there are hundreds of galaxies located at a similar distance, and many of their parameters scale with each other. However, the background galaxies are located at a range of distances, so that their apparent properties do not follow these relations. Useful relations that are commonly used for identifying cluster members are the color-magnitude and luminosity-surface brightness relations (see e.g., \citealp{Misgeld2009}). Already \citet{Binggeli1985} and \citet{Ferguson1989} have used colors, the magnitude-surface brightness relation and galaxy morphology for defining the membership status of their newly found galaxies in the Virgo and Fornax clusters, respectively. Follow-up studies of these surveys based on spectroscopy or surface brightness fluctuations (see e.g., \citealp{Drinkwater2000}, \citealp{Mieske2007}) have proven the photometric classifications to be very robust: more than 90\% of the galaxies selected this manner are confirmed to be cluster members.

\begin{figure*}[!ht]
    \centering
        \includegraphics[width=17cm]{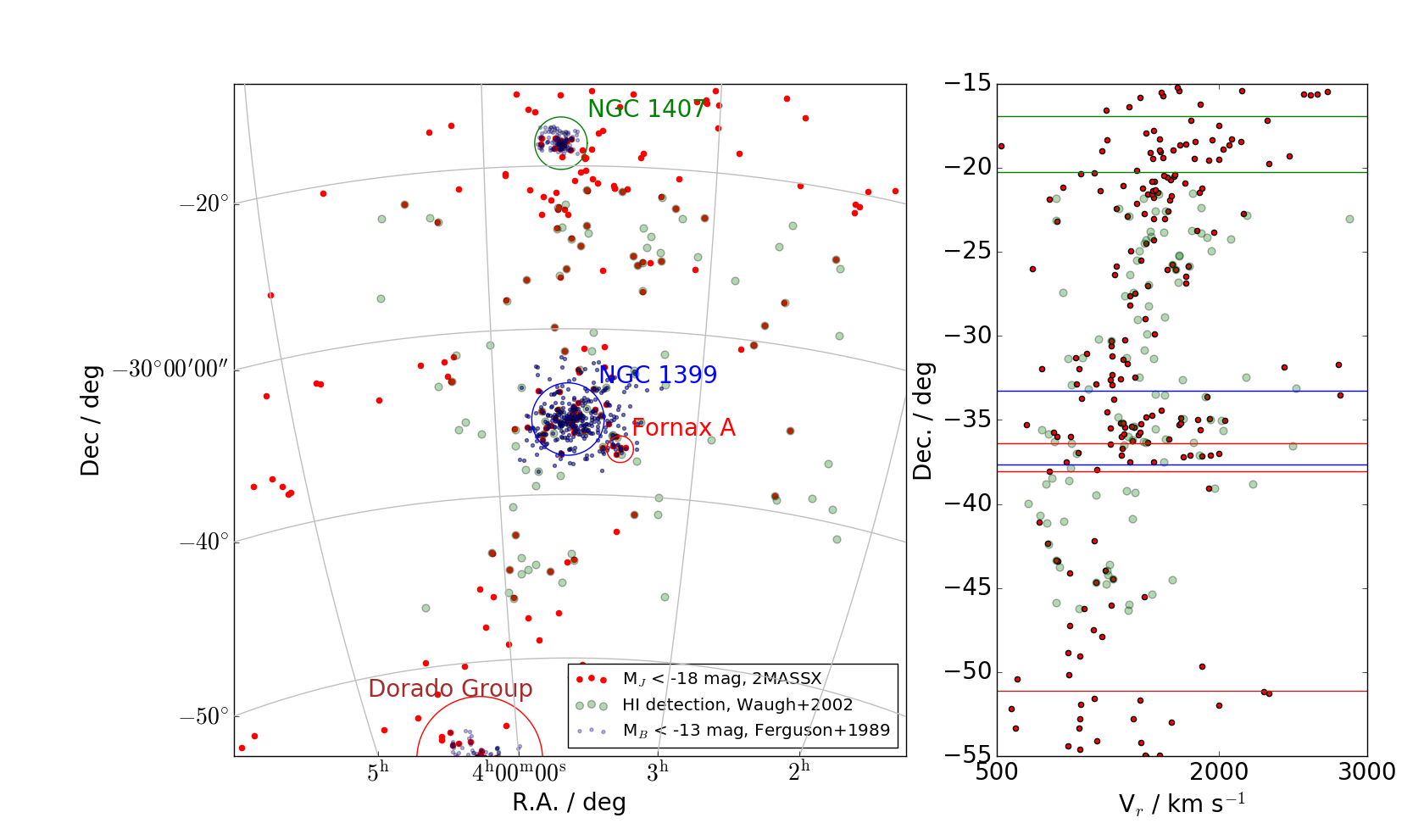}
        \caption{Large scale structure surrounding the Fornax cluster. The left panel shows galaxy right ascension and declination in International Celestial Reference System (ICRS) coordinates, and the right panel shows the recession velocities of the galaxies as a function of declination. At the distance of the Fornax cluster 1 deg corresponds to 0.3 Mpc, and 1000 km/s velocity difference due to Hubble flow corresponds to 14 Mpc (to first order independent of the distance). The galaxies with recession velocities V$_r$ < 4000 km s$^{-1}$ in the 2 Micron All Sky Survey catalog (2MASSX, \citealp{Huchra2012}), are plotted with the red dots, and the galaxies with velocities V$_r$ < 4000 km s$^{-1}$ from \citet{Waugh2002} with the green circles. The FCC galaxies are shown with blue dots. We also indicate the virial radii of the most significant groups in the surroundings of the Fornax cluster with the large circles, and show their names with the corresponding colors. The locations of the circles of the left panel are shown by the horizontal lines in the right panel using the corresponding colors.}
        \label{fig:surroundigs}
\end{figure*}

The Fornax cluster appears on the southern sky centered around the elliptical galaxy NGC 1399 with coordinates R.A. = 54.6209 deg and Dec. = -35.4507 deg \citep{Watson2009}. Its mean recession velocity is 1493$\pm$36 kms$^{-1}$ \citep{Drinkwater2001}, and the mean distance calculated from surface brightness fluctuations of early-type galaxies is 20.0$\pm$0.3$\pm$1.4 Mpc \citep{Blakeslee2009}.  The main cluster is very compact and consists of 22 galaxies brighter than M$_B$ < -18 mag and around 200 fainter galaxies \citep{Ferguson1989}. The Fornax Cluster is part of the larger Fornax-Eridanus structure (see \citealp{Nasonova2011}) located in the Fornax-filament of the cosmic web. Fornax, having a virial mass of M = 7$\times$10$^{13}$ M$_{\odot}$, is the most massive mass concentration (see Fig. \ref{fig:surroundigs}) in the filament. Other significant mass concentrations near the Fornax cluster are the groups around NGC 1316 (Fornax A), NGC 1407 and the Dorado group (see Fig. \ref{fig:surroundigs}). The NGC 1316 group is currently falling into the main group (\citealp{Drinkwater2001}), whereas the other spectroscopically confirmed significant groups are located at least 15 deg ($\approx$ 5 Mpc) away from the Fornax cluster.

\indent The Fornax cluster is an interesting environment to study, since it bridges the mass range of evolved groups to more massive clusters. For instance, \citet{Trentham2009} study dwarf galaxies in the group environments of which the NGC5846 group, with a mass of M = 8.4$\pm$2.0$\times$10$^{13}$ M$_{\odot}$, is more massive than the Fornax cluster. However, regardless of its low mass,  the Fornax cluster has many properties that qualify it as a cluster, such as concentration, X-ray intensity, and evolved galaxy population. Due to its low mass it may also be an interesting test case for simulations: the high-resolution cosmological simulations like Millennium-II \citep{BoylanKolchin2009}, and the ongoing 50 Mpc box simulations of IllustrisTNG \citep{Pillepich2018} have only a handful of Virgo-mass clusters, but many have a Fornax cluster mass, so they provide a great opportunity for both dwarf resolution and good population statistics when using them to interpret observational data.

\indent Due to its southern location the Fornax cluster is not covered by the Sloan Digital Sky Survey (SDSS, \citealp{Alam2015}). The most recent galaxy catalog covering the whole cluster is the Fornax Cluster Catalog (FCC) by \citet{Ferguson1989}. The catalog covers 40 deg$^2$ area centered onto the Fornax cluster, and it contains 2678 galaxies in total. Its given completeness limit in apparent B-magnitude is m$_B$ $\approx$ 19 mag, but it may vary due to visual identification of the galaxies. In the catalog, Ferguson classified galaxies as being either likely cluster galaxies or likely background galaxies using the morphology and surface brightness of the galaxies. The whole catalog contains 340 likely cluster members in the area of the Fornax cluster, and more than two thousand background galaxies. 

\indent Another major effort for mapping the Fornax cluster galaxies with higher resolution was done using the Hubble Space Telescope  \citep{Jordan2007}. In their ACS Fornax Cluster Survey, the authors target the brightest 43 galaxies using two different filters. Their spatial coverage is much smaller than the one of FCC, but the spatial resolution of the observations is superior. The core region of the cluster was also covered with deep observations by \citet{Hilker2003} and \citet{Mieske2007}, who used the 100-inch du Pont telescope and the Inamori-Magellan Areal Camera and Spectrograph - instrument (IMACS, \citealp{Dressler2011}) at Las Campanas Observatory (Chile), respectively. Both observational surveys were performed in V and I bands and they were able to obtain colors and structural parameters of the cluster dwarfs down to M$_V$ = -9 mag. Another ongoing effort to image the Fornax cluster with modern instruments is the Next Generation  Fornax Survey collaboration (NGFS, \citealp{Munoz2015}, \citealp{Eigenthaler2018}). The NGFS aims to cover 30 deg$^2$ area in u', g', i', and Ks bands in the Fornax cluster with similar observations as FDS, using the DECam instrument attached to 4-m telescope Blanco at Cerro Tololo Inter-American Observatory (CTIO) for the optical u', g', and i' bands, and VISTA/VIRCAM (\citealp{Sutherland2015}) for the Ks-band. So far, the NGFS has published their galaxy catalog covering the area within the virial radius of the Fornax cluster (\citealp{Eigenthaler2018}, \citealp{Ordenes-Briceno2018}) with 643 dwarf galaxies altogether.

\indent A major effort for obtaining spectroscopic redshifts for the Fornax cluster galaxies was the 2dF Fornax survey made by \cite{Drinkwater1999}, who obtained spectroscopy for several hundreds of galaxies located in a $\approx$ 9 deg$^2$ area in the main cluster. However, only a few percent of the observed objects were cluster galaxies, since there was no morphological selection for the targets.  Recently, the spectroscopic 2dF observations were extended by additional 12 deg$^2$ (Maddox et al., in prep.), which more than doubles the area with spectroscopic data. The spectroscopic data are limited to relatively high surface brightness objects (B-band central surface brightness $\mu_{0,B}$ < 23 mag arcsc$^{-2}$), which unfortunately excludes most of the dwarf galaxies. Spectroscopic  redshifts are  available for several tens of bright galaxies (m$_J$ < 14 mag) in the Fornax cluster and in its surroundings, made by the 2 Micron All-Sky Survey (2MASS) spectroscopic survey \citep{Huchra2012}. Several spectroscopic redshifts from HI-data were obtained by \citet{Waugh2002}, but most of these galaxies are in the surroundings of the main cluster.

\indent Previous work on the Fornax cluster suggests that the center of the cluster is  dynamically evolved, which means that most of the galaxies have travelled at least once through the cluster center, but there is still ongoing in-fall of subgroups and individual galaxies in the outskirts. The X-ray analysis of the hot intra-cluster gas by \citet{Paolillo2002} shows that there is a concentration of X-ray gas in the center of the cluster that has a mass of M $\approx$ 10$^{11}$ M$_\odot$ within the inner 100 kpc. However, this X-ray gas shows a lopsided distribution toward the northwest, which is a sign of it not being fully virialized. The high concentration of galaxies in the center of the Fornax cluster \citep{Ferguson1989} and the observed mass segregation of the galaxies \citep{Drinkwater2001} are both signs that the galaxies in the center have spent several Gyrs in the cluster environment corresponding to a few crossing times\footnote{
If we consider a galaxy located at half a virial radius from the cluster center ($R$=0.35 Mpc) with a velocity similar to the velocity dispersion of the cluster galaxies, ($V$ =370 km s$^{-1}$), the crossing time is $t_{cross}\approx$1 Gyr. }.  This long standing interaction of galaxies with the cluster potential is possibly the main mechanism that has produced a significant intracluster population of stars, as the one recently traced by globular clusters \citep{Pota2018} and planetary nebulae \citep{Spiniello2018} in the core of the Fornax cluster. This population shows a velocity dispersion which is consistent with the one of the galaxy population in the same area, hence supporting the picture of a cluster core being dynamically evolved.   \citet{Drinkwater2001} analyzed the substructure of the Fornax cluster using the Fornax spectroscopic survey. They discussed that, although showing signs of a relaxed system, the Fornax cluster still has two groups of galaxies with common systematic velocities clearly different from the one of the main cluster. Additionally, the high early-type galaxy fraction in the Fornax cluster  (E+S0+dE+dS0)/all = 0.87 \citep{Ferguson1989} is a sign that the galaxies have spent a long time in the cluster without forming many new stars.

\indent The obtained multiband optical images of FDS extend the previous Fornax surveys with data that cover a large spatial area and are very deep\footnote{Azimuthally averaged profiles can be determined down to $\mu_r$=30 mag arcsec$^{-2}$ \citep{Iodice2016}. See also Section 4.1.}. At the same time their $\approx$ 1 arcsec (100 pc at the distance of the Fornax cluster) resolution allows detailed morphological analysis of dwarf galaxies. The survey has already led to publication of several papers, which have demonstrated the usefulness of this deep high resolution data in various different scientific cases (\citealp{Iodice2016}, \citealp{DAbrusco2016}, \citealp{Iodice2017b}, \citealp{Iodice2017a}, \citealp{Venhola2017}, \citealp{Cantiello2018} ).

\indent In this paper we present the steps necessary to construct the FDS dwarf galaxy catalog containing all the cluster member galaxies with M$_{r'}$ > -18.5 mag. Observations used in this work are described in Section 2. In Sections 3 and 4, we explain the data reduction and calibration, and assess the quality of the final data products, respectively. We then explain the preparation of the galaxy detection images (Section 5), our detection method (Section 6), and the photometric analysis done for the detected galaxies (Section 7). In section 8, we use the photometric parameters of the galaxies to separate the background objects from the cluster galaxies and finally classify the Fornax cluster galaxies into early- and late-type systems.  In Section 9, we compare our catalog with the previous Fornax studies. Throughout the paper we assume a distance of 19.7 Mpc for the Fornax cluster, which corresponds to a distance modulus of 31.51 mag \citep{Blakeslee2009}. Due to the high Galactic latitude of the Fornax cluster (Galactic declination = -53.63 deg) the dust reddening is small\footnote{According to \citet{Schlafly2011} the dust extinction coefficients in the area of the Fornax cluster are 0.05, 0.04, 0.03, and 0.02 mag for u', g', r', and i' filters.} and therefore, if not stated explicitly, we use non-corrected values for magnitudes.

\section{Observations}

The Fornax Deep Survey is a collaboration of the two guaranteed observing time surveys Focus (PI: R. Peletier) and VEGAS (PI: E. Iodice, see also \citealp{Capaccioli2015}) that covers the area of the Fornax cluster and Fornax A subgroup with deep multiband imaging. The FDS is executed using the OmegaCAM \citep{Kuijken2002} instrument attached to the survey telescope of the Very Large Telescope (VST, \citealp{Schipani2012}), which is a 2.6 m telescope located at Cerro Paranal, Chile. The camera consists of 32 $CCD$-chips, has a 0.21 arcsec pixel$^{-1}$ resolution, and a field of view of $\approx$1 deg $\times$ 1 deg. The observations of the FDS were performed between November 2013 and November 2017, and they are listed in Table \ref{table:observations}. All the observations were performed in clear (photometric variations $<$ 10 \%) or photometric conditions  with a typical seeing FWHMs of 1.2, 1.1, 1.0, and 1.0 arcsec in u', g', r', and i'-bands. The u' and g'-band observations were performed in dark time, and the other bands in gray or dark time.

\begin{table}
\caption{Fornax Deep Survey observations used in this work. The columns correspond to the date, ESO observing period, and total exposure times per filter in hours.} 
\label{table:observations} 
\centering  
\begin{tabular}{l c c c c c}       
\hline\hline  
     &      & \multicolumn{4}{c}{Total exposure time [h]} \\  
Date & Period &    u'   &    g'   &   r'   &   i'   \\
\hline      
        Nov, 2013      & P92  &  9.2 & 6.0 & 7.0 & 0.4       \\
    Nov, 2014      & P94  &  16.0 & 15.9 & 11.2 & 12.3  \\   
    Oct, Nov, 2015 & P96  &  6.2 & 20.7 & 21.0 & 2.7   \\
    Oct, Nov, 2016 & P98  &  - & 15.5 & 15.1 & 7.9    \\
    Oct, Nov, 2017 & P100 &  31.8 & - & - & 17.2   \\
\hline       
\end{tabular}
\end{table}

\begin{figure*}[!ht]
    \centering
        \includegraphics[width=17cm]{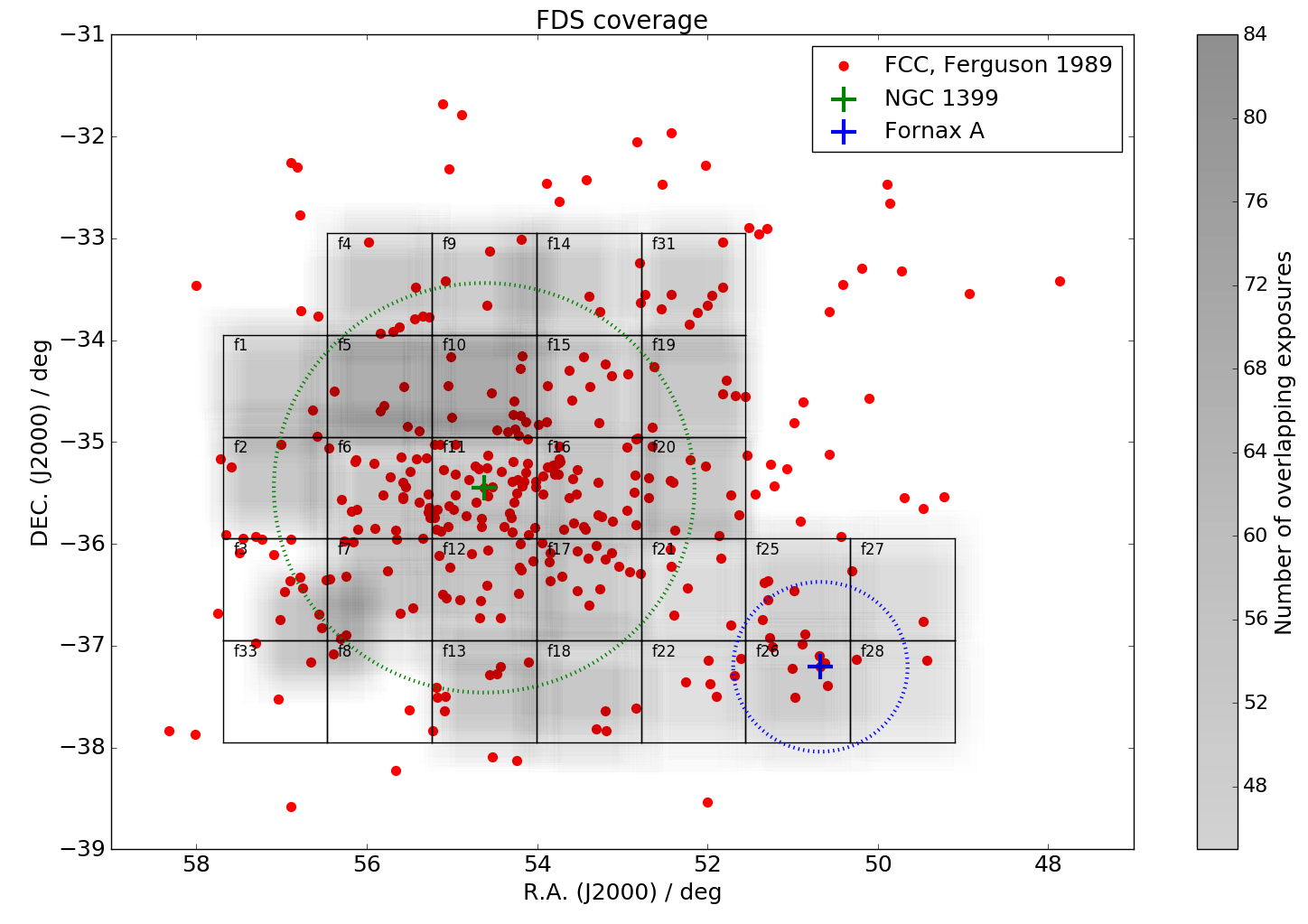}
    \caption{  Locations of the observed 1$^{\circ}$ x 1$^{\circ}$ (corresponding to 325 kpc x 325 kpc at the distance of the Fornax cluster) sized FDS fields are plotted in black. The r'-band weight maps (see Section 3.1) are shown in the gray-scale colors, darker color corresponding to deeper observations.  All the FCC galaxies \citep{Ferguson1989} classified as "likely members" or "definitive members" are shown with red points. We note that compared to FDS, FCC covers a slightly larger area of the cluster. The green dotted circle shows the virial radius of 2.2$^{\circ}$ ($\approx$ 0.7 Mpc, \citealp{Drinkwater2001}), and the green cross shows the central galaxy NGC 1399. The blue cross and the blue dotted line show the peculiar elliptical galaxy NGC 1316 in the center of the Fornax A subgroup, and the 2$\sigma$ galaxy overdensity around it, respectively. }
    \label{fig:fieldlocaitons}
\end{figure*}

\indent The observing strategy of the FDS is described in \citet{Venhola2017} and \citet{Iodice2016} and will be described more comprehensively in the survey paper by Peletier et al. (in preparation), but for completeness, a short description is given also here. The observations were performed using short 3 min exposure times and large $\approx$ 1 deg dithers between the consecutive exposures. The fields were observed in sets of two to three fields in such a way that after visiting all the fields once, an offset of 10 arcmin with respect to the previous observation of a given field was made. Directions of the small $\sim$10 arcmin offsets were randomly chosen around the centers of the fields. The large dithers and offsets ensure that the same objects do not appear twice in the same pixel, and makes it possible to stack consecutive observations as a background model (see Section 3.1).  For reference, the halo of NGC 1399, located in Field 11, extends over an area of 1 deg$^2$ (\citealp{Iodice2016}), which would lead it covering the full field of view of the observations of that field if we did not use the adopted dithering and offset strategy. To obtain the necessary depth in the images each field was visited 75, 55, 55, and 35 times with the u', g', r', and i' filters, respectively. The locations of the observed fields are shown in Fig. \ref{fig:fieldlocaitons}. The observations cover a 20 deg$^2$ area in the main cluster in u', g', r', and i', and additional 6 deg$^2$ in the Fornax A southwest subgroup in g', r', and i'. All observations follow a regular grid of target fields comprising continuous coverage, except in the area of Fields 3, 33, and 8 in which some gaps occur due to bright stars.

\section{Data reduction}

\subsection{Instrumental corrections}

The instrumental corrections applied for each frame include overscan correction, removal of bias, flatfielding, illumination correction, masking of the bad pixels, and subtraction of the background. The data is overscan corrected by subtracting from each pixel row the row-wise median values, read from the {\it CCD} overscan areas. The fine structure of the bias is then subtracted using a master bias frame stacked from ten overscan corrected bias frames.

\indent Flatfielding is done after bias correction using a master flatfield which is combined from eight twilight flatfields and eight dome flatfields. Before combining the different flatfields, the high spatial frequencies are filtered out from the twilight flatfields, and the low  frequency spatial Fourier frequencies from the dome flatfields. This approach is adopted, since the dome flatfields have better signal-to-noise ratios to correct for the pixel-to-pixel sensitivity variations, whereas the twilight flatfields have more similar overall illumination with the science observations. 

\begin{figure*}[!ht]
    \centering
        \includegraphics[width=17cm]{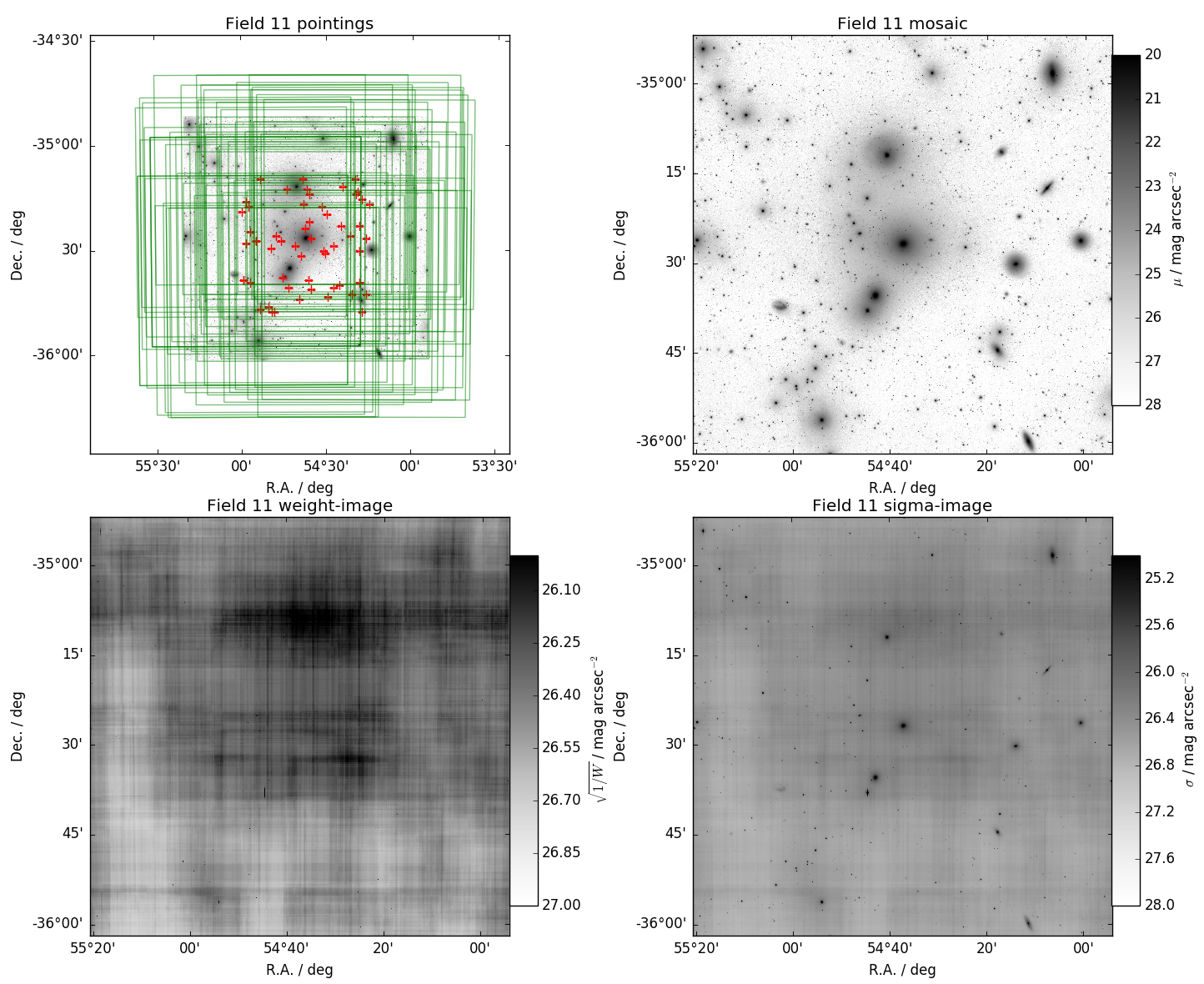}
    \caption{Coverage of the FDS field 11 observations in g'-band are shown with the green squares in the upper left panel, and the centers of the pointings with the red crosses. The median combined mosaic image is shown in the upper right corner, and the corresponding weight- and sigma-images are shown in the lower left and lower right panels, respectively. The color bars in the panels indicate the surface brightness and 1$\sigma$ noise per pixel transformed into surface brightness, respectively.}
    \label{fig:pointings_and_types}
\end{figure*}

\indent During the instrumental reduction, weight maps are also created for each individual frame. Weight maps carry information about the defects or contaminated pixels in the images and also the expected noise associated with each pixel (see lower left panel of Fig. \ref{fig:pointings_and_types}). The hot and cold pixels are detected from the bias and flatfield images, respectively. These pixels are then set to zero in the weight maps. The flatfielded and debiased images are also searched for satellite tracks and cosmic rays, and the values of the pixels in the weight maps corresponding to the contaminated pixels in the science images, are then set to zero. The Hough transformation method \citep{Vandame2001} is applied to the images to pick up the satellite tracks, which are eliminated by masking the lines consisting of more than 1000 pixels that have intensity above the 5-$\sigma$ level relative to the background and are located on the same line. Cosmic rays are detected using SExtractor, and the corresponding pixels are masked from the weight maps. The pixels in the weight maps $W$ have values
\begin{equation}
W = \frac{1}{\sigma^2} \times M_{bad},
\end{equation}
where $\sigma$ is the standard deviation of the background noise and $M_{bad}$ is the combined bad pixel map where the bad pixels have been set to zero and other pixels to one.

\indent The observations contain an additional smooth light component resulting from scattered light. A careful removal of this component is essential for studying the outskirts of the galaxies and the low surface brightness objects. A background model is created first by scaling a set of 12 consecutive exposures of the targets, and then median averaging the stack. The scaling factor $s$ between images A and B is defined by measuring median values within small boxes in image A ($m_A$), and in the same locations in image B ($m_B$), and then taking the median of their ratios:
\begin{equation}
s = \mathrm{median}\left( \frac{m_A}{m_B} \right).
\end{equation} 
For each image among those to be stacked, such a scaling factor is defined with respect to A, and the images are multiplied with these factors before stacking. If there is a large scatter between the ratios of $s$, the chip medians of the exposures are scaled with each other. The scaled images are then median stacked to the background model, and the model is subtracted from image A.  This strategy allows us also to remove the fringe patterns appearing in the OmegaCAM i'-band images, and removes also all the possible residual patterns from the flatfielding.

\indent Systematic photometric residual patterns still remain after flatfielding, which are corrected by applying an illumination correction to the data. We used the correction models made for the Kilo Degree Survey (KiDS, see \citealp{Verdoes-Kleijn2013} for details). The models were made by mapping the photometric residuals across the OmegaCAM's CCD array using a set of dithered observations of Landolt's Selected Area (SA) standard star fields \citep{Landolt1992}, and fitting a linear model to the residuals. The images were multiplied with this illumination correction. The illumination correction is applied after the background removal to avoid producing artificial patterns into the background of images.

\subsection{Astrometric calibration}

The reduced images are calibrated to world coordinates using SCAMP \citep{Bertin2006}. We make the coordinate transformation by applying first the shifts and rotations according to the image headers. The fine tuning of the astrometric calibration is obtained by first associating the source lists extracted from the science images with the 2 Micron All-Sky Survey Point Source Catalog (2MASS PSC, \citealp{Cutri2003}) and fitting the residuals by a second order polynomial plane. This polynomial correction is then applied to the data coordinates, and the pixel size is sampled to 0.2 arcsec pixel$^{-1}$. After applying the astrometric calibration, the remaining differences between the 2MASS PSC objects and the corresponding objects in our data have root mean square of 0.1 arcsec.

\subsection{Flux calibration}

The absolute zeropoint calibration is done by observing standard star fields each night and comparing their OmegaCAM magnitudes with the Sloan Digital Sky Survey Data Release 11 (SDSS DR11, Alam et al. 2015) catalog values. The OmegaCAM point source magnitudes are first corrected for the atmospheric extinction by subtracting a term kX, where X is airmass and k is the atmospheric extinction coefficient with the values of 0.515, 0.182, 0.102 and 0.046 for u', g’, r’ and i’, respectively. The zero-point for a given CCD is the difference between the corrected magnitude of the object measured from a standard star field exposure and the catalog value. The zero-points are defined only once per night, so that for each science observation only the varying airmass was corrected. All magnitudes in the catalog are given in SDSS filters calibrated to AB-system.

\subsection{Making the mosaic images}

The calibrated exposures are median stacked into mosaic images using SWarp \citep{Bertin2010}, and the contaminated pixels are removed using the weight maps. SWarp produces also a mosaic weight map for each mosaic, where the pixel values are inverse of the variance associated to each pixel. We stack the images according to the FDS fields with an extra overlap of 5 arcmin on each side, so that we do not need to cut any large galaxies later in the analysis. As a final result we produce 1.17 deg $\times$ 1.17 deg mosaics and the corresponding weight images. Examples of a g'-band mosaic and the associated weight-images are shown in Fig. \ref{fig:pointings_and_types}. 

\subsection{Sigma-images}

\indent The weight images we produced include the information of the bad pixels and the inverse variance, but do not include the Poisson noise associated to the astronomical objects. For the right weighting of the pixels in the structure analysis of the galaxies (see Section 7.2), we need also sigma-images that include the Poisson noise. We produced the sigma images from the weight images using the equation
\begin{equation}
\sigma = \sqrt{\sigma_{sky,i}^2+\frac{f_i}{GAIN}},
\end{equation}
where $\sigma_{sky,i}=\frac{1}{\sqrt{W}}$, $W$ being the pixel value in the weight image, $f_i$ is the flux in the corresponding pixel in the science image, and $GAIN$ is the ratio of the calibrated flux units to observed electrons as calculated by SWarp during production of the mosaic-images. The lower right panel of Fig. \ref{fig:pointings_and_types} shows an example of a sigma-image of Field 11.

\section{Quality of the mosaics}

To understand the limits of our data and the uncertainties introduced by the calibrations, we made tests for the noise in the images, and the photometric and astrometric accuracy.   

\subsection{Depth}

The image depth (or signal-to-noise ratio, $S/N$) can be calculated theoretically when the telescope size, efficiency of the detector and instrument, brightness of the sky, read-out noise of the instrument, number of exposures, along with the total exposure time, are known. However, in practice there will be also other sources of noise, from the scattered light, reflections between different parts of the images, imperfect background subtraction, and changing the observation conditions. To quantify these effects, we used the final mosaics to measure the actual obtained depth in the images.

\indent To measure the background noise in the images, we defined 500 boxes with 200 $\times$ 200 pixels in size, randomly distributed in the images, and calculate the three times $\sigma$-clipped standard deviations of the pixel values within the boxes. As the final $\sigma$-value of each field we take the median of the calculated standard deviations. The measured $\sigma$s for all fields in the different bands are listed in Table \ref{table:quality}. We find that the obtained depth in the images for 1$\sigma$ signal-to-noise per pixel corresponds to the surface brightness of 26.6, 26.7, 26.1, and 25.5 mag arcsec$^{-2}$ in u', g', r', and i'-bands, respectively. When averaged over 1 arcsec$^2$ area, these values correspond to surface brightness of 28.3, 28.4, 27.8, 27.2 mag arcsec$^{-2}$ in u', g', r', and i', respectively.

\subsection{Photometric accuracy}

As the Fornax cluster is poorly covered with standard star catalogs, a straightforward comparison of the obtained magnitudes with the standard stars to define the photometric accuracy is not possible. However, we can do an internal photometric consistency check by using the fact that the Milky-Way stars form locii in the color-color space that have constant locations and small intrinsic scatters. \citet{Ivezic2004} have performed analysis for the zeropoint  accuracy of the SDSS, using a test which can be used as a comparison.

\begin{figure}[!ht]
    \centering
        \resizebox{\hsize}{!}{\includegraphics{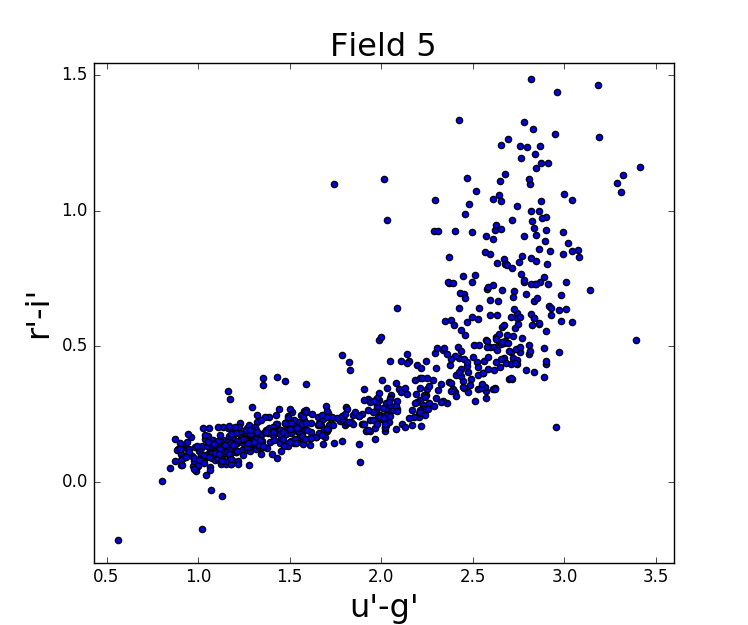}}
        \caption{Colors of the stars in Field 5 are shown in the u'-g' vs. r'-i color-space. The positions of the stars form an inverse L-shaped figure whose two linear parts ({\it locii}) are known to have constant locations and small intrinsic scatter. The locations and scatter of the two apparent locii projected along the principal colors are used for the assessment of the FDS data quality. }
    \label{fig:example_locus}
\end{figure}

\begin{figure}[!ht]
    \centering
        \resizebox{\hsize}{!}{\includegraphics{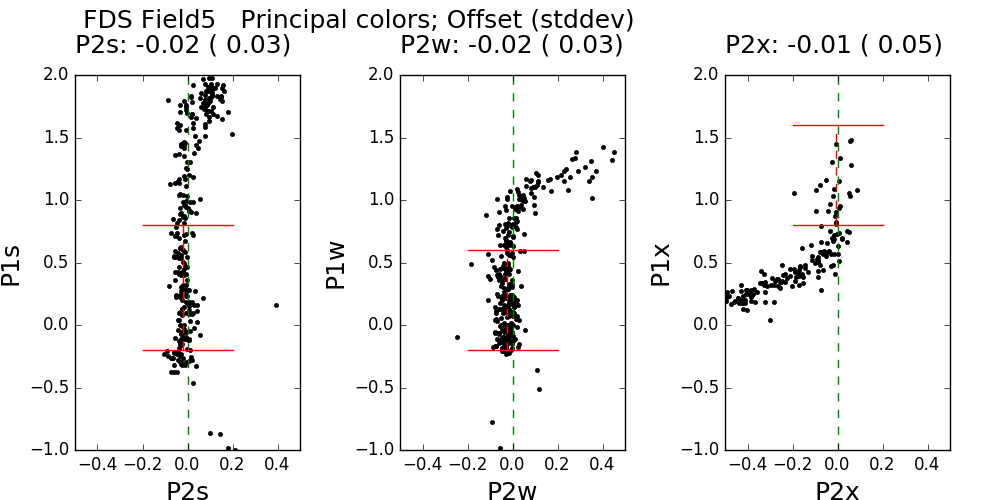}}
        \caption{Stellar locii of the Milky Way stars shown along the three different principal color axes for the FDS Field 5. The dashed vertical lines show the zero offsets, and the horizontal red solid lines show the limits where the offsets and the standard deviations of the locii are measured.}
    \label{fig:single_locus}
\end{figure}

\indent In Fig. \ref{fig:example_locus} we show non-saturated stars of the field 5 in u'-g' versus r'-i' color space. The stars appear in an inverse L-shaped distribution, where two loci are clearly apparent. The scatter in the vertical branch is relatively large in this projection, but reduces considerably when projected along the principal components defined from the full u', g', r', i'- distribution (see Fig. \ref{fig:single_locus}). The principal colors $P1$ and $P2$ as defined by \citet{Ivezic2004} are:

\begin{equation}
\begin{array}{l}
P2s = -0.249\times u' + 0.794 \times g' - 0.555 \times r' + 0.234, \\
P2w = -0.227\times g' + 0.792 \times r' - 0.567 \times i' + 0.050, \\
P2x = 0.707\times g'- 0.707 \times r' -0.988, \\
P1s = 0.910\times u'- 0.495 \times g' - 0.415 \times r' - 1.28, \\
P1w = 0.928\times g' - 0.556 \times r' - 0.372 \times i' - 0.425, \\
P1x = 1.0\times r' - 1.0 \times i',
\end{array}
\end{equation}
\noindent where $u'$, $g'$, $r'$ and $i'$ are apparent magnitudes in the different bands. In our test, we projected stars in our data along these principal colors with the assumption that the stars intrinsically follow these equations with very small intrinsic scatter. To quantify the variations in the offsets of the loci and the scatter around them, we defined the scatter and offsets of the P2 colors with respect to zero projected along the P1 colors.

\begin{figure}[!ht]
    \centering
        \resizebox{\hsize}{!}{\includegraphics{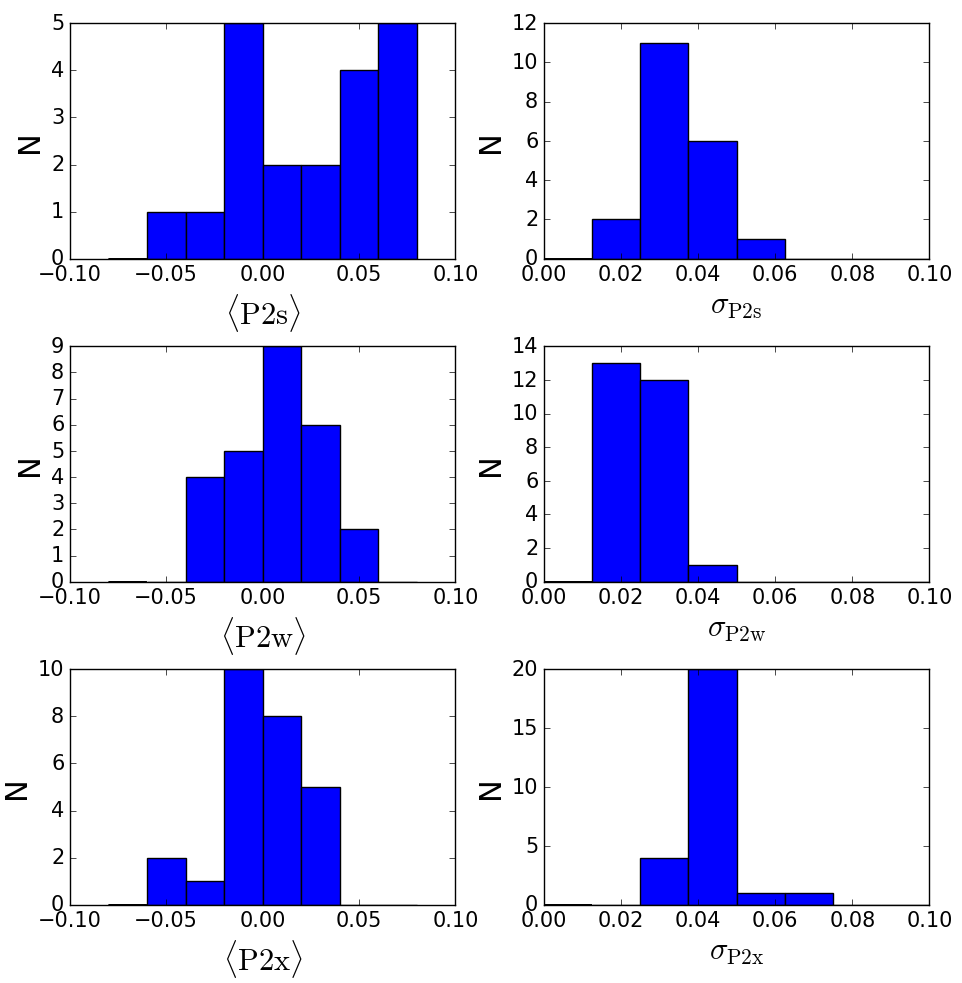}}
        \caption{Left panels: Distributions of the clipped means of the principal colors in the different FDS fields. Right panels: Distributions of the standard deviations of the scatter of the stellar colors around the principal color axes in the different fields.}
        \label{fig:locus_distributions}
\end{figure}

\indent For the stellar locus test we wanted to use bright non-saturated stars. We used SExtractor for the identification of stars, and selected stars that have r'-band apparent magnitudes between 16 mag < m$_{r'}$ < 19 mag (we used $MAG$\_$AUTO$\footnote{$MAG$\_$AUTO$ is the magnitude within 2.5 times the Kron-radius \citep{Kron1980} which is the luminosity weighted first order radial moment from the center of the star.} parameter in SExtractor), and have SExtractor parameter $CLASS$\_$STAR$ > 0.9 ($CLASS$\_$STAR$ tells the probability of an object being a star). The stellar locus test is done for the stars that have $P1$ color components between -0.2 < $P1s$ < 0.8, -0.2 < $P1w$ < 0.6, and 0.8 < $P1x$ < 1.6. In Fig. \ref{fig:single_locus} we show an example of a principal color diagram. The distributions of the measured offsets and scatters in each FDS field are shown in Fig. \ref{fig:locus_distributions}.

\indent The standard deviations of the locus offsets in our data are 0.041, 0.020, 0.024 in s, w, and x, respectively. Within scatter, the deviations are consistent with zero offsets. The median scatters of the stars around the locus are 0.040, 0.025, 0.041 in s, w, and x, respectively.  The corresponding values in SDSS are 0.011, 0.006, and 0.021 for the medians, and 0.031, 0.025, and 0.042 for the standard deviations. This test shows that the errors associated to our zero point definitions are roughly three time as large as for the SDSS images, corresponding to 0.03 mag in g', r', and i' -bands and 0.04 in u'-band.

\subsection{Seeing $FWHM$}

As the images are taken during different epochs with different observing conditions, the point spread function ({\it PSF}) in the images varies. Additionally, when observations with different seeing conditions are stacked into the final mosaics, the radial profile of the {\it PSF} in the stacked images may be different from the original images. Below we describe how the full width at half maximum ($FWHM$) varies between the mosaics, and in Sections 5.1.1 and 5.1.2 we show how we model the {\it PSF} with analytic functions.

\indent The $FWHM$ is straightforward to measure using SExtractor, so we ran it on all the fields to get object lists. From the object list we selected stars (CLASS\_STAR\footnote{CLASS\_STAR parameter requires an input FWHM estimation. We use the median FWHM of the objects in the image that are larger than FWHM>0.25 arcsec. This lower limit was adopted to ensure that the median $FWHM$ is not biased by the false detections consisting of background noise fluctuations.} parameter > 0.95) that have the highest S/N, but are not yet saturated. In our images this corresponds to stars with r'-band aperture magnitudes between 15.5 mag < m$_{r'}$ < 18 mag. The measured median $FWHM$ and their standard deviations within the fields are listed in Table \ref{table:quality}.

\section{Preparation of the detection images}

In the following, we describe the steps for creating the images used for the identification of the galaxies. As a starting point the calibrated stacked mosaics are used (see Section 3.4). We first modelled and subtracted the bright stars (m$_{r'}$  < 15 mag) in the images in the different bands, and then stacked the different bands to make the final detection images.

\subsection{Point-spread function models}

For accurate modeling of the galaxies, it is necessary to take into account the effect of the {\it PSF}. The core of the {\it PSF} ($\lesssim$ 10 arcsec from the center) is determined by the atmospheric turbulence and scattering which vary during the observations. The outer part ($\gtrsim$ 8 arcsec) consists of light scattered from the optical surfaces of the camera that remain constant, apart from the amount of dust in the optics that can slightly alter the outer profile (see \citealp{Sandin2014}).
 
\indent We derived a {\it PSF} model for each of the fields separately. The model is derived in two parts: the inner {\it PSF} is modeled using the brightest non-saturated stars, and the outer {\it PSF} using the outer parts of the saturated stars. Since the number of saturated stars is limited, we use the same outer {\it PSF}-model for all fields, so that only the inner {\it PSF} is modeled in all fields. As shown in Table \ref{table:quality} the {\it PSF} varies also within one field on the order of few tenths of an arcseconds, which means that for a high accuracy modeling of the {\it PSF} one needs to do subfield modeling. This high accuracy is not needed for our S\'ersic profile modeling of extended dwarf galaxies, but is important for compact objects such as ultra compact dwarfs (UCDs) or globular clusters (GCs).

\subsubsection{Inner {\it PSF}}

We followed \citet{Venhola2017} in the creation of the model for the inner 8 arcsec: first we selected stars with r'-band magnitudes between 15.5 mag < m$_r'{}$ < 18 mag, and cut 80 $\times$ 80 pixel areas around the stars. We then more-accurately determined the peaks of the stars by fitting the innermost R < 1 arcsec areas around the centers with a 2D-parabola. We then resampled the images by dividing each pixel into 5$\times$5 subpixels, and recentered the images using the accurate peak coordinates obtained via the parabola fitting. These cuts were then normalized with the flux within the innermost R < 1 arcsec from the center. These normalized stamp images were then median averaged and resampled to the original pixel size to obtain the {\it PSF}-model. 

\indent Theoretically, a Moffat-profile should be sufficient to fit this inner part of the {\it PSF} (\citealp{Moffat1969}, \citealp{Trujillo2001}), but as the mosaics typically consist of images with different seeings, the {\it PSF} of the mosaics is  not well fit by a single Moffat-profile (see \citealp{Venhola2017}). We added a Gaussian to improve the fit in the peak, while leaving the Moffat-profile to dominate for radii R > 2 arcsec. In the combined fitted function
\begin{equation}
I_{inner} = I_{0,Gaus} \exp\left[-\left(\frac{R}{\sqrt{2}\sigma}\right)^2\right] + I_{0,Mof}\left( 1 + \left(\frac{R}{\alpha} \right)^{2} \right) ^{-\beta},
\end{equation}
the first part corresponds to the Gaussian profile, and the second one to the Moffat-profile. $I_{0,Gaus}$ and $ I_{0,Mof}$ correspond to the central intensities of the Gaussian and the Moffat profiles, respectively, $R$ corresponds to radius, $\sigma$ is the standard deviation of the Gaussian, and $\alpha$ and $\beta$ define the extent and the slope of the Moffat profiles, respectively. 

\begin{figure}
        \resizebox{\hsize}{!}{\includegraphics{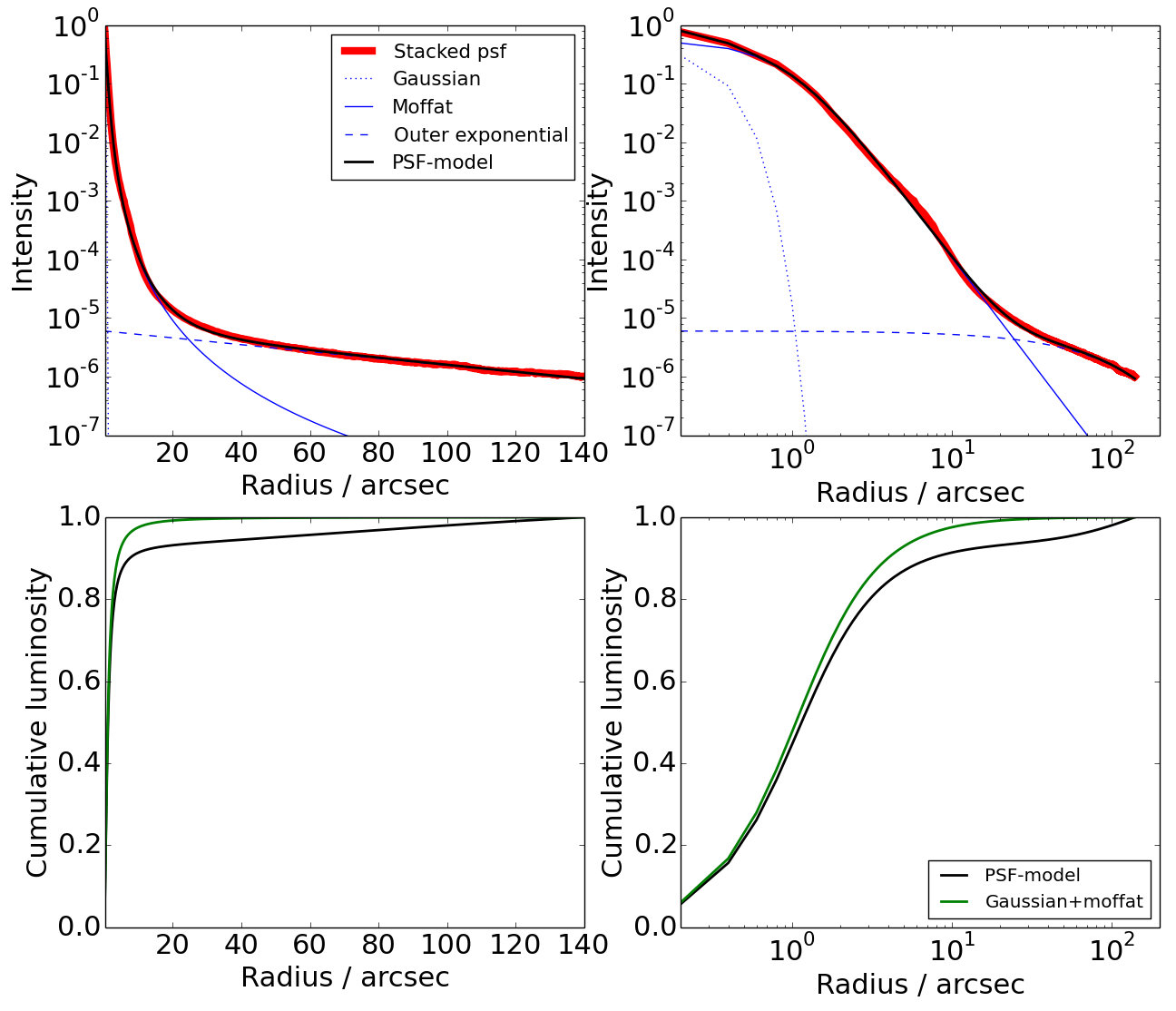}}
  \caption{Upper left and right panels: Stacked intensity profiles (red lines) against linear and logarithmic radius scale, respectively. Shown also are the analytic {\it PSF} model (black lines), the model created from the inner Gaussian (blue dotted line), Moffat model (blue solid line) and the fitted outer exponential function (blue dashed line). The left and right lower panels show the cumulative luminosity fraction within a given radius in linear and logarithmic radial scales, respectively. The green line in the lower panels gives the cumulative flux for only the core part of the {\it PSF} (Gaussian+Moffat).}
  \label{fig:deep_psf}
\end{figure}

\begin{figure*}
        \includegraphics[width=17cm]{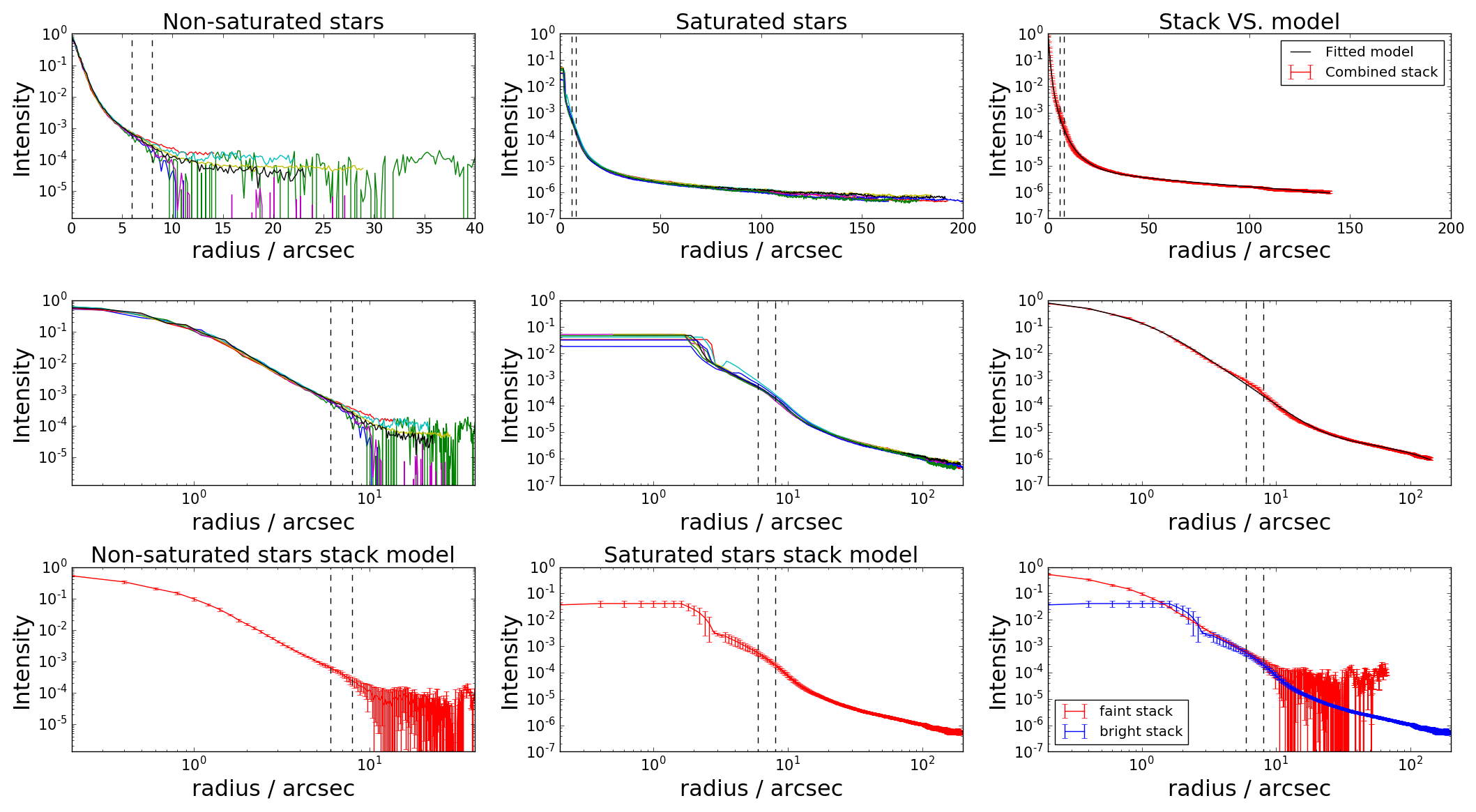}
        \caption{Left panels: Scaled luminosity profiles of the non-saturated stars of field 11 as a function of radius, shown in linear (top left) and logarithmic (middle left) scale. The different colors correspond to different stars. In the lowest panel the stack constructed from the stars is shown. The errorbars indicate the scatter between individual stars. Middle panels: Similar graphs for saturated stars of different fields. Top and middle right panels: Comparisons of the full profiles derived from the observed stars (i.e., combined stack of faint and bright stars), and the corresponding fitted model. Bottom right panel: Bright and faint stacks near the transition zone. The transition zone is shown in all panels using the vertical dashed lines.  }
        \label{fig:psf_stack}
\end{figure*}

\indent The fitting was done so that first the innermost ten arcsec region of the profile was fitted with the Moffat function, and after that the profile within $<$1 arcsec from the center was fitted with the Gaussian function. These initial fits were then used as an input for the second fit where both components were fit simultaneously. The fit results for Field 11 are shown in Fig. \ref{fig:deep_psf}, and the profiles of the individual non-saturated stars are shown in the left panels of the Fig. \ref{fig:psf_stack}. Since the {\it PSF} profiles vary field by field, we list the fit parameters of all the {\it PSF}s in  Table \ref{table:psf_fits}.

\subsubsection{Outer {\it PSF}}

The previously described models of the inner {\it PSF} have low $S/N$ in their outer parts, so that they cannot be used to trace the {\it PSF} down to I < 10$^{-4}$ of the central intensity. To follow the {\it PSF} to fainter levels we have to use the brightest saturated stars.

\indent To model the outer parts of the {\it PSF}, we selected 15 saturated bright stars (m$_{r'}$ < 10 mag) from different fields with no bright galaxies or stars nearby. To scale the flux of the stars we measured the central surface brightnesses of several hundreds of non-saturated stars from the FDS images, and compared the values with the magnitudes of American Association of Variable Star Observers' Photometric All Sky Survey catalog (APASS, \citealp{Henden2012}). These values have a linear relation, which we defined and used to scale the saturated stars (which are not saturated in the APASS-data). Azimuthally averaged radial profiles for the stars spanning up to 3 arcmin distance were made (see Fig. \ref{fig:psf_stack}), which profiles were then combined making an average of them. The outer parts (from R = 40 to 160 arcsec) were fit with an exponential function,
\begin{equation}
I_{exp} = I_{0,exp} \exp\left(-\frac{R}{h}\right),
\end{equation}
where $I_{0,exp}$ is the central intensity and $h$ is the scale length. The exponential profile was selected empirically due to its good fit to the data. From the fits we obtain $h_{g'}$ = 87.38 arcsec and $h_{r'}$ = 74.26 arcsec for the g'- and r'-band scale lengths, respectively, and $I_{0,exp,g'}$ = 1.556$\times10^{-6}$ and $I_{0,exp,r'}$ = 6.022$\times10^{-6}$ for the central intensities in the scaled units ($I_{0}$=1). We use these same parameters in all the fields.

\indent To ensure that the scaling between the faint and bright stars works, we plot the profiles of a set of faint stars and the averaged profile in the left panels of Fig. \ref{fig:psf_stack}, the profiles of the bright saturated stars in the middle panels, and finally show the combined stack model and the fitted model in the right panels. For obtaining the combined stack model, we used an average of the bright and faint profile within 6 arcsec < R < 8 arcsec, the faint star average profile within R < 6 arcsec, and in the outer parts the average profile of the saturated stars. The lower right panel shows the profiles of both bright and faint stars around the area where they are combined showing that their profiles agree well in this area.

\indent  \citet{Sandin2014} analyzed the outer parts of {\it PSF}s of several telescopes up to several hundred arcsecs. They found that at the very large radii (R > 300 arcsec) the {\it PSF} intensity attenuates following $I$ $\propto$ R$^{-2}$ law. In our data, we can follow the {\it PSF} only up to 200 arcsec. In this region our {\it PSF}-profile behaves in a similar manner as most of the {\it PSF}s in Sandin (2014), showing a clear seeing dependent core up to a few tens of arsecs, and an exponential part beyond that. For our purposes it is not necessary to follow the {\it PSF} further than a few arcminutes.

\subsection{Subtraction and masking the fore-ground stars}

Due to the extended {\it PSF} of OmegaCAM, the bright stars contaminate large areas in the images. SExtractor is not designed to find objects in crowded fields, and therefore these outer halos affect the detection efficiency of SExtractor. In particular, the bright stars have spikes and reflection halos, which appear as false detections in the source lists. To prevent the above mentioned bias in the source lists, we subtract the bright stars (m$_{r'}$ < 12 mag) and mask the stars with m$_{r'}$ < 16 mag  in the images before making the combined detection image. To mask the stars in a systematic manner we use the analytic {\it PSF} models described above.

\begin{figure*}
        \includegraphics[width=17cm]{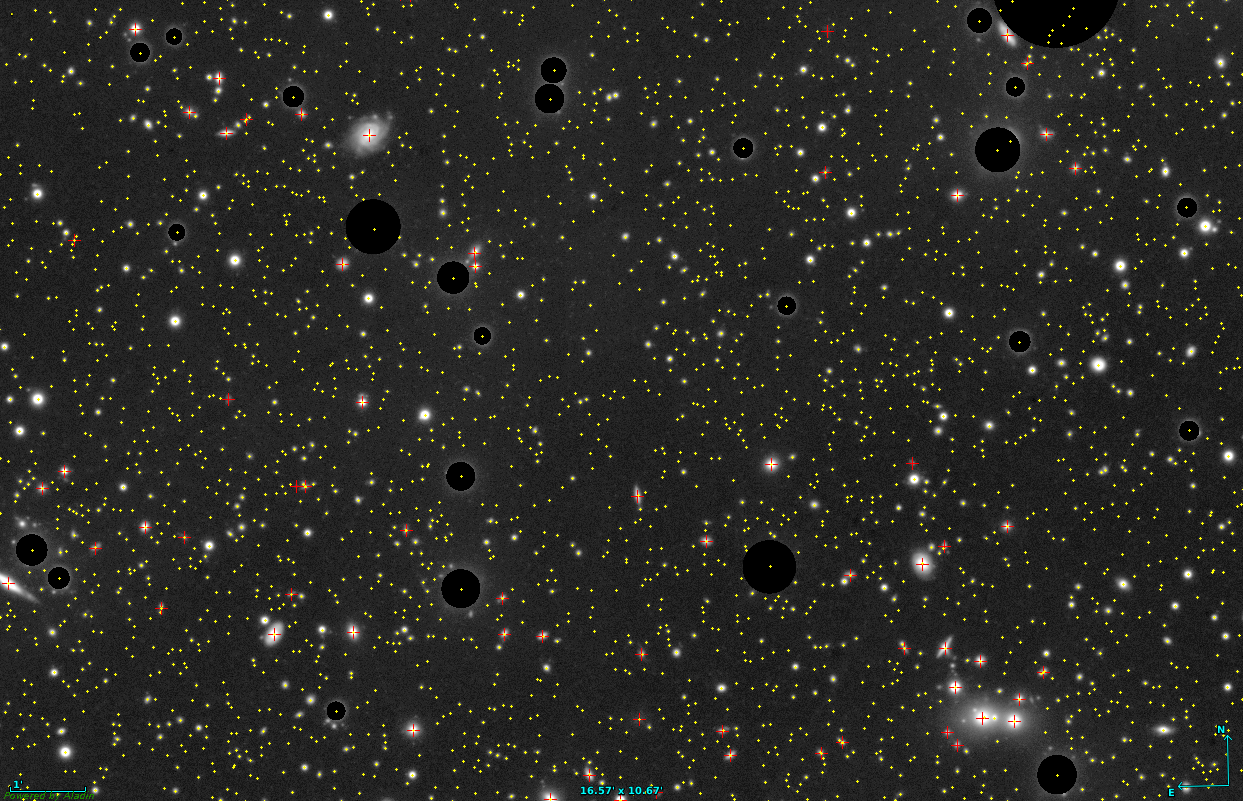}
        \caption{Magnification of Field 5 with the detected objects and masks (black circles) overlaid on the image. The yellow points and red symbols correspond to the initial detections of our detection algorithm, and the objects that pass the A\_IMAGE > 2 arcsec selection limit, respectively. Aladin \citep{Bonnarel2000} was used for generating the image. The image is best viewed in color on-screen.}
                \label{fig:example_detection}
\end{figure*}

\indent In order to decide the masking radius for each star, we need to know their magnitudes. As the bright stars are saturated, we used APASS magnitudes for them. Since APASS includes also galaxies, some of the bright objects in that catalog may be FCC galaxies. To prevent unintentionally masking bright galaxies, we check for FCC galaxies within 5 arcsec around the bright APASS objects before masking or subtracting them. Since we masked only stars that have apparent magnitudes m$_{r'}$ < 16 mag, we do not have to be worried about masking galaxies that are not in the FCC.  We took the coordinates and magnitudes of the stars and subtract the analytic {\it PSF} model from all stars brighter than m$_{r'}$ $<$ 12 mag, as far as the surface brightness level of 29 mag arcsec$^{-2}$. The stars with m$_{r'}$ $<$ 16 mag are masked up to the radius where the analytical model corresponds to the surface brightness of 25.5 mag arcsec$^{-2}$. We find that the spikes and reflections are typically well masked using the selected masking limits (see Fig. \ref{fig:example_detection} for example masks).

\indent As a result, we obtain images where the bright stars are subtracted and the spikes and reflection haloes are masked. The total fraction of the area that could not be used due to these saturated stars is only $\approx$ 3 \%, which will cause incompleteness of the same order into the dwarf galaxy catalog. Moreover, these excluded regions should not cause any systematic bias to our analysis since the stars are randomly distributed in the survey area. After this preprocessing of the images, they may still include some imaging artifacts, which were manually eliminated afterwards. 

\subsection{Creating the final detection images}

To obtain the best image quality for the source detection image, we combine the  star-subtracted g', r' and i'-band images of each field as a g'r'i'-composite image. We calculate"d a weighted average of the frames using the weights 0.4, 0.5 and 0.1 for g', r' and i'-bands, respectively. The weights were selected taking into account the depth of the different bands and the color g'-r' $\approx$ 0.6 of the early-type dwarf (dE) galaxies \citep{Janz2009}.  

\section{Preliminary source lists}

\subsection{Detection algorithm}

In this paper our aim is to detect resolved dwarf galaxies. We used SExtractor for the detection of the objects. An automatic detection method is used instead of a manual one, given the large amount of imaging data. However, SExtractor is not optimal for the detection of low surface brightness galaxies ($\bar{\mu}_{e,r'}$ $\geq$ 24 mag arcsec$^{-2}$), so we test the completeness of our source lists in Section 6.2. An extension dedicated to low surface brightness (LSB) galaxies in the Fornax cluster \citep{Venhola2017}, to be generated with a different detection algorithm, will be added to this catalog in a forthcoming paper (Venhola et al., in prep.). We did not specifically exclude LSB galaxies, but the detection limits in this paper are not very favorable for such galaxies.

\indent SExtractor detects objects by searching for groups of connected pixels that are brighter than a certain detection threshold. In principle the detection can be done with or without subtracting a background model from the detection image. The background model is created by defining a grid of image pixels, and then estimating the background level in each grid box. This is done by iteratively $\sigma$-clipping the pixel distribution within the grid box, and then taking a mean. The grid of means is then interpolated, which makes the background model. In this study the background model is subtracted before detecting the objects.

\indent Some of the bright galaxies are blended with the smaller ones either physically or due to projection. In such cases, we can treat the large galaxies as background and include them into the background model. We can select a background grid size so that it is larger than the sizes of the small galaxies, but smaller than the primary galaxy. While detecting more extended galaxies, both bright and faint, the background grid size should be set to be large enough to prevent introducing false detections, resulting from background maps. For the above reasons, the galaxies have to be detected in several runs aiming for detecting galaxies with different sizes.

\indent We ran SExtractor in three rounds: first for detecting small galaxies, then large galaxies, and finally we tuned the parameters to detect LSB galaxies. For the detection, we used the combined g'r'i'-images (described in Section 5.3) where the bright stars are subtracted and masked. We convolved all the images with a Gaussian kernel with $FWHM$ of ten pixels before the detection, in order to increase the $S/N$ in the images. The SExtractor parameters of the different detection runs are shown in Table \ref{tab:lists}.

\begin{table}
\caption{Parameters used in SExtractor in the different lists. The columns in the table correspond to the name of the list (List), detection threshold (Thresh.) in units of background noise standard deviations (1$\sigma$ corresponds typically to $\mu_{r'}\approx$ 26 mag arcsec$^{-2}$), number of connected pixels above threshold to count as detection (Min. area), and the background model grid size (Back size). In the "LSB" list, an additional 9$\times$9-pixel median filtering was applied to the images before the detection, so the 5$\sigma$ threshold corresponds to 0.55$\sigma$ in the non-filtered image. }
\label{tab:lists}
\centering
\begin{tabular}{c|c|c|c}
\hline\hline
List & Thresh. ($\sigma_{sky}$) & Min. area (pix) & Back size (pix) \\
\hline 
Small & 1  & 10     & 100$\times$100 \\
Large & 50 & 10,000 & 21,000$\times$21,000 \\
LSB   & 5  & 25     & 500$\times$500 \\
\hline
\end{tabular}
\end{table}

\begin{figure}
        \resizebox{\hsize}{!}{\includegraphics{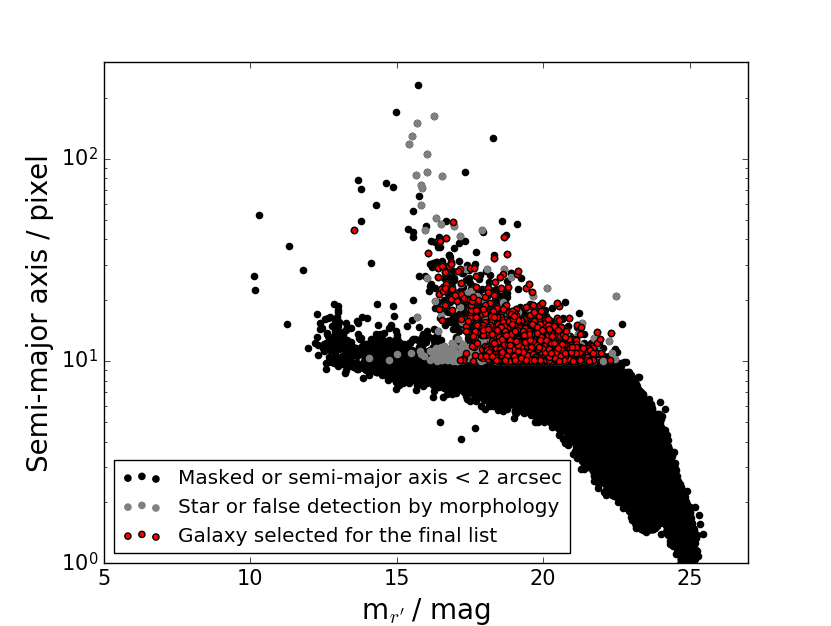}}
        \caption{ Size-magnitude relation of the detected objects in Field 5. Black dots show the objects that have been excluded either for being masked or too small, the gray dots show objects that are morphologically stars or false detections, and the red dots are the galaxies selected for the final catalog. }
        \label{fig:detection}   
\end{figure}

\indent SExtractor outputs object lists with several parameters associated with each object. Most of these detections are Milky-Way stars, false detections or unresolved background galaxies that we want to remove from the lists. First the objects located under the masks generated for the bright stars (described in Section 5.2) were removed from all the lists. Also, the faint stars and unresolved galaxies were removed by excluding the objects with the semi major axis smaller than ten pixels (2 arcsec, see Fig. \ref{fig:example_detection} and Fig. \ref{fig:detection}) measured by SExtractor (A\_IMAGE). This selection based on size excludes also the unresolved Fornax cluster galaxies from our sample (see Section 9.1). However, the 2 arcsec ($\sim$200 pc at the distance of the Fornax cluster) size limit is yet small enough, so that it will not exclude Fornax cluster galaxies similar to the Local Group dSphs that have effective radii between 2 arcsec < R$_e$ < 10 arcsec at the distance of the Fornax cluster (see Fig. \ref{fig:mag-size-LG.png}). On average, this size limit excludes 99.5\% of the detections per field.  The remaining objects in the three lists were then combined. We searched objects within 3 arcsecs from each other. If the same object appeared in several lists, its parameters and coordinates were taken from the list which had the highest detection threshold (in order 1. "Large", 2."Small", 3."LSB"). 

\indent As a result,  cleaned object lists for all fields were obtained. For each target we used the coordinates, magnitudes and semi-major axis lengths obtained with SExtractor as initial values for the photometric pipeline (Section 7). We did not want to make further filtering based on parameters of targets before running the photometric pipeline, since the photometric parameters, like effective radii or magnitudes, given by SExtractor, are not very robust. 

\subsection{Completeness of the detection}

To test the completeness of our detection algorithm, we iteratively embedded 3500 artificial galaxies in sets of 150 galaxies into the Field 10 detection image. As the depth variations in the different fields are only in the order of 0.2 mag (Section 4.1) we assume that the completeness is very similar over the whole survey area. 2D-S\'ersic functions were used as artificial galaxies. The mock galaxies were convolved with the {\it PSF} of OmegaCAM, and the Poisson noise was added into each pixel. The mock galaxies were embedded to the reduced mosaic images with random locations and position angles. We selected a wide range of input parameters to cover the expected parameter space of the dwarf galaxies in the Fornax cluster (m$_{r'}$=16--25 mag, $n$=0.5--3, $b/a$=0.2--1 and R$_e$=1--20 arcsec). We then ran the detection algorithm to test how many of these galaxies we can detect. By detection we required a detection within 3 arcsec from the central coordinates of the embedded galaxy. To also understand the effect of the minimum size-limit, we finally removed the objects with A\_IMAGE < 2 arcsec from the detections.

\begin{figure*}
        \includegraphics[width=17cm]{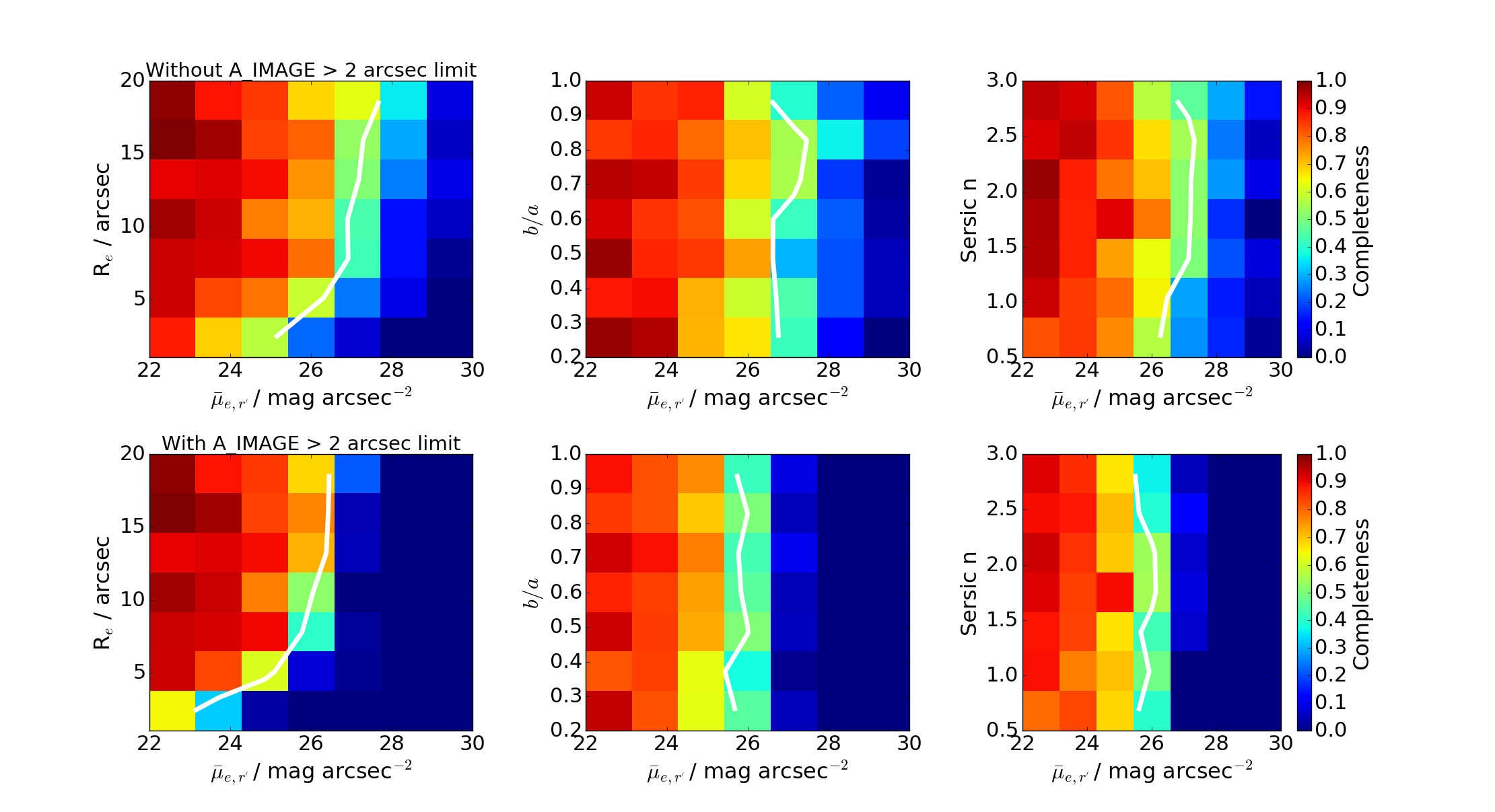}
        \caption{Detection efficiency of our detection algorithm is shown color-coded such that red means more efficient and blue less efficient. The detection efficiency is shown for the effective radius (R$_e$), axis ratio ($b/a$), and Sersic index ($n$), as a function of the galaxy mean effective surface brightness ($\bar{\mu}_{e,r}$). The upper row shows the detection efficiency without applying the minimum size limit of A\_IMAGE of 2 arcsec, and the lower row shows the detection efficiency after applying the limit. The white line shows the 50\% completeness limit.}
   \label{fig:completeness}
\end{figure*}

\indent Figure \ref{fig:completeness} shows the detection efficiency of the galaxies as a function of galaxy magnitude for the different structure parameters with and without using the minimum size limit. We find that the detection efficiency slightly depends on the shape of the galaxy profiles (S\'ersic n) so that more extended and more peaked galaxies are more efficiently detected. Applying the minimum size limit lowers the completeness limit from $\bar{\mu}_{e,r'}$ = 27 mag arcsec$^{-2}$ to $\bar{\mu}_{e,r'}$ = 26 mag arcsec$^{-2}$, and especially it affects the smallest low surface brightness objects. As a result, our detection has the limiting r'-band magnitude with 50\% detection efficiency of m$_{r'}$ = 21 mag and the limiting mean effective surface brightness of $\bar{\mu}_{e,r'}$ = 26 mag arcsec$^{-2}$. In Section 9, we also compare the final detections and completeness with previous galaxy catalogs in the Fornax cluster.

\section{Obtaining the photometric parameters}

\begin{figure*}
\centering
   \includegraphics[width=17cm]{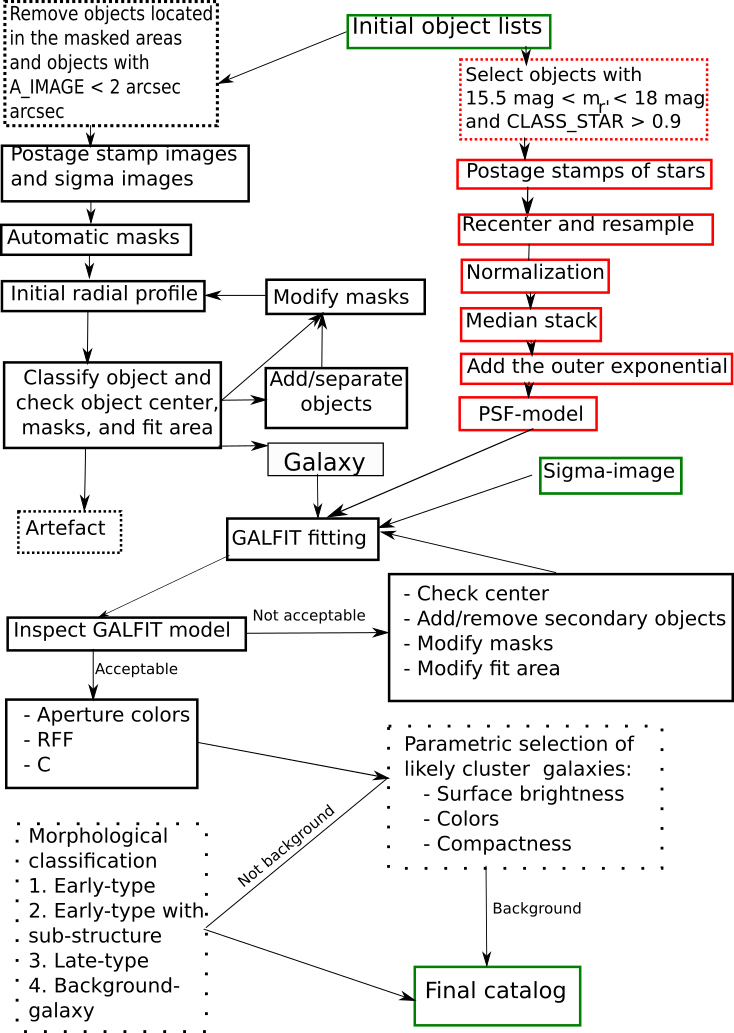}
        \caption{Flow chart of the photometric pipeline.}
   \label{fig:pipe_schematics}
\end{figure*}

Photometric parameters are derived for classification of the galaxies, with the ultimate goal to identify the galaxies that belong to the Fornax cluster. The parameters are obtained for all non-masked galaxies that have (SExtractor) semi-major axis lengths larger than 2 arcsec ($\approx$ 200 pc at the distance of the Fornax cluster). We fit S\'ersic profiles to the 2D flux distributions of the targets using
GALFIT \citep{Peng2002} to obtain the galaxy magnitudes, effective radii and shape properties. We also measured aperture colors, and calculate residual flux fractions ($RFF$, \citealp{Hoyos2011}) and concentration ($C$) for all the objects.  A scheme of the photometry measurements is shown in Fig. \ref{fig:pipe_schematics}, and the steps are described in more detail below. As an input for the photometric pipeline, we use the central coordinates, isophotal magnitudes, and semi-major axis lengths measured with SExtractor.

\subsection{Preparing the images for photometry}

First post-stamp images of the galaxies were made in all bands, limiting the semi-width of images to 10 A\_IMAGE measured by SExtractor. As these semi-major  axis lengths are not always accurate, especially for the low surface brightness objects, some post-stamp images were almost fully covered by the object galaxy. In cases for which too few sky pixels appeared, the image sizes were increased manually. The corresponding sigma-image mosaics were cut in a similar manner.

\indent In the post-stamp images there are also other objects than the primary galaxy like faint stars and other galaxies that need to be masked for not to bias the fitting. We generated initial masks using SExtractor by masking all the sources larger than 100 pixels above the 1$\sigma$-threshold. As the primary galaxy was typically also masked, we removed all the masks within two effective radii from the center of the source. In the inner parts, we wanted to mask only point-like sources, so we identified them using SExtractor (CLASS\_STAR > 0.3), and then used the analytic {\it PSF} model to mask the point sources down to 27 mag arcsec$^{-2}$. These automatically generated masks were then visually inspected and modified (if needed) before fitting.

\subsection{GALFIT models}

\subsubsection{Initial estimation of the parameters}

We estimated the initial input parameters of GALFIT by making an azimuthally averaged radial profile of the galaxy, using circular bins and a bin width of two pixels. We then took the clipped average of each bin and make a cumulative profile up to three semi-major axes lengths (from SExtractor). We then defined the effective radius and  magnitude from the growth curve, which parameter values were used as the input for GALFIT.

\indent The centers of the objects are also defined before running GALFIT. For the objects that have a clear center, we fit the central 10 $\times$ 10 pixel area with a 2d-parabola, and take the peak as the center. For the galaxies that have a flat center, we take the SExtractor coordinates as the center and modify them in the cases where they are obviously wrong. This can happen if the object is split into several parts in the deblending done by SExtractor.

\subsubsection{Partially overlapping objects}

In some cases two galaxies are partially overlapping, so that they cannot be measured robustly separately. This problem can be solved by modeling both galaxies simultaneously with GALFIT. 

\indent Before running GALFIT, we inspected all the post stamp images for close companions. If the two objects were only identified as single object by SExtractor we separated them and ran the whole pipeline for both of them separately. Initial profiles were then generated for both objects, and an additional S\'ersic component was added to the GALFIT model (see next subsection).

\subsubsection{GALFIT modeling}

We used the idl-interface \citep{Salo2015} to run GALFIT. The objects are fitted using either a single S\'ersic function, or a combination of a S\'ersic function and a point source for the nucleus, based on the visual appearance and the radial light profile of the galaxy. In both cases the background is also fitted with a plane of three degrees of freedom (mean intensity, and gradients in x- and y-directions). We left the more complicated multicomponent decompositions for future papers. All the parameters of the S\'ersic component and the background are fitted freely. However, for the nucleus, the center is kept fixed, leaving only the magnitude as a free parameter. In case of nucleated dwarfs, we allowed the S\'ersic component to have a different center than the nucleus, since it is possible to have off-centered nuclei (see \citealp{Bender2005}, but also \citealp{Cote2006}).

\indent We performed the fitting in g' and r' -bands for all galaxies\footnote{u' and i' band were excluded since not all the galaxies have enough signal-to-noise for a robust fit}. The fits are inspected, by looking at the residuals, radial profile with the model overlaid, and the original image with the fitted effective radius (R$_e$) overlaid. For a good fit we required R$_e$ to be within the area that we can see from the galaxy. In the case of a bad fit, (due to imperfect masking or divergence of the model) the masks, center positioning, and the initial radial profile were reiterated.

\subsection{Aperture colors}

We measured colors within the effective radius for all the galaxies using R$_e$, ellipticity and position angle obtained from the r'-band GALFIT model. For the galaxies within the main cluster, we measure u', g', r', and i' aperture magnitudes. For the galaxies in the Fornax A region, we have only g',r', and i', since that area was not observed in the u'-band. We estimated the uncertainty in the aperture magnitudes as

\begin{equation}
\sigma_{\rm aper}^2  = \sigma_{ZP}^2+\left(\frac{2.5}{I_{\rm aper} \ln10}\right)^2(\sigma_{\rm I}+\sigma_{\rm sky})^2,
\end{equation} 

\noindent where $I_{\rm aper}$ is intensity within the aperture, and $\sigma_{\rm I}$, $\sigma_{\rm sky}$ and $\sigma_{\rm ZP}$ are the uncertainties for the surface brightness, the sky, and the photometric zero point, respectively. For the mean intensity we assume Poissonian behavior, so that  $\sigma_{\rm I}=\sqrt{I_{\rm aper} / (GAIN \times n)}\times GAIN$, where $n$ is the number of pixels within the aperture. $I$, $\sigma_{\rm I}$ and $\sigma_{\rm sky}$ are given in flux units, whereas $\sigma_{\rm ZP}$ is in magnitudes.

\subsection{Residual flux fraction ($RFF$)}

The morphological separation of early- and late- type galaxies is done, apart from using the colors, also using the amount of structures in galaxies. Elliptical galaxies are mostly smooth and do not have strongly non-axisymmetric components, S0s have more distinct disk and bulge components, and may have bars, and late-type disk galaxies have star-forming clumps and/or spiral arms. The smoothness parameter is often used to quantify the amount of structures (see \citealp{Conselice2014}). It is calculated by quantifying the residual after subtracting the smoothed image from the original galaxy image. This approach works well when the galaxies are well resolved and are located at similar distances. However, for distant galaxies with small angular sizes the smoothing flattens the radial profiles, so that the residuals increase systematically. Therefore smoothness does not only measure structure, but is somewhat degenerated with steepness of the slope of the radial profiles. 

\indent To overcome the problem related to the smoothness parameter, we decided to use residual flux fraction ($RFF$) that describes how much a galaxy differs from the used model, which in this case is a S\'ersic profile. We measured $RFF$ following \citet{Blakeslee2006}:

\begin{equation}
RFF = \frac{\sum_{i=1}^{n_{r<R_P}}\left( |data_i-model_i|-0.8\sigma_i\right)}{F_{r<R_P}},
\end{equation}

\noindent where $n_{r<R_P}$ is the number of pixels within the Petrosian radius ($R_P$) where the galaxy's surface brightness is 1/5 of the mean surface brightness within that radius. The term $|data_i-model_i|$ corresponds to the absolute value of residual flux at a given point, $F_{r<R_P}$ is the total flux within the $R_P$, and $\sigma_i$ is the pixel value of the sigma-image. The factor 0.8$\sigma$ is the expected mean absolute deviation of the $data_i-model_i$, so that in case of a perfect fit $RFF=0$. The $RFF$ was measured after masking the small background galaxies and point sources that overlap with the galaxy, and in the cases of large overlapping galaxies the large secondary galaxy was modeled and subtracted before calculation of the $RFF$. These steps were done in order to prevent secondary sources biasing the $RFF$ measurements.

\indent However, the $RFF$ parameter is not completely redshift-independent, since seeing blurs more the structures in galaxies at higher redshifts. As shown in Fig. \ref{fig:effect_of_redshift}, late-type galaxies are well separated from early-type systems at low redshift, but it becomes difficult to distinguish the various morphological types as one goes to larger redshifts.

\subsection{Concentration parameter (C)}

\begin{figure}
  \resizebox{\hsize}{!}{\includegraphics{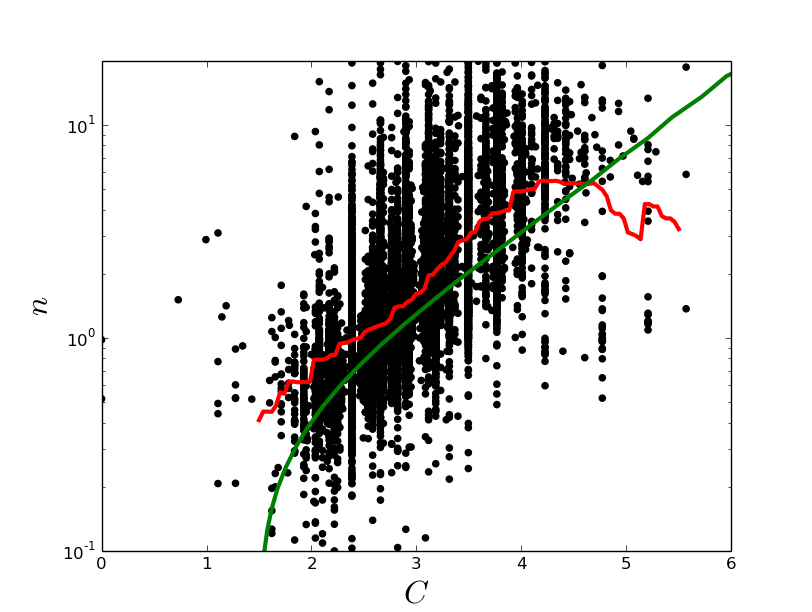}}
        \caption{Values of the S\'ersic index $n$ and the concentration parameter $C$, measured in r'-band for all the objects in our catalog. The red line shows the running mean of the points ($\bar{\log10(n)}$ was used) along the x-axis within intervals of $\Delta C$=0.5, and the green line shows the relation for a pure S\'ersic profile.}
   \label{fig:C_vs_n}
\end{figure}

Galaxies also differ in their concentration; low mass galaxies have low surface brightnesses and approximately exponential radial profiles, whereas high surface brightness galaxies have central mass concentrations. In S\'ersic profiles the parameter $n$ defines the peakedness of the profile, and can be used for morphological classification, in the level that we are interested in this paper. However, we acknowledge that not even for bright elliptical galaxies the S\'ersic profile is an accurate model; for ellipticals NUKER-profiles \citep{Lauer1995} or core-cusp profiles are often used. Therefore, using a non-parametric measure to evaluate the type of profiles is also useful. We used the concentration parameter ($C$) as given in \citet{Conselice2014}
\begin{equation}
C = 5* \log\left( \frac{R_{80\%}}{R_{20\%}} \right),
\end{equation}
where $R_{20\%}$ and $R_{80\%}$ are the radii that enclose 20\% and 80\%, respectively, of the galaxy's total light. The $R_{20\%}$ and $R_{80\%}$ are obtained by first measuring Petrosian magnitude\footnote{Petrosian magnitude is measured using elliptical aperture with size of 1.5$\times\rho_P$, where $\rho_P$ is the radius where the local surface brightness is one fifth of the mean surface brightness within the radius.} for the galaxy, and defining these radii from the growth curve derived from the radial profile. The lower right panel in Fig.\ref{fig:effect_of_redshift} shows how early-type galaxies of a given luminosity have higher concentration than late-type galaxies. In Fig. \ref{fig:C_vs_n} we show how the non-parametric concentration relates with the S\'ersic index $n$ obtained via one-component fit. We also show in Fig. \ref{fig:C_vs_n} how $C$ and $n$ are related for a S\'ersic profile (see also \citep{Janz2014}). Average S\'ersic indices of the galaxies follow a similar trend to the pure S\'ersic index, but with a large scatter and a small offset so that the real galaxies have higher S\'ersic index at a given $C$. This offset is likely explained by the fact that the effects of the {\it PSF} are taken into account in the S\'ersic $n$ (obtained from the GALFIT models) but not in $C$.

\subsection{Uncertainties of the GALFIT models}

Our photometric measurements have uncertainties arising from two different sources: at the low surface brightness end of the galaxy distribution we are limited by the signal-to-noise, and at the bright end the galaxies have typically more structure than our simple S\'ersic models assume.

\begin{figure*}
        \includegraphics[width=17cm]{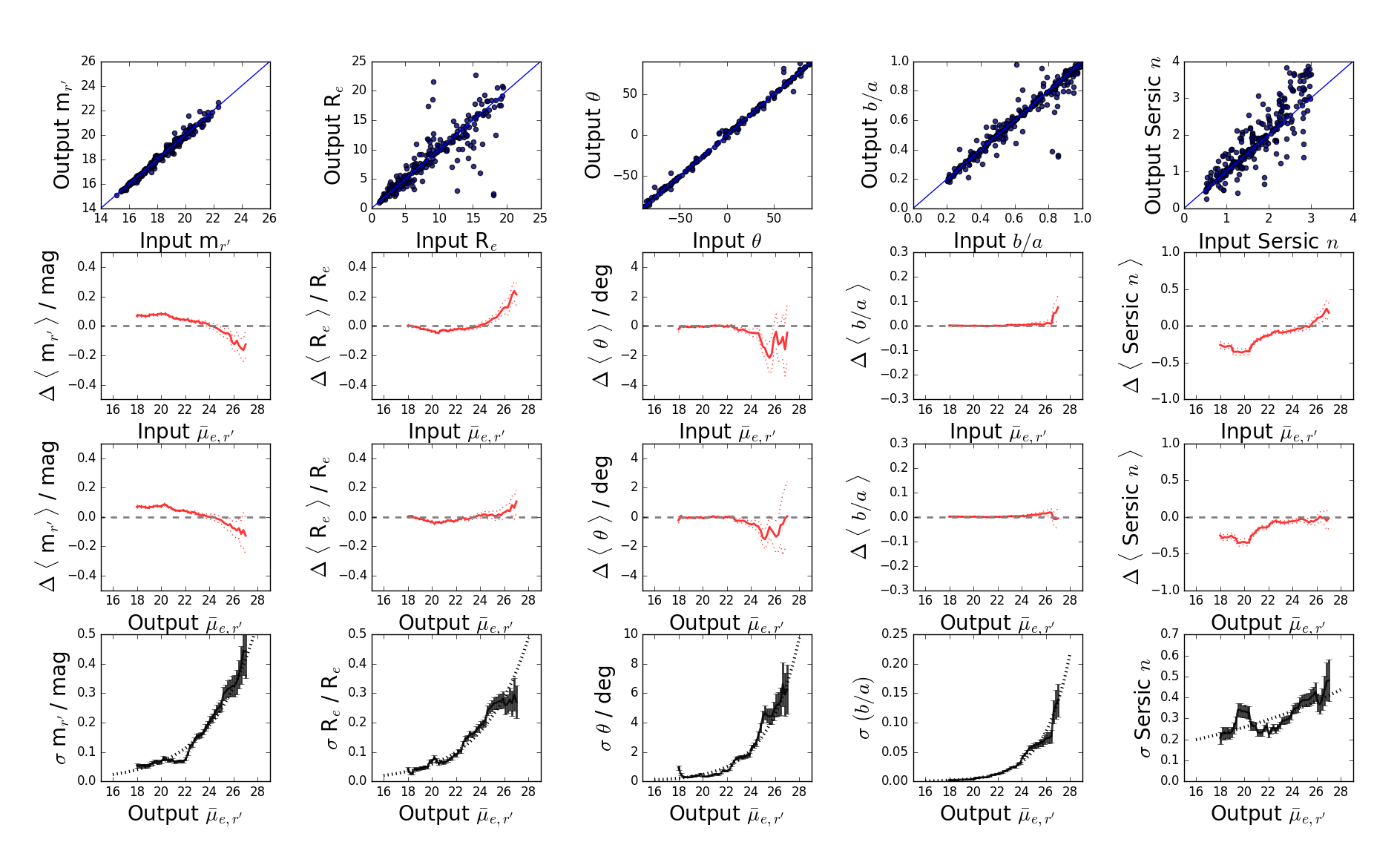}
     \caption{ Top row panels: Comparison of the input structural parameters of the mock galaxies to the values measured by our photometric pipeline. The shown parameters  are  apparent magnitude (m$_{r'}$),  effective radius in arcsec (R$_e$),  position angle ($\theta$),  axis ratio ($b/a$), and S\'ersic index ($n$), and the blue diagonal lines represent the 1:1 ratio. Second and third row panels: Mean differences between the input and output parameters (input - output) as a function of their input and output mean effective r'-band surface brightness, $\bar{\mu}_{e,r'}$, respectively. Bottom row: Standard deviations of the input - output parameter differences as a function of the input mean $\bar{\mu}_{e,r'}$. The dotted lines in the bottom row panels show the fits to the standard deviations as defined in Eq. 10.}
     \label{fig:accuracy}
\end{figure*}

\indent We quantify the fit uncertainty in the low surface brightness end using the mock galaxies embedded in the r'-band images (described in Section 6.2). We made photometric measurements for ~400 detected mock galaxies having a large range of  structural properties. The differences between the input and output values, and the systematic shifts and standard deviations between the input and output values as a function of surface brightness, are shown in Fig. \ref{fig:accuracy}. As expected, the uncertainties  in the parameters increase toward the fainter (lower surface brightness) galaxies. Slight systematic trends also appear in the total magnitudes and S\'ersic indices, but are smaller than the uncertainties of those parameters.

\indent Similarly to \citet{Hoyos2011} and \citet{Venhola2017} we fit the standard deviations of the input-output residuals\footnote{By "standard deviation of residual" we mean $\sigma = \sqrt{\Sigma^N_{j=1}(input_j-ouput_j)^2/(N-1)}$, where $N$ is the number of mock galaxies in a given $\bar{\mu}_{e,r'}$ bin.} We fit the $\sigma$ with the function
\begin{equation}
\log_{10}(\sigma) = \alpha\times\bar{\mu}_{e,r'} + \beta,
\end{equation}
where $\alpha$ and $\beta$ are free parameters, and $\bar{\mu}_{e,r'}$ is the measured mean effective surface brightness. The fit results are listed in Table \ref{tab:fitparams}, and the fits to the standard deviations are shown in Fig. \ref{fig:accuracy}. We use these functions to estimate the measurement uncertainties for the galaxy parameters given by the photometric pipeline. These uncertainties are given with the galaxy parameters in the catalog. We note that these empirically measured uncertainties are significantly larger than the formal uncertainties given by GALFIT that only take in account the statistical uncertainty due to the pixel noise.

\begin{table}
\caption{Fit parameters from the Eq. 10. The first column shows the parameter, and second and third columns show the constant ($\beta$) and the slope  ($\alpha$) in Eq. 10 for the given parameter.  }
\label{tab:fitparams}
\centering
\begin{tabular}{c|c|c}
\hline\hline
Parameter & $\alpha$ & $\beta$ \\
\hline 
m$_r$        & 0.107 & -3.309 \\
R$_e$        & 0.111 & -3.443 \\
$\theta$     & 0.175 & -3.854 \\
$b/a$        & 0.211 & -6.535 \\
S\'ersic $n$ & 0.030 & -1.194 \\
\end{tabular}
\end{table}

\begin{figure}
  \resizebox{\hsize}{!}{\includegraphics{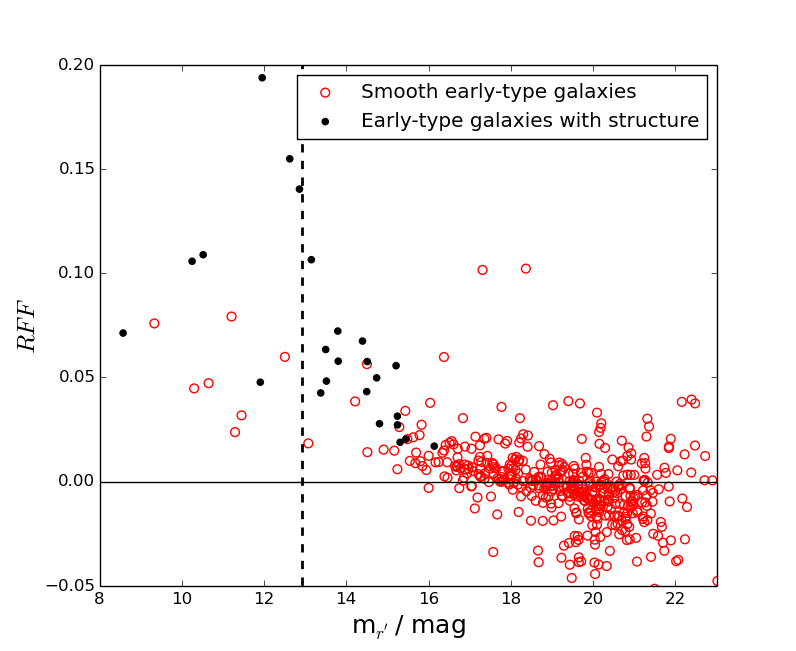}}
  \caption{RFF shown as a function of the r'-band apparent magnitude for the Fornax cluster early-type galaxies in our catalog. The black points show the galaxies which have clear structure (for example a bar or an inner disk) differing from the single S\'ersic model, and the red points are smooth early-type galaxies. The points on the left side of the vertical dashed line are giant galaxies (M$_{r'}$< -18.5 mag) not included in our final catalog. The horizontal line shows the $RFF$=0 level. The systematic shift of $RFF$ from zero in the high luminosity end can be understood as galaxies starting to differ from S\'ersic profiles, whereas the increasing scatter in the low luminosity end is mostly explained by decreasing signal-to-noise.
  }
  \label{fig:model_limits}
\end{figure}

\indent The uncertainty arising from the difference between the intrinsic profile of the galaxy and the fitted model is important for bright galaxies, which typically need several components to adequately fit their light distribution. Also, the models we use cannot fit star-forming clumps of the dwarf irregular galaxies (dIrr), which introduces some additional uncertainty for their fits. The bias introduced by the star-formation areas could be reduced slightly by doing the fits in i'-band, but as the signal-to-noise of the i'-band is significantly lower\footnote{Fornax dwarfs are bluer than r'-i' $\leq$ 0.3, and r'-band is 0.6 mag arcsec$^{-2}$ deeper than i'-band, which makes r'-band at least 0.3 mag deeper than i'-band for those galaxies.  } for the faintest Fornax cluster galaxies than in r'-band, we use r'-  and g'-band data for fitting. We used $RFF$ to quantify how well the S\'ersic models fit the galaxies. In Fig. \ref{fig:model_limits} we show how the $RFF$ is near zero for the early-type galaxies with m$_{r'}$ > 15 mag, and then rises for the galaxies brighter than that indicating increasing amount of structure. It is difficult to quantify the uncertainties associated to the model, but in Section 9 we show that even for the most massive dwarfs, our measurements of magnitudes and effective radii agree well with the values from the literature. 

\section{Separation of the cluster and background galaxies}

Optical photometry alone is not optimal for defining the cluster membership of the objects, since some degeneracy exists in the projected structural and color properties of cluster galaxies and those at higher redshift (see next subsections for details).  A reliable separation requires spectroscopic redshifts, but using the known scaling relations between the properties of the galaxies, we can separate the likely cluster members from the background objects. In the following, we calibrate our selection limits using archival spectroscopic data, select the likely cluster members, and finally test the purity of the selections. 

\indent The Fornax spectroscopic survey of \citet{Drinkwater2000} and its extension (Maddox et al., in prep.) provide spectra for some galaxies with r'-band magnitudes m$_{r'}$<18, but is generally limited to relatively high surface brightness objects ($\mu_{0,r'}$ $\lesssim$ 23 mag arcsec$^{-2}$). It has also a smaller spatial extent than the FDS. According to \citet{Drinkwater2001}, the mean  recession velocity of the Fornax cluster galaxies is $\langle V \rangle$ = 1493 $\pm$ 36 km s$^{-1}$ and the standard deviation of the velocity distribution is $\sigma_V$ = 374 $\pm$ 26 km s$^{-1}$. In what follows, we assume that the galaxies belong to the cluster if they have recession velocities within 2$\sigma_v$ of the mean corresponding to 745 kms$^{-1}$ < V < 2241 kms$^{-1}$. As we are interested in identifying galaxies at the distance of the Fornax cluster rather than identifying the galaxies physically bound to the cluster we do not use varying velocity limits at different cluster-centric radii. For the galaxies with no spectroscopic data available, we can use several other criteria to separate them from background galaxies, as explained below. These criteria are tested using the galaxies with spectroscopic data.

\subsection{Effect of redshift on the morphological and structural parameters}

It is well known that when the distance of a galaxy increases, its angular size decreases, but the surface brightness stays almost constant\footnote{The contribution of the redshift dimming of the surface brightness by factor of $1/(1+z)^4$, is small for the low- to mid-redshift galaxies, with redshift z<0.1 (at z=0.1, there is 50\% dimming).}. This makes intrinsically bright galaxies at large distance to have a low total apparent luminosities but high surface brightness. On the other hand, cluster galaxies follow the magnitude-surface brightness relation \citep{Binggeli1984}, so that the cluster dwarf galaxies with low total luminosity also have low surface brightness. This means that most of the background galaxies should have a brighter surface brightness for a given total magnitude, than the cluster galaxies. Additionally, the intrinsically large background galaxies should have more structure (such as bars or spiral arms) than the low-mass cluster galaxies of a similar apparent size, and also to be more centrally concentrated. However, although we understand well the expected differences between the background and cluster galaxies, it is not trivial how these differences appear in our structural and morphological parameters, once the effects of seeing, $S/N$, and the use of simple decomposition models are taken into account. In the Appendix B we show quantitatively how the parameters of the galaxies change as they get redshifted. To set the local group dwarf galaxies in the context of Fornax cluster we also show them in Fig. \ref{fig:effect_of_redshift}.

\subsection{Preliminary selection cuts}

We identified the cluster galaxies using the following criteria: firstly they become bluer with decreasing luminosity (e.g., \citealp{Roediger2017}), secondly the surface brightness of the cluster galaxies decreases with decreasing total luminosity, and thirdly the faint cluster galaxies are less concentrated than the background galaxies (e.g., \citealp{Misgeld2009}). 

\subsubsection{Color cut}

\begin{figure}
        \resizebox{\hsize}{!}{\includegraphics{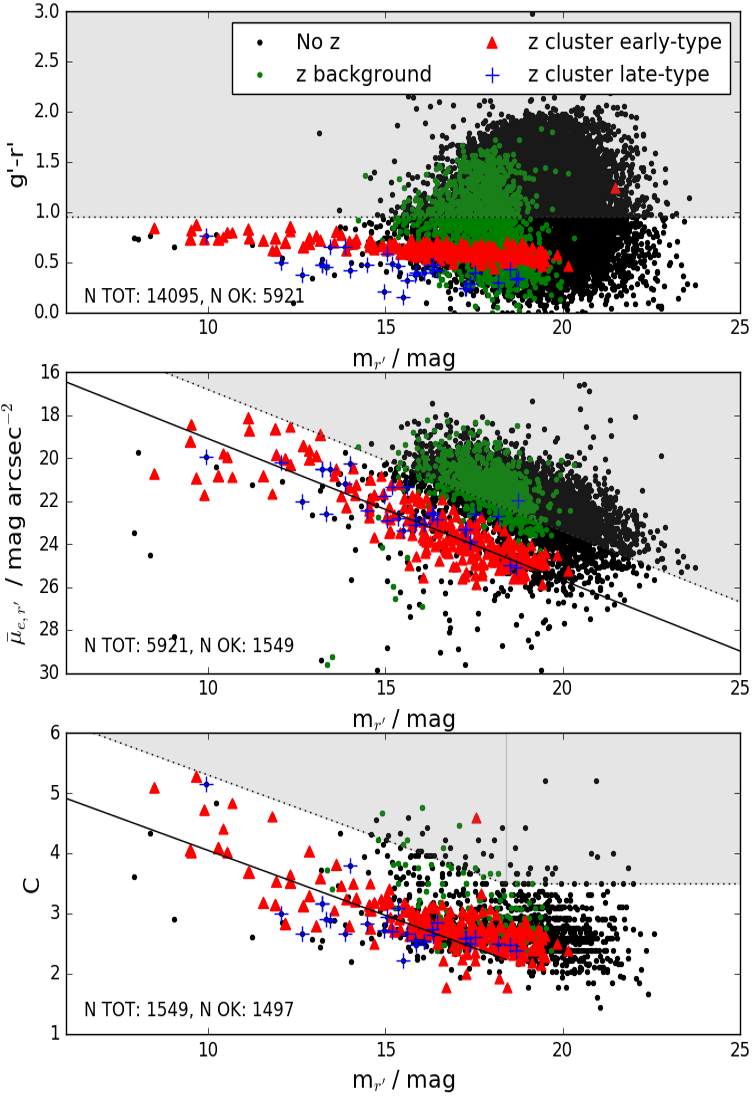}}
    \caption{Illustration of our main criteria for distinguishing the cluster and background galaxies from each other. The panels from top to bottom show how the g'-r' color (also g'-i' cut was used which looks very similar), the mean effective surface brightness $\bar{\mu}_{e,r'}$, and the concentration parameter $C$ of the spectrally confirmed \citep{Drinkwater2000} early- (red symbols) and late-type (blue symbols) cluster and background galaxies (green symbols), scale with the r'-band apparent magnitude (m$_{r'}$). The solid lines show the fits to the early-type Fornax cluster galaxies, and the dotted lines show the selection limits. The excluded areas are shaded with gray. The black dots correspond to objects with no spectra available. The numbers in each plot correspond to the total number of galaxies before the cut, and the number of galaxies that remain after the cut. The two lower panels show only the galaxies that have not been excluded in the previous steps.} 
        \label{fig:selection}
\end{figure}

\indent To calibrate our selection limits, we used the cluster and background galaxies with spectroscopic data. For the color selection we selected the brightest spectroscopically confirmed cluster galaxies\footnote{NGC1399 has g'-r' $\approx$ 0.8 and g'-i' $\approx$ 1.2 in the central parts \citep{Iodice2016}.} and exclude all the galaxies that are at least 0.15 mag redder than that, which corresponds to g'-r' > 0.95 and g'-i' > 1.35 (see the top panel in Fig. \ref{fig:selection}). These limits are $\approx$3$\sigma$ of the calibration uncertainties toward red from the color of NGC1399, which means that by this selection limit we are not likely to remove any galaxies with intrinsic colors bluer than that from our sample. This selection excludes more than half (N$\approx$8200) of the detected galaxies. The excluded galaxies include most of the large background ellipticals and spirals, as their intrinsic colors are similar to the largest cluster galaxies, and their apparent colors are even redder due to redshift. However, after this cut our sample still includes a significant amount of low and mid-redshift background field spirals, and possibly also lower redshift moderate mass ellipticals from the background clusters. These galaxies are bluer than the largest ellipticals of the Fornax cluster. 

\indent Previous studies have suggested the existence of very red dwarf galaxies in clusters, including both low surface brightness \citep{Conselice2003} and compact dwarf galaxies \citep{Price2009}. They appear as red outliers from the red sequence. However, no compact elliptical galaxies of \citet{Price2009} would have been excluded with our color cut, since they are still bluer than the most massive ellipticals. Five of the 53 galaxies in the sample by \citet{Conselice2003} would have been excluded from our sample but as \citet{Penny2008} showed later using spectroscopic subsample, those red outliers in the sample of Conselice were background galaxies.  In principle, the color cut would also remove galaxies that appear red due to their internal dust extinction, but for the evolved nature of the Fornax cluster, it is very unlikely that such galaxies exist in this environment.

\subsubsection{Surface brightness cut}

\begin{figure}
        \resizebox{\hsize}{!}{\includegraphics{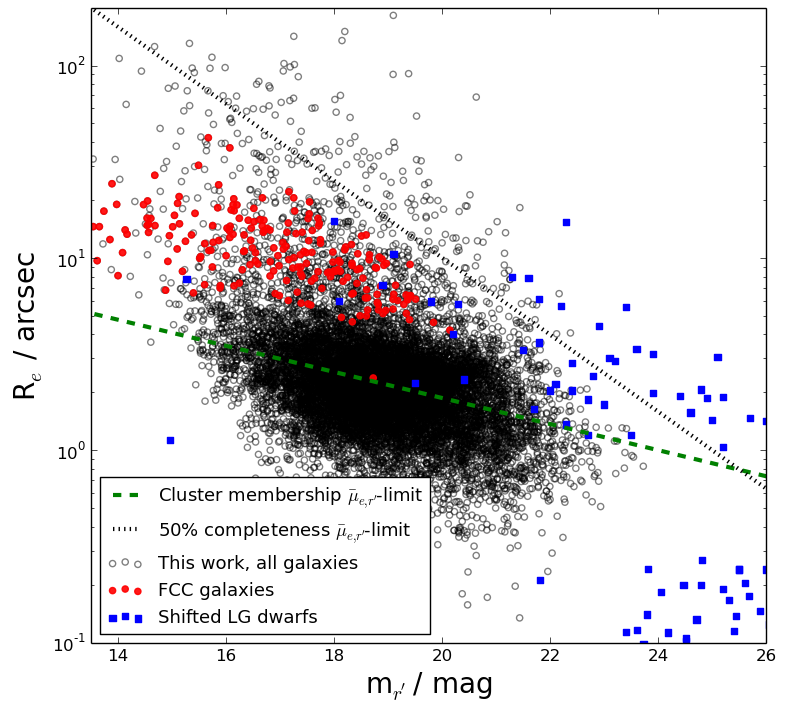}}
    \caption{Comparison of our surface brightness selection cut (green dashed line) described in Section 8.2.2 with the size-magnitude relation of the galaxies in our sample (black and red symbols), and in the local group galaxies of \citet{Brodie2011} after shifting them to the distance of the Fornax cluster (blue squares). The red points show the galaxies in our sample that are classified as likely cluster members by \citet{Ferguson1989}. The gray dotted line shows our 50\% surface brightness completeness limit of $\bar{\mu}_e$ = 26 mag arcsec$^{-2}$. As the blue squares are mostly above the surface brightness selection limit, we would not exclude similar galaxies in the Fornax cluster from our sample by applying the selection limit. The blue squares appearing in the bottom right corner are star cluster like objects that would appear as point sources at the distance of the Fornax cluster. } 
        \label{fig:mag-size-LG.png}
\end{figure}

In order to separate the background galaxies that have higher surface brightness for a given apparent magnitude than the cluster galaxies, we made a linear fit for the cluster galaxies in the magnitude-surface brightness space. For the confirmed cluster galaxies this is shown in Fig. \ref{fig:selection} (red dots in the middle panel). It appears that the slope of the relation between m$_{r'}$ and $\mu_{e,r'}$ changes at  m$_{r'}$ $\approx$ 12 mag so that the galaxies fainter and brighter than that have different slopes (\citealp{Binggeli1984}, \citealp{Misgeld2011}, \citealp{Eigenthaler2018}). Since we are interested in dwarf galaxies in this work, we fit the galaxies only in the faint part, that is, with m$_{r'}$ > 12 mag, and use this fit for the classification. We then defined the mean deviations of the galaxies around the fit and exclude the galaxies that have brighter surface brightness than three standard deviations from the cluster sequence (gray area in Fig. \ref{fig:selection} mid panel). This selection aims to exclude massive high surface brightness background galaxies, but as shown in Fig. \ref{fig:effect_of_redshift}, will not exclude many bright galaxies with z < 0.04. This selection excludes three quarters of the remaining galaxies  leaving only  N = 1549 galaxies. We are aware that there exist compact galaxies in the Fornax cluster that might be excluded due to this criterion. We discuss these galaxies in Section 9. However, as shown in Fig. \ref{fig:mag-size-LG.png} this surface brightness cut would not exclude galaxies similar to Local Group dSphs from our sample.

\subsubsection{Concentration cut}

Finally, we used the concentration parameter to exclude the remaining background elliptical galaxies, which are otherwise difficult to separate morphologically from the cluster dwarfs.  We fit the magnitude-$C$ relation for cluster galaxies with m$_{r'}$ < 16 mag, and classify the galaxies that are located more than 2$\sigma$ above that relation, and have $C$ > 3.5, as background galaxies. The latter criterion is adopted to make sure that we did not exclude exponential disks, or galaxies with even flatter luminosity distribution from our sample. By this condition we exclude additional $\approx$50 galaxies from the cluster sample, leaving 1497 galaxies as likely cluster members ($\approx$10\% of the total sample). Even after this cut, there remains some fraction of spectroscopically confirmed background galaxies, which are later removed from the cluster sample according to their visual morphological appearance (see Sect. 8.3).

\subsection{Visual classification of the selected sample}

\begin{figure*}[!ht]
        \centering
        \includegraphics[height=24cm]{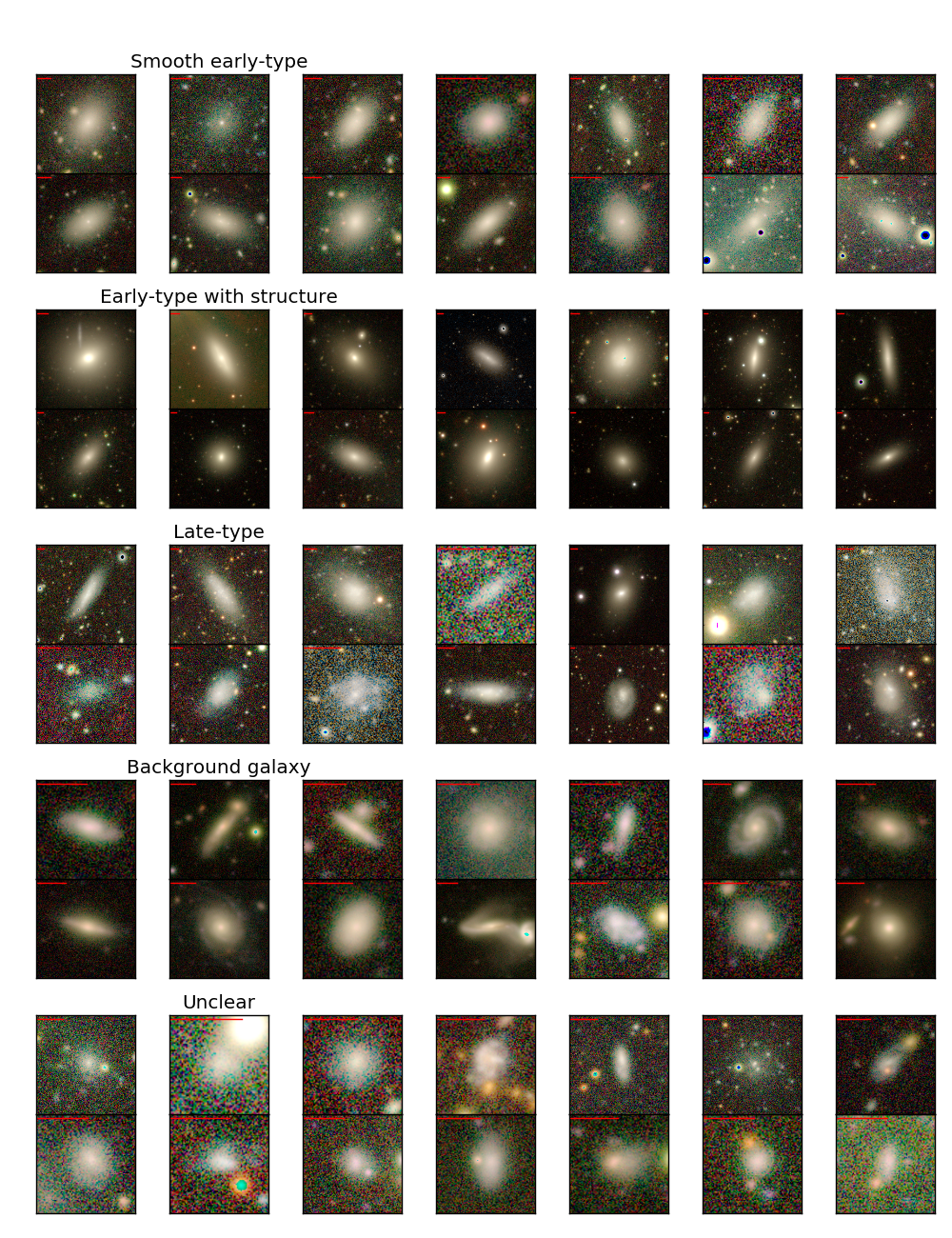}
        \caption{Color images of the galaxies with different morphological classifications as defined in Section 8.3. g', r', and i' bands of the FDS data are used as the blue, green, and red channels in the images, respectively. The galaxies have very different sizes on the sky, so that we have added bars with the length of 10 arcsec to each image.}
        \label{fig:morphological_examples}
\end{figure*}

After selecting the likely cluster member galaxies by their photometric parameters, we make also a first order visual morphological classification of the galaxies.  This classification is not meant for the use of any detailed morphological analysis, but rather to further identify the galaxies that belong to the Fornax cluster. To do the classification in practice, we generate color images of all the likely cluster galaxies and inspect the color and residual images (data$-$model) simultaneously. We separate the galaxies into five groups according to  their morphology, with examples shown in Fig. \ref{fig:morphological_examples}:

\begin{itemize}

\item {\bf Smooth early-type:} Galaxies that have a smooth red appearance, and do not have structures like clearly distinguishable bars or spiral arms. If a dwarf galaxy has an unresolved point-like nucleus, it will be classified into this group as well. This group therefore includes giant early-type galaxies with no clear structure, and nucleated and non-nucleated dwarf ellipticals.

\item {\bf Early-type with structure:} Galaxies that are red and have no star-forming clumps, but have structures such as bulge, bar, or spiral arms. They are not well modeled by a single S\'ersic function. This group includes S0s and dEs with prominent disk features. 

\item {\bf Late-type:} Galaxies that are blue, and have star-forming clumps. This group includes spirals, blue compact dwarfs and dwarf irregular galaxies.

\item {\bf Background:} Small galaxies that show features like bars or spiral arms. Since such features are not likely to appear in low-mass cluster dwarfs (see \citealp{Janz2014}), we conclude them to be background galaxies. We are aware that some dwarf galaxies have also spiral structure and bars \citep{Lisker2006}, but those are mostly found in the most massive dwarfs, and even in them, the fraction of light in the disk structures compared to the smooth spherical component of the galaxy is so low, that there is no danger to mix these dwarfs to the background spirals.

\item {\bf Unclear:} Galaxies whose morphological type is not clear. For example, galaxies that have low surface brightness, but possess some weak structures resembling a bar or a central bulge.

\end{itemize}

\indent In summary, of the 1497 identified galaxies there were 577 likely cluster members, of which 453 were classified as smooth early-types, 24 as early-types with structure, and 100 as late-types. Of the parametrically selected galaxies, 897 were classified as background systems and 22 as uncertain cases.

\subsection{Parametric classification of the uncertain objects}

\begin{figure*}
        \includegraphics[width=17cm]{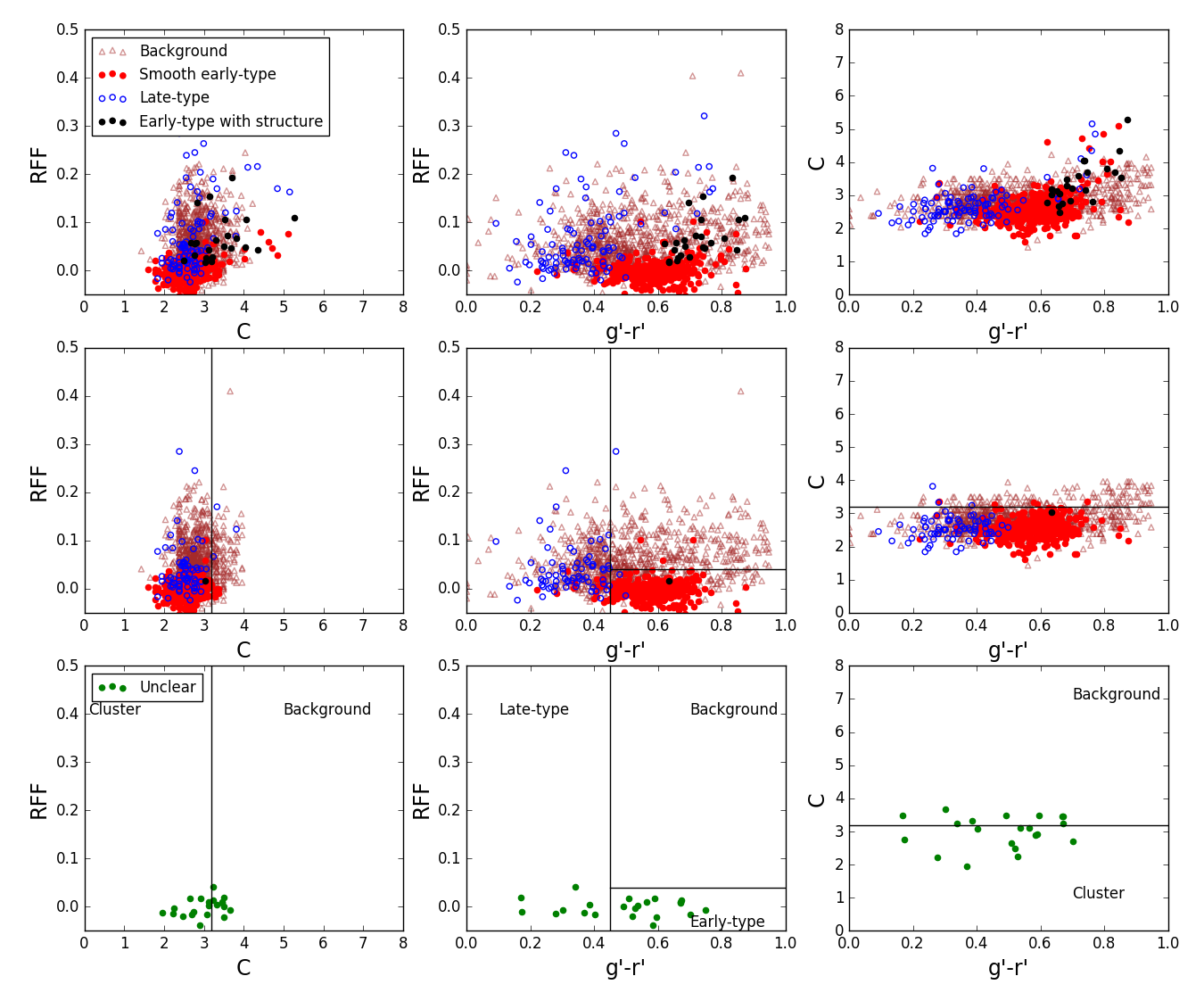}
    \caption{Evaluation of the remaining uncertain cluster memberships, after applying the main selection criteria.  The top row of panels shows how the g'-r' color, concentration ($C$), and $RFF$ are related for the galaxies that we selected using the preliminary selection cuts (see Section 8.2) and visual classification (see Section 8.3). The different colors correspond to visual morphological classifications of these galaxies, and the colors are explained in the legend of the upper left panel.  The mid row shows the same parameters as the top row but limiting to galaxies with m$_{r'}$ > 15 mag. This is the range of the galaxies that we could not classify morphologically with certainty. In the bottom row we show the parameters for these uncertain cases with the green points. The limits that we used for classifying the uncertain cases into early- and late-type cluster, and background galaxies are shown with black lines. Altogether nine of galaxies with uncertain classifications were classified as background, four as cluster late-types, and nine as cluster early-type dwarfs.} 
        \label{fig:morphology_groups}
\end{figure*}

The 22 galaxies, which we were not able to classify morphologically with certainty, may be Fornax cluster galaxies according to their colors, surface brightnesses and concentrations. To give them classifications, we can compare their morphological parameters with those galaxies that we were able to classify morphologically. In Fig. \ref{fig:morphology_groups} we show how the different morphological classes correlate with these parameters, including color, $RFF$ and the $C$ parameters. Such correlations help us to evaluate which of the uncertain cases may still form part of the Fornax cluster. It is clear that in the color-$RFF$ plane a simple color cut is not enough to explain most of the division of galaxies between early-type and late-type systems.

\indent If we concentrate only on the low luminosity galaxies with m$_{r'}$ > 15 mag, the separation of early-type and late-type galaxies is simpler (see lower panels in Fig. \ref{fig:morphology_groups}). All of them have low concentration parameters and $RFF$-values, and the g'-r' colors show a straight forward division between early-type and late-type galaxies. Using these properties, we give parametric classifications for the uncertain galaxies. The galaxies with g'-r' < 0.45 and $C$ < 3.2 are classified as late-types, the ones with  g'-r' > 0.45 and $RFF$ < 0.05 and $C$ < 3.2 are classified as early-types, and the others are classified as background objects. Applying these criteria, 13 of the 22 galaxies with uncertain classifications appeared to be real cluster members, of which four are late-types and nine early-types. Adding these 13 galaxies into our sample of likely cluster galaxies, our total galaxy number increases to 590 galaxies.

\subsection{Final catalog}

Of the 590 cluster members, 564 are dwarf galaxies (m$_{r'}$ $\geq$ 12.5 mag) presented in our Fornax cluster dwarf galaxy catalog. The number of background objects is 13,505. Our catalog has a minimum semi-major axis size limit of 2 arcsec, and it reaches a 50\% completeness at limiting surface brightness of $\bar{\mu}_{e,r'}$ < 26 mag arcsec$^{-2}$. Of the cluster dwarf galaxies 470 are early-types, of which 24 have substructure, and the remaining 94 galaxies are classified as late-type systems. The LSB galaxies and the bright galaxies will appear in separate catalogs to be published by Venhola et al. (in prep.) and \citet{Iodice2018}, respectively. Producing separate catalogs makes sense because different analysis methods are used to obtain the parameters of the bright galaxies, and in case of the LSBs are also used when identifying the galaxies. For example, with the simple S\'ersic fitting used in this study, it is not possible to derive any reliable physical parameters of the bright galaxies \citep{Spavone2017}. However, for completeness in this study we provide also the parameters of all the detected background objects independent of the galaxy magnitude. An extract of the catalog is given  in Table \ref{tab:cat_example}. The full dwarf galaxy catalog with the measured parameters and classifications, and a separate catalog with the background objects, are available as online material at the CDS via anonymous ftp to cdsarc.u-strasbg.fr (130.79.128.5) or via \url{http://cdsweb.u-strasbg.fr/cgi-bin/gcat?J/A+A/}.

\begin{sidewaystable*}
\caption{An example page of our dwarf galaxy catalog including likely cluster galaxies. The columns from left to right correspond to: target, right ascension (R.A.) and declination (Dec.) in ICRS coordinates, axis-ratio ($b/a$), position angle measured from north toward east ($\theta$), apparent r'-band total magnitude (m$_{r'}$), r'-band effective radius in arcseconds (R$_{e}$), S\'ersic index ($n$), aperture magnitudes within the effective radius in the u', g', r', and i' filters, the concentration index ($C$), and the residual flux fraction ($RFF$). The values after the $\pm$-signs correspond to the 1$\sigma$ uncertainties in the parameters. The final column gives the morphological type of the galaxy according to our classifications in Section 8.3: 'e' corresponding to smooth early-type, 'e(s)' to early-type with structure, and 'l' to late-type galaxy, and '*' at the end of the morphological classification indicating that the object has a nucleus.}
\label{tab:cat_example}
\centering\tiny
\begin{tabular}{lcccccccccccccc}
\hline\hline
Target & R.A. & Dec. & $b/a$ & $\theta$ & m$_{r'}$ & R$_{e}$ & $n$ & u' & g' & r' & i' & $C$ & $RFF$ & Morphology  \\
\hline 

F12D327 & 54.3038 & -36.2579 & 0.54 $\pm$ 0.08 & -84.3 $\pm$ 4.6 & 20.0 $\pm$ 0.3 & 5.9 $\pm$ 1.6 & 0.8 $\pm$ 0.4 & 22.91 $\pm$ 0.07 & 21.58 $\pm$ 0.05 & 21.15 $\pm$ 0.04 & 21.27 $\pm$ 0.04  & 2.0 & 0.01 & e \\ 
F12D349 & 54.2052 & -36.2297 & 0.33 $\pm$ 0.05 & 85.3 $\pm$ 3.2 & 17.3 $\pm$ 0.2 & 13.4 $\pm$ 2.8 & 1.0 $\pm$ 0.4 & 20.44 $\pm$ 0.04 & 19.41 $\pm$ 0.03 & 18.77 $\pm$ 0.03 & 18.39 $\pm$ 0.04  & 2.3 & 0.00 & e \\ 
F12D366 & 55.1832 & -36.1853 & 0.84 $\pm$ 0.10 & 30.9 $\pm$ 5.5 & 17.8 $\pm$ 0.3 & 19.6 $\pm$ 5.8 & 1.0 $\pm$ 0.4 & 20.42 $\pm$ 0.06 & 19.21 $\pm$ 0.04 & 18.62 $\pm$ 0.03 & 18.71 $\pm$ 0.04  & 2.8 & -0.02 & e \\ 
F12D367 & 54.0536 & -36.1665 & 0.55 $\pm$ 0.02 & 66.7 $\pm$ 1.2 & 15.5 $\pm$ 0.1 & 9.9 $\pm$ 1.1 & 1.5 $\pm$ 0.3 & 18.52 $\pm$ 0.04 & 17.12 $\pm$ 0.03 & 16.49 $\pm$ 0.03 & 16.17 $\pm$ 0.04  & 3.0 & 0.02 & e(s) \\ 
F12D396 & 54.9691 & -36.1419 & 0.56 $\pm$ 0.08 & 85.5 $\pm$ 4.4 & 21.0 $\pm$ 0.3 & 3.4 $\pm$ 0.9 & 0.6 $\pm$ 0.4 & 24.02 $\pm$ 0.07 & 22.93 $\pm$ 0.06 & 22.25 $\pm$ 0.05 & 21.84 $\pm$ 0.04  & 2.1 & -0.01 & e \\ 
F12D399 & 55.1571 & -36.1211 & 0.67 $\pm$ 0.06 & -43.0 $\pm$ 3.6 & 18.3 $\pm$ 0.2 & 9.4 $\pm$ 2.1 & 0.7 $\pm$ 0.4 & 20.76 $\pm$ 0.04 & 19.83 $\pm$ 0.03 & 19.24 $\pm$ 0.03 & 18.84 $\pm$ 0.04  & 2.4 & -0.01 & e \\ 
F12D406 & 54.7741 & -36.0989 & 0.73 $\pm$ 0.02 & -3.6 $\pm$ 1.5 & 17.0 $\pm$ 0.1 & 6.4 $\pm$ 0.8 & 0.8 $\pm$ 0.3 & 19.78 $\pm$ 0.06 & 18.39 $\pm$ 0.03 & 17.82 $\pm$ 0.03 & 17.56 $\pm$ 0.04  & 2.6 & 0.01 & e* \\ 
F12D429 & 54.5801 & -36.0656 & 0.78 $\pm$ 0.04 & -15.9 $\pm$ 2.6 & 18.3 $\pm$ 0.2 & 6.5 $\pm$ 1.2 & 0.7 $\pm$ 0.3 & 21.11 $\pm$ 0.04 & 19.80 $\pm$ 0.03 & 19.22 $\pm$ 0.03 & 19.00 $\pm$ 0.04  & 2.4 & -0.00 & e \\ 
F12D433 & 54.8846 & -36.0994 & 0.90 $\pm$ 0.08 & 79.5 $\pm$ 4.3 & 20.3 $\pm$ 0.3 & 4.7 $\pm$ 1.2 & 0.6 $\pm$ 0.4 & 23.04 $\pm$ 0.07 & 21.81 $\pm$ 0.05 & 21.29 $\pm$ 0.04 & 21.27 $\pm$ 0.04  & 2.3 & -0.02 & e \\ 
F12D436 & 55.0014 & -36.0936 & 0.59 $\pm$ 0.09 & -35.2 $\pm$ 5.0 & 20.5 $\pm$ 0.3 & 5.2 $\pm$ 1.4 & 0.8 $\pm$ 0.4 & 23.17 $\pm$ 0.10 & 21.94 $\pm$ 0.06 & 21.43 $\pm$ 0.05 & 21.14 $\pm$ 0.04  & 2.0 & -0.01 & e \\ 
F12D463 & 54.5262 & -36.0501 & 0.89 $\pm$ 0.06 & -83.9 $\pm$ 3.6 & 19.9 $\pm$ 0.2 & 4.6 $\pm$ 1.0 & 0.8 $\pm$ 0.4 & 22.33 $\pm$ 0.05 & 21.34 $\pm$ 0.04 & 20.78 $\pm$ 0.04 & 20.57 $\pm$ 0.04  & 2.6 & -0.00 & e \\ 
F12D491 & 54.4249 & -35.9553 & 0.46 $\pm$ 0.11 & -13.0 $\pm$ 5.8 & 19.5 $\pm$ 0.3 & 9.4 $\pm$ 2.9 & 0.7 $\pm$ 0.4 & 21.86 $\pm$ 0.05 & 21.16 $\pm$ 0.04 & 20.56 $\pm$ 0.04 & 20.29 $\pm$ 0.04  & 2.2 & -0.03 & e \\ 
F12D492 & 54.3645 & -35.9629 & 0.90 $\pm$ 0.07 & 10.0 $\pm$ 4.1 & 19.7 $\pm$ 0.3 & 5.9 $\pm$ 1.4 & 1.0 $\pm$ 0.4 & 21.99 $\pm$ 0.12 & 21.03 $\pm$ 0.07 & 20.57 $\pm$ 0.05 & 20.38 $\pm$ 0.05  & 2.4 & -0.03 & e \\ 
F12D608 & 54.7168 & -36.3010 & 0.89 $\pm$ 0.04 & 56.8 $\pm$ 2.5 & 20.8 $\pm$ 0.2 & 2.0 $\pm$ 0.4 & 1.1 $\pm$ 0.3 & 23.50 $\pm$ 0.05 & 22.39 $\pm$ 0.05 & 21.90 $\pm$ 0.04 & 21.71 $\pm$ 0.04  & 2.2 & 0.01 & e \\ 
F13D004 & 55.0100 & -37.9414 & 0.40 $\pm$ 0.08 & -38.2 $\pm$ 4.6 & 21.1 $\pm$ 0.3 & 3.5 $\pm$ 0.9 & 0.4 $\pm$ 0.4 & 24.15 $\pm$ 0.07 & 23.54 $\pm$ 0.08 & 23.12 $\pm$ 0.07 & 22.88 $\pm$ 0.05  & 2.1 & -0.02 & l \\ 
F13D042 & 55.2303 & -37.8376 & 0.62 $\pm$ 0.02 & 36.5 $\pm$ 1.4 & 15.7 $\pm$ 0.1 & 10.9 $\pm$ 1.4 & 1.1 $\pm$ 0.3 & 18.63 $\pm$ 0.04 & 17.25 $\pm$ 0.03 & 16.54 $\pm$ 0.03 & 16.24 $\pm$ 0.04  & 2.8 & 0.01 & e \\ 
F13D044 & 54.4815 & -37.8560 & 0.72 $\pm$ 0.05 & -45.1 $\pm$ 3.0 & 17.6 $\pm$ 0.2 & 10.6 $\pm$ 2.1 & 1.0 $\pm$ 0.4 & 19.98 $\pm$ 0.04 & 19.20 $\pm$ 0.03 & 18.51 $\pm$ 0.03 & 18.15 $\pm$ 0.04  & 2.9 & 0.01 & e \\ 
F13D054 & 55.1427 & -37.8518 & 0.76 $\pm$ 0.04 & 45.0 $\pm$ 2.6 & 18.5 $\pm$ 0.2 & 6.1 $\pm$ 1.1 & 1.4 $\pm$ 0.3 & 21.18 $\pm$ 0.05 & 19.80 $\pm$ 0.03 & 19.34 $\pm$ 0.03 & 19.09 $\pm$ 0.04  & 3.1 & 0.01 & e \\ 
F13D058 & 55.1262 & -37.8280 & 0.38 $\pm$ 0.02 & 10.0 $\pm$ 1.2 & 16.2 $\pm$ 0.1 & 7.3 $\pm$ 0.8 & 1.4 $\pm$ 0.3 & 18.08 $\pm$ 0.04 & 17.43 $\pm$ 0.03 & 17.11 $\pm$ 0.03 & 16.95 $\pm$ 0.04  & 2.8 & 0.24 & l \\ 
F13D074 & 54.2367 & -37.8208 & 0.64 $\pm$ 0.08 & -9.4 $\pm$ 4.6 & 18.8 $\pm$ 0.3 & 9.9 $\pm$ 2.6 & 0.7 $\pm$ 0.4 & 23.19 $\pm$ 0.11 & 20.52 $\pm$ 0.04 & 19.86 $\pm$ 0.03 & 19.44 $\pm$ 0.04  & 2.4 & -0.00 & e \\ 
F13D130 & 54.9101 & -37.7097 & 0.95 $\pm$ 0.03 & 12.7 $\pm$ 1.9 & 19.5 $\pm$ 0.2 & 2.7 $\pm$ 0.4 & 1.3 $\pm$ 0.3 & 22.06 $\pm$ 0.05 & 21.01 $\pm$ 0.04 & 20.51 $\pm$ 0.03 & 20.30 $\pm$ 0.04  & 2.8 & 0.00 & e \\ 
F13D162 & 55.0871 & -37.6449 & 0.57 $\pm$ 0.05 & 88.4 $\pm$ 3.1 & 17.2 $\pm$ 0.2 & 13.8 $\pm$ 2.9 & 0.8 $\pm$ 0.4 & 20.39 $\pm$ 0.04 & 18.91 $\pm$ 0.03 & 18.23 $\pm$ 0.03 & 18.00 $\pm$ 0.04  & 2.5 & 0.00 & e \\ 
F13D165 & 55.0854 & -37.6441 & 0.58 $\pm$ 0.05 & -88.3 $\pm$ 3.0 & 17.1 $\pm$ 0.2 & 13.6 $\pm$ 2.8 & 0.8 $\pm$ 0.4 & 20.13 $\pm$ 0.04 & 18.76 $\pm$ 0.03 & 18.06 $\pm$ 0.03 & 17.81 $\pm$ 0.04  & 2.5 & 0.04 & e \\ 
F13D176 & 54.0200 & -37.6152 & 0.86 $\pm$ 0.07 & 14.6 $\pm$ 4.0 & 20.3 $\pm$ 0.3 & 4.4 $\pm$ 1.1 & 0.8 $\pm$ 0.4 & 23.84 $\pm$ 0.10 & 21.81 $\pm$ 0.05 & 21.16 $\pm$ 0.04 & 20.97 $\pm$ 0.04  & 2.6 & -0.04 & e \\ 
F13D221 & 55.1205 & -37.5228 & 0.66 $\pm$ 0.02 & -53.6 $\pm$ 1.5 & 19.2 $\pm$ 0.1 & 2.3 $\pm$ 0.3 & 1.1 $\pm$ 0.3 & 22.11 $\pm$ 0.04 & 20.98 $\pm$ 0.03 & 20.35 $\pm$ 0.03 & 20.06 $\pm$ 0.04  & 2.6 & 0.01 & e \\ 
F13D224 & 55.2271 & -37.5166 & 0.91 $\pm$ 0.07 & -63.7 $\pm$ 4.3 & 19.4 $\pm$ 0.3 & 7.1 $\pm$ 1.8 & 0.7 $\pm$ 0.4 & 23.54 $\pm$ 2.04 & 20.70 $\pm$ 0.16 & 20.14 $\pm$ 0.10 & 19.88 $\pm$ 0.08  & 2.4 & -0.01 & e* \\ 
F13D230 & 55.0774 & -37.4994 & 0.45 $\pm$ 0.04 & 58.0 $\pm$ 2.7 & 17.9 $\pm$ 0.2 & 8.4 $\pm$ 1.6 & 0.9 $\pm$ 0.3 & 21.15 $\pm$ 0.04 & 19.68 $\pm$ 0.03 & 19.00 $\pm$ 0.03 & 18.69 $\pm$ 0.04  & 2.7 & 0.00 & e \\ 
F13D258 & 55.1850 & -37.4083 & 0.76 $\pm$ 0.02 & -82.1 $\pm$ 1.6 & 16.4 $\pm$ 0.2 & 9.2 $\pm$ 1.3 & 0.8 $\pm$ 0.3 & 19.32 $\pm$ 0.04 & 18.01 $\pm$ 0.03 & 17.29 $\pm$ 0.03 & 17.02 $\pm$ 0.04  & 2.5 & 0.02 & e \\ 
F13D284 & 54.1532 & -37.3653 & 0.52 $\pm$ 0.06 & -72.3 $\pm$ 3.7 & 19.4 $\pm$ 0.2 & 6.1 $\pm$ 1.4 & 0.8 $\pm$ 0.4 & 22.22 $\pm$ 0.04 & 21.01 $\pm$ 0.04 & 20.42 $\pm$ 0.03 & 20.12 $\pm$ 0.04  & 2.5 & -0.00 & e \\ 
F13D299 & 54.5557 & -37.2899 & 0.88 $\pm$ 0.03 & 78.1 $\pm$ 2.1 & 14.7 $\pm$ 0.2 & 26.9 $\pm$ 4.3 & 1.3 $\pm$ 0.3 & 17.49 $\pm$ 0.04 & 16.11 $\pm$ 0.03 & 15.45 $\pm$ 0.03 & 14.73 $\pm$ 0.04  & 2.5 & 0.02 & e(s) \\ 

... & & & & & & & & & & & & & \\

\end{tabular}
\end{sidewaystable*}

\section{Comparison with the literature}

\subsection{Detections}

To assess its quality, it is important to compare the completeness of the FDS catalog described in this paper, to the previous most complete Fornax Cluster Catalog by Ferguson (1989). Our galaxy catalog is known to miss some galaxies: those overlapping with the bright stars, galaxies projected on top of the bright galaxies, galaxies that are small, and galaxies which have very low surface brightnesses (such as UDGs). These biases were quantified in Section 6.2 using mock galaxies, which showed that we are able to detect galaxies down to $\bar{\mu}_{e,r}$ $\approx$ 26 mag arcsec$^{-2}$ with more than a 50\% completeness and miss $\approx$3\% of the galaxies due to overlapping bright objects. Here we make comparisons with the FCC and the LSB galaxy sample of \citet{Venhola2017}, latter of which (when extended to the whole cluster) will form part of the complete FDS dwarf galaxy catalog\footnote{We also miss UCDs and cEs that appear (nearly) unresolved in our data.}.  Additionally, we show a comparison with the visually identified samples of \citet{Mieske2007} and \citet{Eigenthaler2018} that are limited to the central parts of the cluster.

\begin{figure}
        \resizebox{\hsize}{!}{\includegraphics{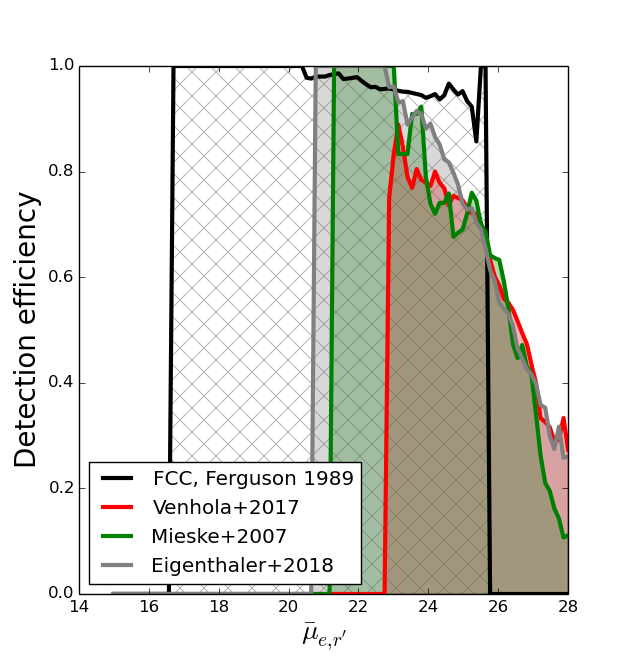}}
        \caption{Completeness of the cluster galaxies with our detection algorithm compared to the previous Fornax samples. The black, red, green, and gray lines show the detection efficiency of our algorithm when compared with the galaxies from FCC, \citet{Venhola2017}, \citet{Mieske2007}, and \citet{Eigenthaler2018}, respectively.}
        \label{fig:real_det}
\end{figure}

\indent We took the above mentioned catalogs and searched for FDS dwarf galaxies within 5 arcsec from those objects. The black, red, green, and gray lines in the Fig. \ref{fig:real_det} correspond to the detection efficiency while comparing our sample with the FCC galaxies, the LSB galaxies by \citet{Venhola2017}, the galaxies in \citet{Mieske2007}, and the galaxy sample of \citet{Eigenthaler2018}. This comparison with other existing catalogs is in good agreement with the mock galaxy tests: at the LSB-end ($\bar{\mu}_{e,r}$ > 24 mag arcsec$^{-2}$) the detection efficiency drops as a function of surface brightness reaching 50\% completeness between $\bar{\mu}_{e,r}$ = 26-27 mag arcsec$^{-2}$. Close to 100\% completeness is obtained for the non-LSB galaxies. 

\begin{figure}
        \resizebox{\hsize}{!}{\includegraphics{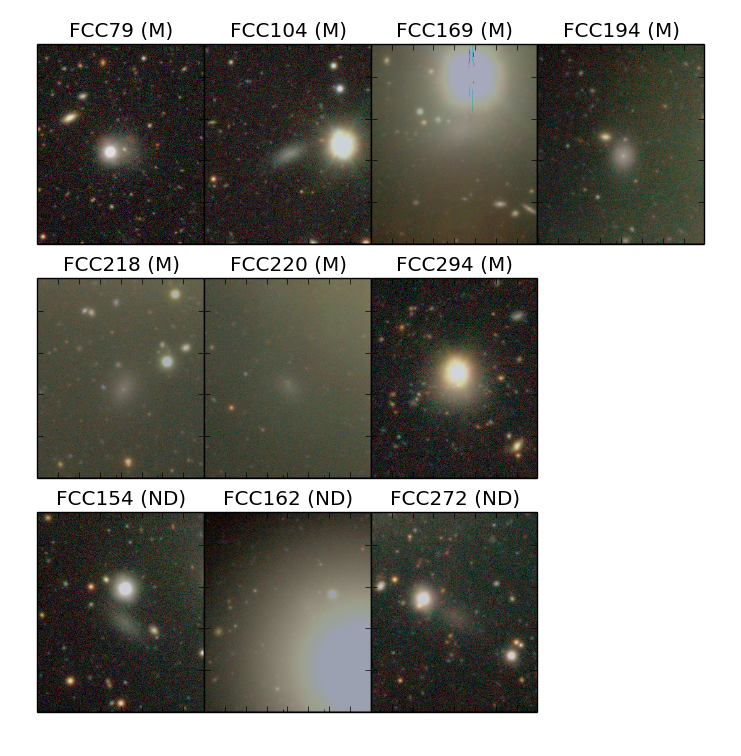}}
        \caption{ i', r', g',- color composite images of the FCC galaxies that are missing from our catalog since they are either located under the masks that were used to exclude objects in the areas contaminated by bright stars, or were not detected. After each object's name we indicate if the object was excluded due to the masks (M) or was not detected (ND). }
        \label{fig:missing_fcc}
\end{figure}

\indent Ten galaxies are classified as likely Fornax cluster members in the FCC, but not found with our detection method (see Fig. \ref{fig:missing_fcc} for thumbnails): seven of these overlap with saturated stars and are therefore excluded due to the masks, and three were missed due to their low surface brightness. The number of missed galaxies due to the masking is consistent with our estimation in Section 5.2.  In the case of FCC162 we could not detect any object in the location indicated in the FCC (see also \citealp{Eigenthaler2018}). The other two missing galaxies appear morphologically to be cluster members, and they will be included in the LSB extension of this catalog.  We have not analyzed here in detail the differences in object detections between our catalog and the catalogs other than FCC, since possible differences are mostly due to our incompleteness at the LSB-end. Our next paper concentrating on the detection of LSB galaxies in the Fornax area (Venhola et al., in prep.) will include a more detailed comparison with respect to completeness.

\begin{figure}
        \resizebox{\hsize}{!}{\includegraphics{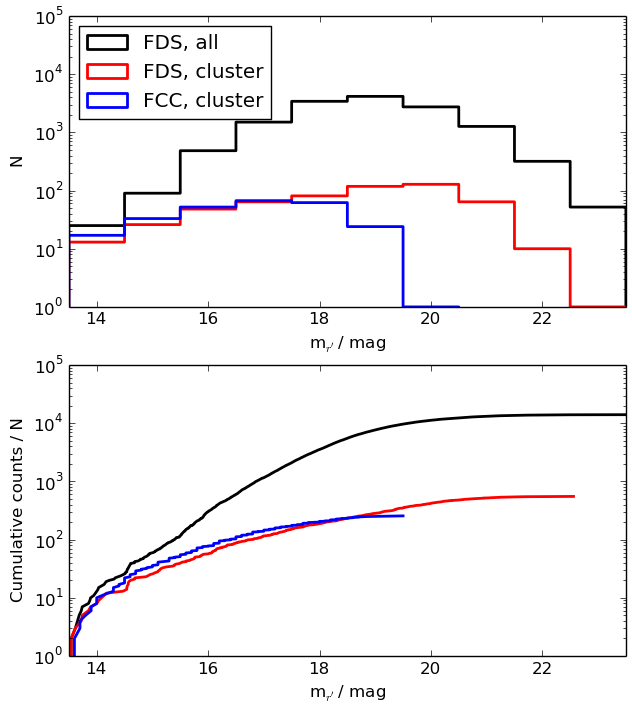}}
        \caption{Upper panel: Distribution of the apparent r'-band magnitudes of in FCC (the blue histogram), compared to the ones detected in this work. We show the distribution of the galaxies classified as likely cluster dwarfs with the red histogram and the one of the likely background galaxies with the black histogram. Lower panel: Same distributions in a cumulative profile. The FCC magnitudes are transformed from the B-band to r'-band using the mean color difference of $\langle B-r' \rangle$ = 1 mag defined comparing our values with the ones of FCC. }
        \label{fig:mag_hist}
\end{figure}

\indent Figure \ref{fig:mag_hist} compares the total dwarf galaxy counts of FCC \citep{Ferguson1989} and our catalog as a function of the total galaxy magnitude. It is clear that our catalog extends three to four magnitudes deeper than FCC. The numbers of likely cluster members in the magnitude bins match well between the two catalogs, within the magnitude range where the FCC is complete (m$_{r'}$ $\lesssim$ 18 mag). FCC has slightly more cluster members in the two brightest bins due to its larger spatial extent.

\begin{figure}
        \resizebox{\hsize}{!}{\includegraphics{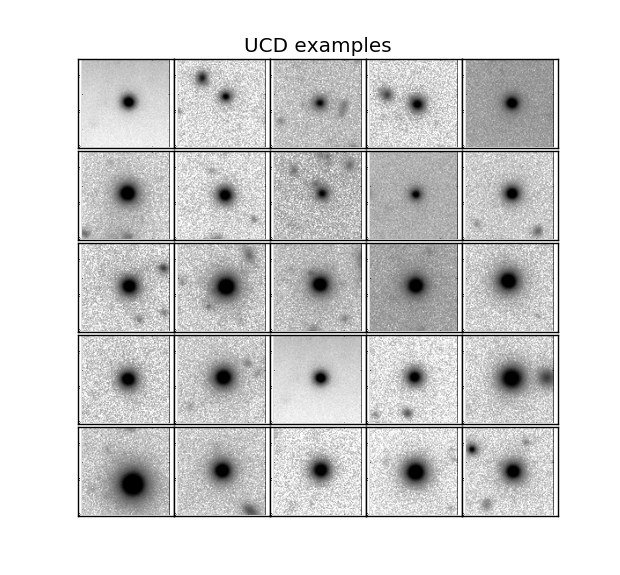}}
        \caption{ r'-band images of UCD galaxies in Fornax, for which the cluster membership has been confirmed with spectroscopic redshifts of \citet{Drinkwater2000}, but which do not appear in our catalog due to the selection limit related to size. The size of the postage stamp images is 20 arcsec $\times$ 20 arcsec. }
\label{fig:ucd_examples}
\end{figure}

The spectroscopically confirmed sample of \citet{Drinkwater2000} contains also compact galaxies that are not classified as likely cluster members in the FCC. These compact galaxies are ultra compact dwarf galaxies (UCDs) that have small sizes and high surface brightnesses. To test whether some of these galaxies should be in our catalog according to their size, but were excluded due to their high surface brightness, we cross-matched our initial detection lists with the objects of Drinkwater et al. We find that there are 93 objects in common. We visually checked these objects, and they all appear unresolved in our data, and due to the A\_IMAGE > 2 arcsec selection limit are not in our catalog. We show examples of these objects in Fig. \ref{fig:ucd_examples}.

\subsection{Magnitudes and effective radii}

In the FCC, effective radii and magnitudes are defined from IIIa-J-bandbass photometric plates using growth curves, which could lead to a systematical difference compared to our r'-band measurements. However, a comparison with their parameter values can be used as a sanity check for the parameters obtained by us. To match the magnitude system with ours, we use the FCC magnitudes transformed to the B-band\footnote{We also applied the correction that is required to transform the magnitudes measured from photometric plates into CCD magnitudes: B$_{CCD}$=1.10*B$_{photo}$-1.37, defined by \citet{Ferguson1989} when the FCC magnitudes were compared  to the ones by \citet{Caldwell1987}.} as given by \citet{Ferguson1989}.  To take into account the different photometric filters used in these works we transformed our g'-band magnitudes to the Johnson B-band using the transformation formula defined by Lupton (2005)\footnote{From the SDSS website \url{http://www.sdss3.org/dr8/algorithms/sdssUBVRITransform.php}}, B = g' + 0.3130$\times$(g'-r') + 0.2271. 

\begin{figure}
        \resizebox{\hsize}{!}{\includegraphics{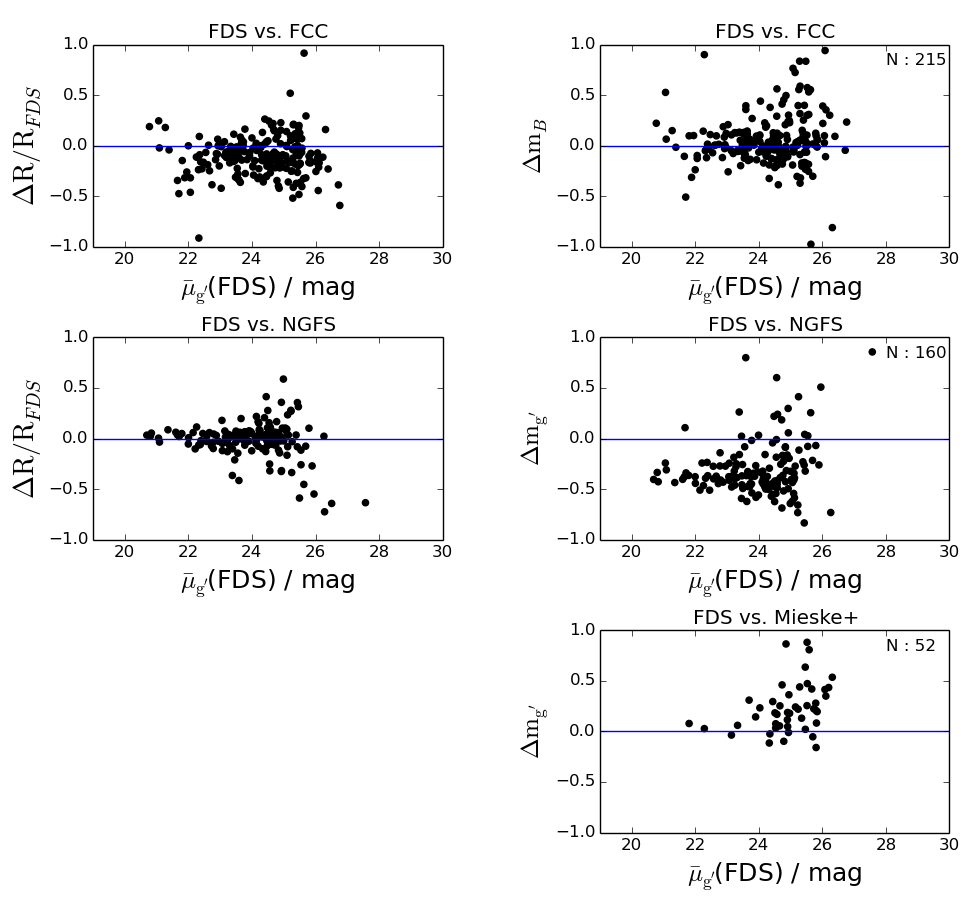}}
        \caption{Effective radii (R$_e$, {\it left panels}) and apparent magnitudes (m$_{B/g'}$, {\it right panels}) of the galaxies obtained in this work are compared to those given in FCC, NGFS \citep{Eigenthaler2018}, and \citet{Mieske2007}, from the top to bottom rows, respectively, using the galaxies common in the studies. The differences are with respect to FDS values, i.e., $\Delta X= X_{FDS} - X_{Ref.}$. The x-axis shows the mean effective surface brightness ($\bar{\mu}_{e,g'}$) of the galaxies, and the blue lines show the zero offsets. \citet{Mieske2007} does not include effective radii for the galaxies, and therefore the comparison is not shown. The number of galaxies in common between the studies are marked into the upper right corners of the right-side panels.}
\label{fig:params_vs_FCC}
\end{figure}

\indent The upper panels in Fig. \ref{fig:params_vs_FCC} show the comparison between our effective radii and magnitudes, and the ones of FCC as a function of mean effective surface brightness for the 215 dwarf galaxies common between the studies. We find that the values agree well with small offsets,$\Delta$(R$_e$(FCC)/R$_e$(FDS)) = -0.11 and $\Delta$(m$_B$(FCC)-m$_{B*}$(FDS)) = 0.01 mag, and a relatively small scatter, $\sigma$(R$_e$(FCC)/R$_e$(FDS)) = 0.18 and $\sigma$(m$_B$(FCC)-m$_{B*}$(FDS)) = 0.25 mag that increases toward lower surface brightness. For the effective radii the offset likely results from FCC using a bluer filter and a different method when defining R$_e$, resulting to slightly smaller values as they miss light in the outskirts of the galaxies. As FCC uses magnitudes based on growth curves they miss an increasing fraction of galaxies' light with decreasing surface brightness. This appears as an increasing difference in the apparent magnitude toward the lower luminosity galaxies, between the two studies.

\indent To avoid the caveats included in the filter transformations and methodological differences in the measurements of the magnitudes and effective radii, we also compare our values with the DECAM g'-band magnitudes of \citet{Eigenthaler2018} for the 160 dwarf galaxies common between the works. Eigenthaler et al. use also GALFIT to fit S\'ersic profiles to the 2D-light distribution of the galaxies in g'-band, similarly to us. We show the comparisons in the second row panels of Fig. \ref{fig:params_vs_FCC} for the galaxies common in these studies. This comparison gives the offsets of $\Delta$(R$_e$(NGFS)/R$_e$(FDS)) = 0.02 and $\Delta$(m$_g$(NGFS)-m$_g$(FDS)) = -0.33 mag, and scatters of $\sigma$(R$_e$(NGFS)/R$_e$(FDS)) = 0.12 and $\sigma$(m$_g$(NGFS)-m$_g)$(FDS)) = 0.23 mag. The observed offset in the magnitudes is surprising given the very good agreement between the effective radii of these two samples. If we neglected the offset, the magnitudes are in good agreement.

\indent To confirm that the observed offset with the NGFS galaxies is not due to a bias in our data we also show a comparison with the magnitudes of \citet{Mieske2007} for the 52 galaxies common in the studies. Their V-band magnitudes, based on the curve of growth analysis, were transformed to g'-band using the transformations given in Appendix C. The magnitudes are in agreement but a slight trend is apparent toward low surface brightness end, so that their magnitudes become fainter than ours. 

\indent These tests show that our effective radii and magnitudes are in good agreement with the other Fornax galaxy samples in the literature, with the exception of the magnitudes of Eigenthaler et al. which are significantly offset. Since we did not observe such an offset in the photometric quality assessments of our data (Section 4.2), GALFIT model quality assessments (Section 7.2.3), nor in comparison with other samples, we conclude that most probably this offset is due to a bias in the calibration of Eigenthaler et al..

\subsection{Assessment of the galaxy colors}

In Section 4.2, we quantify the uncertainty associated with the photometric calibration of our data. In addition to that, we also test the galaxy colors for possible biases when compared with other samples. The color-magnitude relation of early-type cluster galaxies, in other words the red sequence (RS), shows only minor variations between the different nearby clusters (\citealp{Hamraz2018}) and a comparison with those relations can be therefore used as a sanity check for the obtained colors. Here we make the comparisons just for the sake of assessing the colors for possible off-sets. A more detailed analysis with a physical interpretation will follow in the upcoming paper (Venhola et al., in prep.).

\indent Due to a lack of other samples in the Fornax cluster, except for the one of the NGFS that has an offset in the g'-band magnitudes with respect to our sample, and \citet{Mieske2007} that uses V-I colors, we also include works for the Virgo cluster into our comparison. \citet{Roediger2017} have measured the colors of the RS in the core parts of the Virgo cluster with MegaCAM in a luminosity range similar to this work. As Roediger et al. use MegaCAM filters that differ from the SDSS filters, we transformed their colors into the SDSS system using the transformations given in Appendix C. We also make comparisons with the work of \citet{Janz2009} who measure the RS within a larger area in the Virgo cluster using the SDSS filters, but have a smaller luminosity range. To match the sample of Janz et al. with the other samples, we select only their galaxies from the same area as Roediger et al. For the comparison within the Fornax cluster, we use those of \citet{Eigenthaler2018} (see previous subsection for details) and \citet{Mieske2007}, of which the latter measure V-I colors that we have transformed to g'-i' colors using the transformation formulas shown in Appendix C.

In Fig. \ref{fig:cmr_comparison}, we show the color-magnitude relation of the early-type galaxies. To match the samples of NGFS and NGVS, we selected only the galaxies located within 1.4 deg (corresponding to two core radii) from the center of the cluster. We find that there are offsets with respect to the sample of Eigenthaler et al. in the u'-r' and u'-g' colors, so that their colors are significantly bluer by 0.3-0.4 mag. When compared with the samples of Mieske et al. or Roediger et al., there are no such offsets. When compared with Janz et al., a small offset appears in the colors of the brightest dwarfs so that their colors are slightly bluer, but since the colors are otherwise very similar this difference is likely explained by the small number of galaxies in that luminosity range.

\indent As a conclusion, our galaxy colors are in good agreement with the previous measurements done in the Virgo and Fornax clusters, except with the sample of Eigenthaler et al. As the galaxy colors are measured similarly in these two samples, using the same filters and there are no significant differences in the measured effective radii, the difference in the colors is likely be due to a calibration bias in the sample of Eigenthaler et al.

\begin{figure*}[!ht]
        \centering
        \includegraphics[width=17cm]{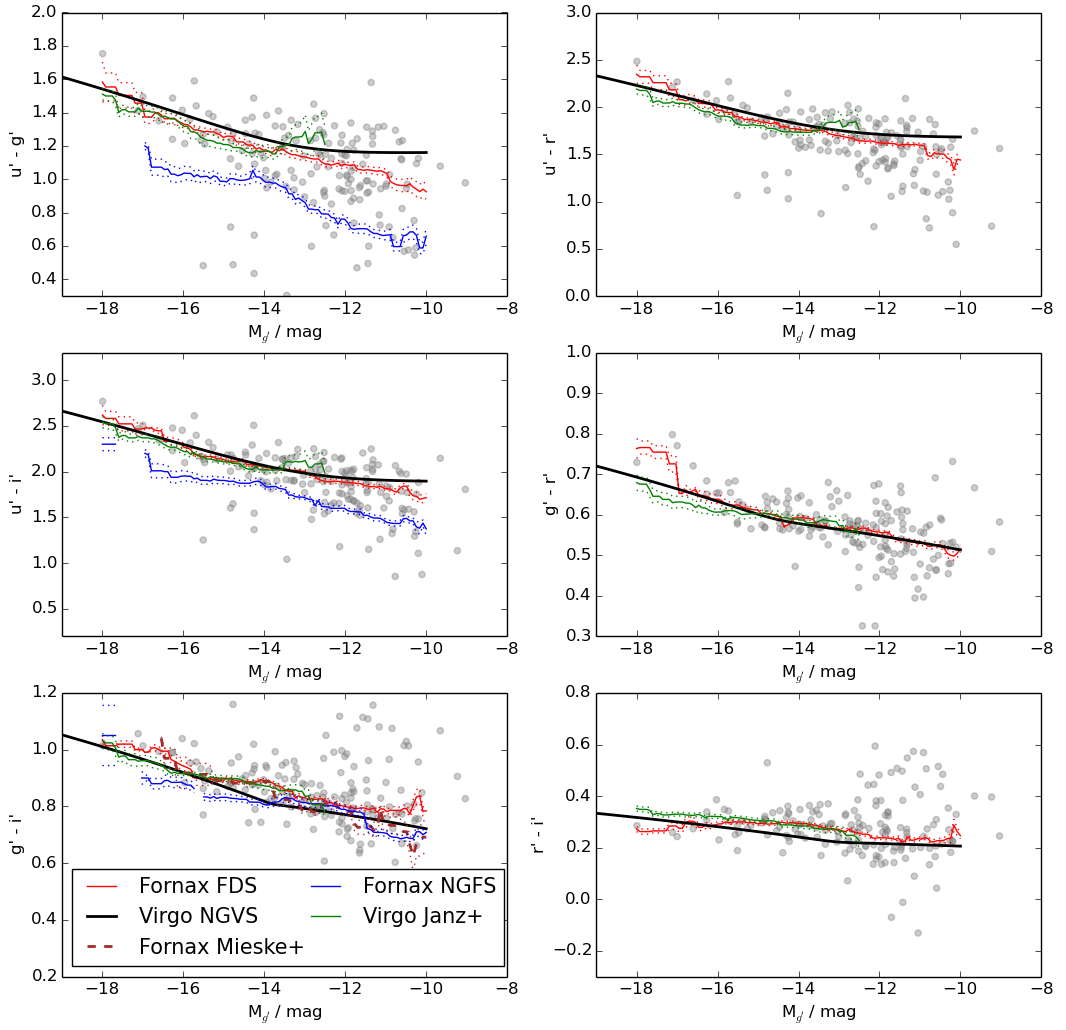}
        \caption{Color-magnitude relations of the early-type galaxies in our sample, within two core radii from the cluster center, are shown with the gray dots. The solid red lines show the running means of the colors measured with an interval of $\Delta\mathrm{M}_{g'}$=1 mag, and the red dotted lines show the uncertainty of the mean. The blue solid line, the brown dashed line, and the green solid line show similarly the samples of NGFS \citep{Eigenthaler2018}, \citet{Mieske2007}, and \citet{Janz2009}, respectively. The solid black lines show the color-magnitude relations of the Virgo early-type dwarf galaxies by \citet{Roediger2017}.}
        \label{fig:cmr_comparison}
\end{figure*}

\subsection{Parametric selection accuracy and contamination from the background objects}

In Section 8.2 above, we apply the parametric cuts to separate most of the background galaxies from the cluster galaxies. To understand the number of possible background galaxies that remain in the sample after applying the parametric selection cuts, we compared our classifications (cluster member or background galaxy) with the objects that have spectroscopic redshifts in \citet{Drinkwater2000}.

\indent In their sample there are 53 cluster galaxies that are not UCDs and 1782 background galaxies, that are also present in our list of detections. We find that only two of the 53 cluster galaxies with known redshifts are excluded from our sample. On the other hand, 194 background galaxies of the original 1782 spectroscopically confirmed background objects remain in our sample after the initial cuts. 

\indent After the initial parametric selection of cluster galaxies, we made a visual morphological classification (Section 8.3) to further exclude background galaxies that remained in our sample. We find that all the spectroscopically confirmed cluster galaxies that remained in our sample after the initial cuts, were correctly associated as cluster members according to our morphological classifications. Conversely, 
of the 194 spectroscopically confirmed background galaxies that remained after applying our cuts, eight were erroneously associated as cluster galaxies by us. We then also removed these eight galaxies from our catalog.

\indent In summary, we find that, using our method (including the initial cuts and morphological classifications of the remaining objects), among the galaxies with spectroscopic redshifts there is a 96.2$\pm$2.7\% ($\frac{53-2}{53}$) chance for a cluster galaxy to be classified as such. For a background galaxy, we get a 99.6$\pm$0.2\% ($\frac{1782-8}{1782}$) chance for it to be classified as a background galaxy.

\indent If we assume that the classification accuracy holds also for the galaxies without spectroscopic redshifts, we can estimate the number of possibly erroneously classified background galaxies using our method. However, this estimation probably overestimates the contamination, because in the low-luminosity end the cluster galaxies have low surface brightnesses and they are relatively blue, which makes it easier to separate them from the background objects. We have 14,095 galaxies in our sample, of which 564 galaxies are classified as cluster dwarf galaxies and 13,505 as background galaxies. With our accuracy of 99.6\% $\pm$ 0.2\% that corresponds to $\approx$ 30--80 false positives corresponding to $\approx$10\% of our final cluster dwarf sample identifications.

\subsection{Cluster membership classifications compared to Eigenthaler}

\citet{Eigenthaler2018} also use morphological classifications for separating cluster and background galaxies. Since they do not provide a list of background galaxies, here we inspect only the galaxies that have been classified as background galaxies by us and as cluster members by them.

\begin{figure*}
        \includegraphics[width=17cm]{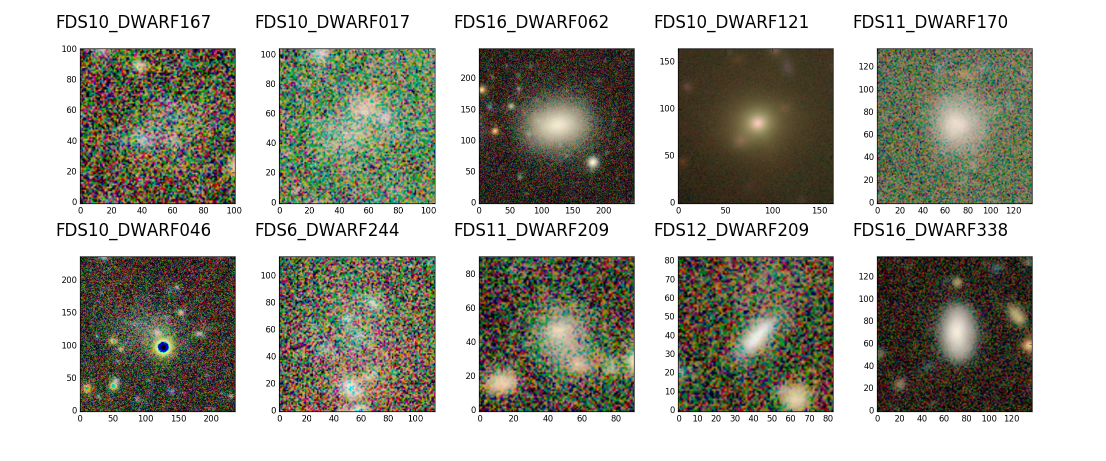}
        \caption{Post-stamp images of the galaxies classified as likely cluster members by \citet{Eigenthaler2018} but classified by us as being likely background galaxies.}
\label{fig:dif_eig_fds}
\end{figure*}

\indent Two known cluster galaxies were falsely classified as background objects by us when the parametric selection cuts were applied (see Fig. \ref{fig:selection}). These two galaxies are included in the sample of Eigenthaler. Other than that, there are ten galaxies in their sample that we classified as background galaxies according to morphology. These galaxies are shown in Fig. \ref{fig:dif_eig_fds}. As seen from Fig. \ref{fig:dif_eig_fds} these galaxies are either very small and very faint or show weak central components. From the photometric data alone it is impossible to robustly decide possible cluster membership of these objects. As so few objects of these samples were classified differently, we conclude that the cluster membership classifications of these samples are very similar.

\subsection{Comparison of our morphological classifications with FCC}

We also compared the consistency between the morphological classifications of \citet{Ferguson1989} and ours. Ferguson uses the full de Vaucouleurs' classification system instead of our simple division into early and late-type galaxies. In our division, dE's, S0's and E's are early-type galaxies whereas S(B)a/b/c/d's, ImV's, and BCD's are late-type systems. Ferguson has many galaxies that have uncertain classifications, which we have not included in our comparison.

\indent As a result, of the 190 early-type galaxies in FCC we classified
188 as early-types, and two as late-types. The two galaxies with differing classifications (FDS22DWARF244/FCC46 and FDS16DWARF000/FCC148) are both dominated by reddish spheroid component in their outer parts, but have embedded bluish disks. Their colors within the effective radii are also clearly bluer than the ones of the red sequence galaxies. These galaxies clearly have properties of both late-type and early-type galaxies. Since they have signs of recent or ongoing star formation in their central parts we classify them as late types, consistently with BCDs. We also classified all the 15 late-types in the classification of Ferguson as late-types.

\section{Summary and conclusions}

Optical data covering 26 deg$^2$ area in the Fornax cluster and the Fornax A subgroup were obtained using the OmegaCAM instrument, attached to the VLT Survey Telescope, located at Cerro Paranal in Chile. A new dwarf galaxy catalog was created, which goes three magnitudes deeper than the previous most complete Fornax cluster catalog (Ferguson 1989). The 26 deg$^2$ area is fully covered in g', r', and i' -bands, and a 20 deg$^2$ area of the main cluster was also covered in u'-band. In this paper we present the observations, data reduction, and the quality assessment of the data. We also present a galaxy detection algorithm and photometric pipeline, which were tested and used to create the dwarf galaxy catalog. We first used colors, concentration, and surface brightness to select the likely Fornax cluster galaxies. We classified the selected galaxies into different classes using visual and parametric morphological classifications. Our main results are:

\begin{itemize}

\item We generated a new catalog of the resolved galaxies for the 26 deg$^2$ area in the Fornax cluster. The catalog includes 14,095 galaxies. It reaches 50\% completeness limits at major axis length $a$ > 2 arcsec,  total r'-band apparent magnitude 13 mag < m$_{r'}$ < 21.1 mag (corresponding to -18.5 mag < M$_{r'}$ < -10.5 mag at the distance of the Fornax cluster), and at the mean effective surface brightness $\bar{\mu}_{r'}$ < 26 mag arcsec$^{-2}$.   
\vskip 0.3cm

\item We used GALFIT to fit all the galaxies in the catalog using either a single S\'ersic function or a S\'ersic function with an additional {\it PSF} component as a nucleus. We used mock galaxies to define the uncertainties in the parameters obtained with these models. The photometric parameters of all 14,095 galaxies are given in electronic form.
\vskip 0.3cm

\item We used cuts in the color-magnitude, luminosity-surface brightness, and luminosity-concentration relations to separate cluster galaxies from the background objects. We then inspected the selected likely cluster galaxies, and classified them according to their visual morphology and morphological parameters. As a result we classify 13,505 galaxies as likely background galaxies and 564 dwarf galaxies as likely cluster members. Of the cluster members 470 galaxies are early-type, and 94 late-type systems. Additionally there are 22 cluster galaxies that are not dwarfs and are not therefore included int our catalogs. 
\vskip 0.3cm

\item We compared the galaxies in our catalog with literature, and found that the cluster membership and morphological classifications are consistent with the previous works in the Fornax cluster. 10 galaxies of the FCC within the FDS area are missing from our catalog due to known selection effects, but in general our catalog extends three magnitudes deeper than the FCC. 
\vskip 0.3cm

\item Extrapolating from the spectroscopic redshift samples in the bright luminosity regime of our sample, we estimate that we are able to obtain the correct separation between the cluster members and background galaxies with a 0.4\% probability of assigning a true background galaxy to the cluster sample and 96.2\% for real cluster galaxies. This implies a background galaxy contamination rate of $\approx$ 10\% in our final cluster dwarf galaxy sample. 
\vskip 0.3cm

\item We compared our photometric parameters with the works of \citet{Ferguson1989}, \citet{Eigenthaler2018}, and \citet{Mieske2007} and find good agreement for the effective radii and magnitudes, with the exception of comparison with Eigenthaler et al, where we find an offset of 0.3 mag between their and our g'-band magnitudes. 
\vskip 0.3cm

\item We assessed the quality of our galaxy colors by comparing them with other samples in the Fornax and Virgo clusters. We showed that our colors are in a good agreement with other previous works. We report and offset between the u'-g' and u'-i' colors of ours and the ones of \citet{Eigenthaler2018}.
\vskip 0.3cm

\end{itemize}

\indent In summary, together with the catalogs of \citet{Iodice2018} containing massive Fornax cluster galaxies, and Venhola et al. (2017, and in prep.) containing the low surface brightness galaxies, our catalog comprises a complete set of resolved galaxies in the Fornax cluster.

{\bf Acknowledgements : } A.V.\ would like to thank the Vilho, Yrjö, and Kalle Väisälä Foundation of the Finnish Academy of Science and Letters for the financial support during the writing of this paper. GvdV acknowledges funding from the European Research Council (ERC) under the European Union's Horizon 2020 research and innovation program under grant agreement No 724857 (Consolidator Grant ArcheoDyn). R.F.P., T.L., E.L., H.S., E.I., and J.J.\ acknowledge financial support from the European Union's Horizon 2020 research and innovation program under the Marie Sk\l{}odowska-Curie grant agreement No.\ 721463 to the SUNDIAL ITN network. H.S.\ and E.L.\ are also supported by the Academy of Finland grant n:o 297738. C.W.\ is supported by the Deutsche Forschungsgemeinschaft (DFG, German
Research Foundation) through project 394551440.

\bibliographystyle{aa}
\bibliography{dwarf_paper}

\begin{thebibliography}{78}
\expandafter\ifx\csname natexlab\endcsname\relax\def\natexlab#1{#1}\fi

\bibitem[{{Alam} {et~al.}(2015){Alam}, {Albareti}, {Allende Prieto}, {Anders},
  {Anderson}, {Anderton}, {Andrews}, {Armengaud}, {Aubourg}, {Bailey}, \&
  et~al.}]{Alam2015}
{Alam}, S., {Albareti}, F.~D., {Allende Prieto}, C., {et~al.} 2015, \apjs, 219,
  12

\bibitem[{{Bender} {et~al.}(2005){Bender}, {Kormendy}, {Bower}, {Green},
  {Thomas}, {Danks}, {Gull}, {Hutchings}, {Joseph}, {Kaiser}, {Lauer},
  {Nelson}, {Richstone}, {Weistrop}, \& {Woodgate}}]{Bender2005}
{Bender}, R., {Kormendy}, J., {Bower}, G., {et~al.} 2005, \apj, 631, 280

\bibitem[{{Bertin}(2006)}]{Bertin2006}
{Bertin}, E. 2006, in Astronomical Society of the Pacific Conference Series,
  Vol. 351, Astronomical Data Analysis Software and Systems XV, ed.
  C.~{Gabriel}, C.~{Arviset}, D.~{Ponz}, \& S.~{Enrique}, 112

\bibitem[{{Bertin}(2010)}]{Bertin2010}
{Bertin}, E. 2010, {SWarp: Resampling and Co-adding FITS Images Together},
  Astrophysics Source Code Library

\bibitem[{{Bilicki} {et~al.}(2018){Bilicki}, {Hoekstra}, {Brown}, {Amaro},
  {Blake}, {Cavuoti}, {de Jong}, {Georgiou}, {Hildebrandt}, {Wolf}, {Amon},
  {Brescia}, {Brough}, {Costa-Duarte}, {Erben}, {Glazebrook}, {Grado},
  {Heymans}, {Jarrett}, {Joudaki}, {Kuijken}, {Longo}, {Napolitano},
  {Parkinson}, {Vellucci}, {Kleijn}, \& {Wang}}]{Bilicki2018}
{Bilicki}, M., {Hoekstra}, H., {Brown}, M.~J.~I., {et~al.} 2018, \aap, 616, A69

\bibitem[{{Binggeli} {et~al.}(1985){Binggeli}, {Sandage}, \&
  {Tammann}}]{Binggeli1985}
{Binggeli}, B., {Sandage}, A., \& {Tammann}, G.~A. 1985, \aj, 90, 1681

\bibitem[{{Binggeli} {et~al.}(1984){Binggeli}, {Sandage}, \&
  {Tarenghi}}]{Binggeli1984}
{Binggeli}, B., {Sandage}, A., \& {Tarenghi}, M. 1984, \aj, 89, 64

\bibitem[{{Blakeslee} {et~al.}(2006){Blakeslee}, {Holden}, {Franx}, {Rosati},
  {Bouwens}, {Demarco}, {Ford}, {Homeier}, {Illingworth}, {Jee}, {Mei},
  {Menanteau}, {Meurer}, {Postman}, \& {Tran}}]{Blakeslee2006}
{Blakeslee}, J.~P., {Holden}, B.~P., {Franx}, M., {et~al.} 2006, \apj, 644, 30

\bibitem[{{Blakeslee} {et~al.}(2009){Blakeslee}, {Jord{\'a}n}, {Mei},
  {C{\^o}t{\'e}}, {Ferrarese}, {Infante}, {Peng}, {Tonry}, \&
  {West}}]{Blakeslee2009}
{Blakeslee}, J.~P., {Jord{\'a}n}, A., {Mei}, S., {et~al.} 2009, \apj, 694, 556

\bibitem[{{Bonnarel} {et~al.}(2000){Bonnarel}, {Fernique}, {Bienaym{\'e}},
  {Egret}, {Genova}, {Louys}, {Ochsenbein}, {Wenger}, \&
  {Bartlett}}]{Bonnarel2000}
{Bonnarel}, F., {Fernique}, P., {Bienaym{\'e}}, O., {et~al.} 2000, \aaps, 143,
  33

\bibitem[{{Boylan-Kolchin} {et~al.}(2009){Boylan-Kolchin}, {Springel}, {White},
  {Jenkins}, \& {Lemson}}]{BoylanKolchin2009}
{Boylan-Kolchin}, M., {Springel}, V., {White}, S.~D.~M., {Jenkins}, A., \&
  {Lemson}, G. 2009, \mnras, 398, 1150

\bibitem[{{Brodie} {et~al.}(2011){Brodie}, {Romanowsky}, {Strader}, \&
  {Forbes}}]{Brodie2011}
{Brodie}, J.~P., {Romanowsky}, A.~J., {Strader}, J., \& {Forbes}, D.~A. 2011,
  \aj, 142, 199

\bibitem[{{Caldwell} \& {Bothun}(1987)}]{Caldwell1987}
{Caldwell}, N. \& {Bothun}, G.~D. 1987, \aj, 94, 1126

\bibitem[{{Cantiello} {et~al.}(2018){Cantiello}, {D'Abrusco}, {Spavone},
  {Paolillo}, {Capaccioli}, {Limatola}, {Grado}, {Iodice}, {Raimondo},
  {Napolitano}, {Blakeslee}, {Brocato}, {Forbes}, {Hilker}, {Mieske},
  {Peletier}, {van de Ven}, \& {Schipani}}]{Cantiello2018}
{Cantiello}, M., {D'Abrusco}, R., {Spavone}, M., {et~al.} 2018, \aap, 611, A93

\bibitem[{{Capaccioli} {et~al.}(2015){Capaccioli}, {Spavone}, {Grado},
  {Iodice}, {Limatola}, {Napolitano}, {Cantiello}, {Paolillo}, {Romanowsky},
  {Forbes}, {Puzia}, {Raimondo}, \& {Schipani}}]{Capaccioli2015}
{Capaccioli}, M., {Spavone}, M., {Grado}, A., {et~al.} 2015, \aap, 581, A10

\bibitem[{{Conselice}(2014)}]{Conselice2014}
{Conselice}, C.~J. 2014, \araa, 52, 291

\bibitem[{{Conselice} {et~al.}(2003){Conselice}, {Gallagher}, \&
  {Wyse}}]{Conselice2003}
{Conselice}, C.~J., {Gallagher}, III, J.~S., \& {Wyse}, R.~F.~G. 2003, \aj,
  125, 66

\bibitem[{{C{\^o}t{\'e}} {et~al.}(2006){C{\^o}t{\'e}}, {Piatek}, {Ferrarese},
  {Jord{\'a}n}, {Merritt}, {Peng}, {Ha{\c s}egan}, {Blakeslee}, {Mei}, {West},
  {Milosavljevi{\'c}}, \& {Tonry}}]{Cote2006}
{C{\^o}t{\'e}}, P., {Piatek}, S., {Ferrarese}, L., {et~al.} 2006, \apjs, 165,
  57

\bibitem[{{Cutri} {et~al.}(2003){Cutri}, {Skrutskie}, {van Dyk}, {Beichman},
  {Carpenter}, {Chester}, {Cambresy}, {Evans}, {Fowler}, {Gizis}, {Howard},
  {Huchra}, {Jarrett}, {Kopan}, {Kirkpatrick}, {Light}, {Marsh}, {McCallon},
  {Schneider}, {Stiening}, {Sykes}, {Weinberg}, {Wheaton}, {Wheelock}, \&
  {Zacarias}}]{Cutri2003}
{Cutri}, R.~M., {Skrutskie}, M.~F., {van Dyk}, S., {et~al.} 2003, VizieR Online
  Data Catalog, 2246

\bibitem[{{D'Abrusco} {et~al.}(2016){D'Abrusco}, {Cantiello}, {Paolillo},
  {Pota}, {Napolitano}, {Limatola}, {Spavone}, {Grado}, {Iodice}, {Capaccioli},
  {Peletier}, {Longo}, {Hilker}, {Mieske}, {Grebel}, {Lisker}, {Wittmann}, {van
  de Ven}, {Schipani}, \& {Fabbiano}}]{DAbrusco2016}
{D'Abrusco}, R., {Cantiello}, M., {Paolillo}, M., {et~al.} 2016, \apjl, 819,
  L31

\bibitem[{{Dressler}(1980)}]{Dressler1980}
{Dressler}, A. 1980, \apj, 236, 351

\bibitem[{{Dressler} {et~al.}(2011){Dressler}, {Bigelow}, {Hare}, {Sutin},
  {Thompson}, {Burley}, {Epps}, {Oemler}, {Bagish}, {Birk}, {Clardy},
  {Gunnels}, {Kelson}, {Shectman}, \& {Osip}}]{Dressler2011}
{Dressler}, A., {Bigelow}, B., {Hare}, T., {et~al.} 2011, \pasp, 123, 288

\bibitem[{{Drinkwater} {et~al.}(2001){Drinkwater}, {Gregg}, \&
  {Colless}}]{Drinkwater2001}
{Drinkwater}, M.~J., {Gregg}, M.~D., \& {Colless}, M. 2001, \apjl, 548, L139

\bibitem[{{Drinkwater} {et~al.}(1999){Drinkwater}, {Phillipps}, \&
  {Jones}}]{Drinkwater1999}
{Drinkwater}, M.~J., {Phillipps}, S., \& {Jones}, J.~B. 1999, in Astronomical
  Society of the Pacific Conference Series, Vol. 170, The Low Surface
  Brightness Universe, ed. J.~I. {Davies}, C.~{Impey}, \& S.~{Phillips}, 120

\bibitem[{{Drinkwater} {et~al.}(2000){Drinkwater}, {Phillipps}, {Jones},
  {Gregg}, {Deady}, {Davies}, {Parker}, {Sadler}, \& {Smith}}]{Drinkwater2000}
{Drinkwater}, M.~J., {Phillipps}, S., {Jones}, J.~B., {et~al.} 2000, \aap, 355,
  900

\bibitem[{{Eigenthaler} {et~al.}(2018){Eigenthaler}, {Puzia}, {Taylor},
  {Ordenes-Brice{\~n}o}, {Mu{\~n}oz}, {Ribbeck}, {Alamo-Mart{\'{\i}}nez},
  {Zhang}, {{\'A}ngel}, {Capaccioli}, {C{\^o}t{\'e}}, {Ferrarese}, {Galaz},
  {Grebel}, {Hempel}, {Hilker}, {Lan{\c c}on}, {Mieske}, {Miller}, {Paolillo},
  {Powalka}, {Richtler}, {Roediger}, {Rong}, {S{\'a}nchez-Janssen}, \&
  {Spengler}}]{Eigenthaler2018}
{Eigenthaler}, P., {Puzia}, T.~H., {Taylor}, M.~A., {et~al.} 2018, \apj, 855,
  142

\bibitem[{{Ferguson}(1989)}]{Ferguson1989}
{Ferguson}, H.~C. 1989, \aj, 98, 367

\bibitem[{{Ferrarese} {et~al.}(2012){Ferrarese}, {C{\^o}t{\'e}}, {Cuillandre},
  {Gwyn}, {Peng}, {MacArthur}, {Duc}, {Boselli}, {Mei}, {Erben}, {McConnachie},
  {Durrell}, {Mihos}, {Jord{\'a}n}, {Lan{\c c}on}, {Puzia}, {Emsellem},
  {Balogh}, {Blakeslee}, {van Waerbeke}, {Gavazzi}, {Vollmer}, {Kavelaars},
  {Woods}, {Ball}, {Boissier}, {Courteau}, {Ferriere}, {Gavazzi},
  {Hildebrandt}, {Hudelot}, {Huertas-Company}, {Liu}, {McLaughlin}, {Mellier},
  {Milkeraitis}, {Schade}, {Balkowski}, {Bournaud}, {Carlberg}, {Chapman},
  {Hoekstra}, {Peng}, {Sawicki}, {Simard}, {Taylor}, {Tully}, {van Driel},
  {Wilson}, {Burdullis}, {Mahoney}, \& {Manset}}]{Ferrarese2012}
{Ferrarese}, L., {C{\^o}t{\'e}}, P., {Cuillandre}, J.-C., {et~al.} 2012, \apjs,
  200, 4

\bibitem[{{Hamraz} {et~al.}(2018){Hamraz}, {Peletier}, {Khosroshahi},
  {Valentijn}, \& {Venhola}}]{Hamraz2018}
{Hamraz}, E., {Peletier}, R.~F., {Khosroshahi}, H.~G., {Valentijn}, E.~A.
  amd~{den Brok}, M., \& {Venhola}, A. 2018, \mnras

\bibitem[{{Henden} {et~al.}(2012){Henden}, {Levine}, {Terrell}, {Smith}, \&
  {Welch}}]{Henden2012}
{Henden}, A.~A., {Levine}, S.~E., {Terrell}, D., {Smith}, T.~C., \& {Welch}, D.
  2012, Journal of the American Association of Variable Star Observers
  (JAAVSO), 40, 430

\bibitem[{{Hilker} {et~al.}(2003){Hilker}, {Mieske}, \& {Infante}}]{Hilker2003}
{Hilker}, M., {Mieske}, S., \& {Infante}, L. 2003, \aap, 397, L9

\bibitem[{{Hoyos} {et~al.}(2011){Hoyos}, {den Brok}, {Verdoes Kleijn},
  {Carter}, {Balcells}, {Guzm{\'a}n}, {Peletier}, {Ferguson}, {Goudfrooij},
  {Graham}, {Hammer}, {Karick}, {Lucey}, {Matkovi{\'c}}, {Merritt}, {Mouhcine},
  \& {Valentijn}}]{Hoyos2011}
{Hoyos}, C., {den Brok}, M., {Verdoes Kleijn}, G., {et~al.} 2011, \mnras, 411,
  2439

\bibitem[{{Huchra} {et~al.}(2012){Huchra}, {Macri}, {Masters}, {Jarrett},
  {Berlind}, {Calkins}, {Crook}, {Cutri}, {Erdo{\v g}du}, {Falco}, {George},
  {Hutcheson}, {Lahav}, {Mader}, {Mink}, {Martimbeau}, {Schneider},
  {Skrutskie}, {Tokarz}, \& {Westover}}]{Huchra2012}
{Huchra}, J.~P., {Macri}, L.~M., {Masters}, K.~L., {et~al.} 2012, \apjs, 199,
  26

\bibitem[{{Iodice} {et~al.}(2016){Iodice}, {Capaccioli}, {Grado}, {Limatola},
  {Spavone}, {Napolitano}, {Paolillo}, {Peletier}, {Cantiello}, {Lisker},
  {Wittmann}, {Venhola}, {Hilker}, {D'Abrusco}, {Pota}, \&
  {Schipani}}]{Iodice2016}
{Iodice}, E., {Capaccioli}, M., {Grado}, A., {et~al.} 2016, \apj, 820, 42

\bibitem[{{Iodice} {et~al.}(2017{\natexlab{a}}){Iodice}, {Spavone},
  {Cantiello}, {D'Abrusco}, {Capaccioli}, {Hilker}, {Mieske}, {Napolitano},
  {Peletier}, {Limatola}, {Grado}, {Venhola}, {Paolillo}, {Van de Ven}, \&
  {Schipani}}]{Iodice2017b}
{Iodice}, E., {Spavone}, M., {Cantiello}, M., {et~al.} 2017{\natexlab{a}},
  \apj, 851, 75

\bibitem[{{Iodice} {et~al.}(2017{\natexlab{b}}){Iodice}, {Spavone},
  {Capaccioli}, {Peletier}, {Richtler}, {Hilker}, {Mieske}, {Limatola},
  {Grado}, {Napolitano}, {Cantiello}, {D'Abrusco}, {Paolillo}, {Venhola},
  {Lisker}, {Van de Ven}, {Falcon-Barroso}, \& {Schipani}}]{Iodice2017a}
{Iodice}, E., {Spavone}, M., {Capaccioli}, M., {et~al.} 2017{\natexlab{b}},
  \apj, 839, 21

\bibitem[{{Iodice} {et~al.}(2018){Iodice}, {Spavone}, {Capaccioli}, {Peletier},
  {van de Ven}, {Napolitano}, {Hilker}, {Mieske}, {Limatola}, {Grado},
  {Venhola}, {Cantiello}, {Paolillo}, {Falcon-Barroso}, {D’Abrusco}, \&
  {Schipani}}]{Iodice2018}
{Iodice}, E., {Spavone}, M., {Capaccioli}, M., {et~al.} 2018, \aap, submitted

\bibitem[{{Ivezi{\'c}} {et~al.}(2004){Ivezi{\'c}}, {Lupton}, {Schlegel},
  {Boroski}, {Adelman-McCarthy}, {Yanny}, {Kent}, {Stoughton}, {Finkbeiner},
  {Padmanabhan}, {Rockosi}, {Gunn}, {Knapp}, {Strauss}, {Richards},
  {Eisenstein}, {Nicinski}, {Kleinman}, {Krzesinski}, {Newman}, {Snedden},
  {Thakar}, {Szalay}, {Munn}, {Smith}, {Tucker}, \& {Lee}}]{Ivezic2004}
{Ivezi{\'c}}, {\v Z}., {Lupton}, R.~H., {Schlegel}, D., {et~al.} 2004,
  Astronomische Nachrichten, 325, 583

\bibitem[{{Jaff{\'e}} {et~al.}(2018){Jaff{\'e}}, {Poggianti}, {Moretti},
  {Gullieuszik}, {Smith}, {Vulcani}, {Fasano}, {Fritz}, {Tonnesen}, {Bettoni},
  {Hau}, {Biviano}, {Bellhouse}, \& {McGee}}]{Jaffe2018}
{Jaff{\'e}}, Y.~L., {Poggianti}, B.~M., {Moretti}, A., {et~al.} 2018, \mnras,
  476, 4753

\bibitem[{{Janz} {et~al.}(2014){Janz}, {Laurikainen}, {Lisker}, {Salo},
  {Peletier}, {Niemi}, {Toloba}, {Hensler}, {Falc{\'o}n-Barroso}, {Boselli},
  {den Brok}, {Hansson}, {Meyer}, {Ry{\'s}}, \& {Paudel}}]{Janz2014}
{Janz}, J., {Laurikainen}, E., {Lisker}, T., {et~al.} 2014, \apj, 786, 105

\bibitem[{{Janz} \& {Lisker}(2009)}]{Janz2009}
{Janz}, J. \& {Lisker}, T. 2009, \apjl, 696, L102

\bibitem[{{Jord{\'a}n} {et~al.}(2007){Jord{\'a}n}, {Blakeslee}, {C{\^o}t{\'e}},
  {Ferrarese}, {Infante}, {Mei}, {Merritt}, {Peng}, {Tonry}, \&
  {West}}]{Jordan2007}
{Jord{\'a}n}, A., {Blakeslee}, J.~P., {C{\^o}t{\'e}}, P., {et~al.} 2007, \apjs,
  169, 213

\bibitem[{{Jordi} {et~al.}(2006){Jordi}, {Grebel}, \& {Ammon}}]{Jordi2006}
{Jordi}, K., {Grebel}, E.~K., \& {Ammon}, K. 2006, \aap, 460, 339

\bibitem[{{Kron}(1980)}]{Kron1980}
{Kron}, R.~G. 1980, \apjs, 43, 305

\bibitem[{{Kuijken} {et~al.}(2002){Kuijken}, {Bender}, {Cappellaro},
  {Muschielok}, {Baruffolo}, {Cascone}, {Iwert}, {Mitsch}, {Nicklas},
  {Valentijn}, {Baade}, {Begeman}, {Bortolussi}, {Boxhoorn}, {Christen},
  {Deul}, {Geimer}, {Greggio}, {Harke}, {H{\"a}fner}, {Hess}, {Hess}, {Hopp},
  {Ilijevski}, {Klink}, {Kravcar}, {Lizon}, {Magagna}, {M{\"u}ller},
  {Niemeczek}, {de Pizzol}, {Poschmann}, {Reif}, {Rengelink}, {Reyes},
  {Silber}, \& {Wellem}}]{Kuijken2002}
{Kuijken}, K., {Bender}, R., {Cappellaro}, E., {et~al.} 2002, The Messenger,
  110, 15

\bibitem[{{Landolt}(1992)}]{Landolt1992}
{Landolt}, A.~U. 1992, \aj, 104, 340

\bibitem[{{Lauer} {et~al.}(1995){Lauer}, {Ajhar}, {Byun}, {Dressler}, {Faber},
  {Grillmair}, {Kormendy}, {Richstone}, \& {Tremaine}}]{Lauer1995}
{Lauer}, T.~R., {Ajhar}, E.~A., {Byun}, Y.-I., {et~al.} 1995, \aj, 110, 2622

\bibitem[{{Lisker} {et~al.}(2006){Lisker}, {Glatt}, {Westera}, \&
  {Grebel}}]{Lisker2006}
{Lisker}, T., {Glatt}, K., {Westera}, P., \& {Grebel}, E.~K. 2006, \aj, 132,
  2432

\bibitem[{{Mieske} {et~al.}(2007){Mieske}, {Hilker}, {Infante}, \& {Mendes de
  Oliveira}}]{Mieske2007}
{Mieske}, S., {Hilker}, M., {Infante}, L., \& {Mendes de Oliveira}, C. 2007,
  \aap, 463, 503

\bibitem[{{Misgeld} \& {Hilker}(2011)}]{Misgeld2011}
{Misgeld}, I. \& {Hilker}, M. 2011, \mnras, 414, 3699

\bibitem[{{Misgeld} {et~al.}(2009){Misgeld}, {Hilker}, \&
  {Mieske}}]{Misgeld2009}
{Misgeld}, I., {Hilker}, M., \& {Mieske}, S. 2009, \aap, 496, 683

\bibitem[{{Moffat}(1969)}]{Moffat1969}
{Moffat}, A.~F.~J. 1969, \aap, 3, 455

\bibitem[{{Mu{\~n}oz} {et~al.}(2015){Mu{\~n}oz}, {Eigenthaler}, {Puzia},
  {Taylor}, {Ordenes-Brice{\~n}o}, {Alamo-Mart{\'{\i}}nez}, {Ribbeck},
  {{\'A}ngel}, {Capaccioli}, {C{\^o}t{\'e}}, {Ferrarese}, {Galaz}, {Hempel},
  {Hilker}, {Jord{\'a}n}, {Lan{\c c}on}, {Mieske}, {Paolillo}, {Richtler},
  {S{\'a}nchez-Janssen}, \& {Zhang}}]{Munoz2015}
{Mu{\~n}oz}, R.~P., {Eigenthaler}, P., {Puzia}, T.~H., {et~al.} 2015, \apjl,
  813, L15

\bibitem[{{Nasonova} {et~al.}(2011){Nasonova}, {de Freitas Pacheco}, \&
  {Karachentsev}}]{Nasonova2011}
{Nasonova}, O.~G., {de Freitas Pacheco}, J.~A., \& {Karachentsev}, I.~D. 2011,
  \aap, 532, A104

\bibitem[{{Ordenes-Brice{\~n}o} {et~al.}(2018){Ordenes-Brice{\~n}o},
  {Eigenthaler}, {Taylor}, {Puzia}, {Alamo-Mart{\'{\i}}nez}, {Ribbeck},
  {Mu{\~n}oz}, {Zhang}, {Grebel}, {{\'A}ngel}, {C{\^o}t{\'e}}, {Ferrarese},
  {Hilker}, {Lan{\c c}on}, {Mieske}, {Miller}, {Rong}, \&
  {S{\'a}nchez-Janssen}}]{Ordenes-Briceno2018}
{Ordenes-Brice{\~n}o}, Y., {Eigenthaler}, P., {Taylor}, M.~A., {et~al.} 2018,
  \apj, 859, 52

\bibitem[{{Paolillo} {et~al.}(2002){Paolillo}, {Fabbiano}, {Peres}, \&
  {Kim}}]{Paolillo2002}
{Paolillo}, M., {Fabbiano}, G., {Peres}, G., \& {Kim}, D.-W. 2002, \apj, 565,
  883

\bibitem[{{Peng} {et~al.}(2002){Peng}, {Ho}, {Impey}, \& {Rix}}]{Peng2002}
{Peng}, C.~Y., {Ho}, L.~C., {Impey}, C.~D., \& {Rix}, H.-W. 2002, \aj, 124, 266

\bibitem[{{Peng} {et~al.}(2012){Peng}, {Lilly}, {Renzini}, \&
  {Carollo}}]{Peng2012}
{Peng}, Y.-j., {Lilly}, S.~J., {Renzini}, A., \& {Carollo}, M. 2012, \apj, 757,
  4

\bibitem[{{Peng} {et~al.}(2014){Peng}, {Lilly}, {Renzini}, \&
  {Carollo}}]{Peng2014}
{Peng}, Y.-j., {Lilly}, S.~J., {Renzini}, A., \& {Carollo}, M. 2014, \apj, 790,
  95

\bibitem[{{Penny} \& {Conselice}(2008)}]{Penny2008}
{Penny}, S.~J. \& {Conselice}, C.~J. 2008, \mnras, 383, 247

\bibitem[{{Pillepich} {et~al.}(2018){Pillepich}, {Springel}, {Nelson}, {Genel},
  {Naiman}, {Pakmor}, {Hernquist}, {Torrey}, {Vogelsberger}, {Weinberger}, \&
  {Marinacci}}]{Pillepich2018}
{Pillepich}, A., {Springel}, V., {Nelson}, D., {et~al.} 2018, \mnras, 473, 4077

\bibitem[{{Pota} {et~al.}(2018){Pota}, {Napolitano}, {Hilker}, {Spavone},
  {Schulz}, {Cantiello}, {Tortora}, {Iodice}, {Paolillo}, {D'Abrusco},
  {Capaccioli}, {Puzia}, {Peletier}, {Romanowsky}, {van de Ven}, {Spiniello},
  {Norris}, {Lisker}, {Munoz}, {Schipani}, {Eigenthaler}, {Taylor},
  {S{\'a}nchez-Janssen}, \& {Ordenes-Brice{\~n}o}}]{Pota2018}
{Pota}, V., {Napolitano}, N.~R., {Hilker}, M., {et~al.} 2018, \mnras
  [\eprint[arXiv]{1803.03275}]

\bibitem[{{Price} {et~al.}(2009){Price}, {Phillipps}, {Huxor}, {Trentham},
  {Ferguson}, {Marzke}, {Hornschemeier}, {Goudfrooij}, {Hammer}, {Tully},
  {Chiboucas}, {Smith}, {Carter}, {Merritt}, {Balcells}, {Erwin}, \&
  {Puzia}}]{Price2009}
{Price}, J., {Phillipps}, S., {Huxor}, A., {et~al.} 2009, \mnras, 397, 1816

\bibitem[{{Roediger} {et~al.}(2017){Roediger}, {Ferrarese}, {C{\^o}t{\'e}},
  {MacArthur}, {S{\'a}nchez-Janssen}, {Blakeslee}, {Peng}, {Liu}, {Munoz},
  {Cuillandre}, {Gwyn}, {Mei}, {Boissier}, {Boselli}, {Cantiello}, {Courteau},
  {Duc}, {Lan{\c c}on}, {Mihos}, {Puzia}, {Taylor}, {Durrell}, {Toloba},
  {Guhathakurta}, \& {Zhang}}]{Roediger2017}
{Roediger}, J.~C., {Ferrarese}, L., {C{\^o}t{\'e}}, P., {et~al.} 2017, \apj,
  836, 120

\bibitem[{{Salo} {et~al.}(2015){Salo}, {Laurikainen}, {Laine}, {Comer{\'o}n},
  {Gadotti}, {Buta}, {Sheth}, {Zaritsky}, {Ho}, {Knapen}, {Athanassoula},
  {Bosma}, {Laine}, {Cisternas}, {Kim}, {Mu{\~n}oz-Mateos}, {Regan}, {Hinz},
  {Gil de Paz}, {Menendez-Delmestre}, {Mizusawa}, {Erroz-Ferrer}, {Meidt}, \&
  {Querejeta}}]{Salo2015}
{Salo}, H., {Laurikainen}, E., {Laine}, J., {et~al.} 2015, \apjs, 219, 4

\bibitem[{{Sandin}(2014)}]{Sandin2014}
{Sandin}, C. 2014, \aap, 567, A97

\bibitem[{{Schipani} {et~al.}(2012){Schipani}, {Capaccioli}, {Arcidiacono},
  {Argomedo}, {Dall'Ora}, {D'Orsi}, {Farinato}, {Magrin}, {Marty}, {Ragazzoni},
  \& {Umbriaco}}]{Schipani2012}
{Schipani}, P., {Capaccioli}, M., {Arcidiacono}, C., {et~al.} 2012, in
  \procspie, Vol. 8444, Ground-based and Airborne Telescopes IV, 84441C

\bibitem[{{Schlafly} \& {Finkbeiner}(2011)}]{Schlafly2011}
{Schlafly}, E.~F. \& {Finkbeiner}, D.~P. 2011, \apj, 737, 103

\bibitem[{{Spavone} {et~al.}(2017){Spavone}, {Capaccioli}, {Napolitano},
  {Iodice}, {Grado}, {Limatola}, {Cooper}, {Cantiello}, {Forbes}, {Paolillo},
  \& {Schipani}}]{Spavone2017}
{Spavone}, M., {Capaccioli}, M., {Napolitano}, N.~R., {et~al.} 2017, \aap, 603,
  A38

\bibitem[{{Spiniello} {et~al.}(2018){Spiniello}, {Napolitano}, {Arnaboldi},
  {Tortora}, {Coccato}, {Capaccioli}, {Gerhard}, {Iodice}, {Spavone},
  {Cantiello}, {Peletier}, {Paolillo}, \& {Schipani}}]{Spiniello2018}
{Spiniello}, C., {Napolitano}, N.~R., {Arnaboldi}, M., {et~al.} 2018, \mnras,
  477, 1880

\bibitem[{{Sutherland} {et~al.}(2015){Sutherland}, {Emerson}, {Dalton},
  {Atad-Ettedgui}, {Beard}, {Bennett}, {Bezawada}, {Born}, {Caldwell}, {Clark},
  {Craig}, {Henry}, {Jeffers}, {Little}, {McPherson}, {Murray}, {Stewart},
  {Stobie}, {Terrett}, {Ward}, {Whalley}, \& {Woodhouse}}]{Sutherland2015}
{Sutherland}, W., {Emerson}, J., {Dalton}, G., {et~al.} 2015, \aap, 575, A25

\bibitem[{{Trentham} \& {Tully}(2009)}]{Trentham2009}
{Trentham}, N. \& {Tully}, R.~B. 2009, \mnras, 398, 722

\bibitem[{{Trujillo} {et~al.}(2001){Trujillo}, {Aguerri}, {Cepa}, \&
  {Guti{\'e}rrez}}]{Trujillo2001}
{Trujillo}, I., {Aguerri}, J.~A.~L., {Cepa}, J., \& {Guti{\'e}rrez}, C.~M.
  2001, \mnras, 328, 977

\bibitem[{{Vandame}(2001)}]{Vandame2001}
{Vandame}, B. 2001, in Mining the Sky, ed. A.~J. {Banday}, S.~{Zaroubi}, \&
  M.~{Bartelmann}, 595

\bibitem[{{Venhola} {et~al.}(2017){Venhola}, {Peletier}, {Laurikainen}, {Salo},
  {Lisker}, {Iodice}, {Capaccioli}, {Verdois Kleijn}, {Valentijn}, {Mieske},
  {Hilker}, {Wittmann}, {van de Ven}, {Grado}, {Spavone}, {Cantiello},
  {Napolitano}, {Paolillo}, \& {Falc{\'o}n-Barroso}}]{Venhola2017}
{Venhola}, A., {Peletier}, R., {Laurikainen}, E., {et~al.} 2017, \aap, 608,
  A142

\bibitem[{{Verdoes Kleijn} {et~al.}(2013){Verdoes Kleijn}, {Kuijken},
  {Valentijn}, {Boxhoorn}, {Begeman}, {Deul}, {Helmich}, \&
  {Rengelink}}]{Verdoes-Kleijn2013}
{Verdoes Kleijn}, G.~A., {Kuijken}, K.~H., {Valentijn}, E.~A., {et~al.} 2013,
  Experimental Astronomy, 35, 103

\bibitem[{{Watson} {et~al.}(2009){Watson}, {Schr{\"o}der}, {Fyfe}, {Page},
  {Lamer}, {Mateos}, {Pye}, {Sakano}, {Rosen}, {Ballet}, {Barcons}, {Barret},
  {Boller}, {Brunner}, {Brusa}, {Caccianiga}, {Carrera}, {Ceballos}, {Della
  Ceca}, {Denby}, {Denkinson}, {Dupuy}, {Farrell}, {Fraschetti}, {Freyberg},
  {Guillout}, {Hambaryan}, {Maccacaro}, {Mathiesen}, {McMahon}, {Michel},
  {Motch}, {Osborne}, {Page}, {Pakull}, {Pietsch}, {Saxton}, {Schwope},
  {Severgnini}, {Simpson}, {Sironi}, {Stewart}, {Stewart}, {Stobbart}, {Tedds},
  {Warwick}, {Webb}, {West}, {Worrall}, \& {Yuan}}]{Watson2009}
{Watson}, M.~G., {Schr{\"o}der}, A.~C., {Fyfe}, D., {et~al.} 2009, \aap, 493,
  339

\bibitem[{{Waugh} {et~al.}(2002){Waugh}, {Drinkwater}, {Webster},
  {Staveley-Smith}, {Kilborn}, {Barnes}, {Bhathal}, {de Blok}, {Boyce},
  {Disney}, {Ekers}, {Freeman}, {Gibson}, {Henning}, {Jerjen}, {Knezek},
  {Koribalski}, {Marquarding}, {Minchin}, {Price}, {Putman}, {Ryder}, {Sadler},
  {Stootman}, \& {Zwaan}}]{Waugh2002}
{Waugh}, M., {Drinkwater}, M.~J., {Webster}, R.~L., {et~al.} 2002, \mnras, 337,
  641

\end{thebibliography}

\begin{appendix}
\section{Quality of the FDS fields}
We give the quality parameters of all the FDS fields in Table \ref{table:quality}. The tests made to obtain the parameters are described in Sections 4.1 and 4.3. We also give the parameters of the {\it PSF} models (Section 5.1) we used for all the fields in Table \ref{table:psf_fits}.

\begin{table*} 
\caption{ Image quality of the FDS fields. The first column gives the name of the field, the four next columns give the mean $FWHM$ and $RMS$ of the $FWHM$ within the field in the different bands, and the four last columns show the surface brightness corresponding to 1$\sigma$ $S/N$ per pixel for a given field in the different photometric bands.} 
\label{table:quality} 
\centering\begin{tabular}{l c c c c c c c c} 
\hline\hline 
 & $FWHM\pm \sigma_{FWHM}$ / arcsec & & & & depth / mag arcsec$^{-2}$& & & \\ 
field & u' & g' & r' & i' & u' & g' & r' & i' \\ 
\hline 

Field1  & 1.17$\pm$0.09 & 1.35$\pm$0.09 & 1.14$\pm$0.15 & 0.69$\pm$0.07 & 25.16 & 26.76 & 26.05 & 25.24  \\ 
Field2  & 1.21$\pm$0.04 & 1.11$\pm$0.07 & 0.90$\pm$0.07 & 0.79$\pm$0.08 & 25.14 & 26.60 & 26.03 & 25.00  \\ 
Field4  & 1.18$\pm$0.11 & 1.39$\pm$0.05 & 1.19$\pm$0.12 & 0.70$\pm$0.07 & 25.20 & 26.70 & 26.01 & 25.26  \\ 
Field5  & 1.33$\pm$0.07 & 1.15$\pm$0.10 & 1.39$\pm$0.14 & 1.08$\pm$0.15 & 25.58 & 26.79 & 26.10 & 25.22  \\ 
Field6  & 1.11$\pm$0.05 & 0.84$\pm$0.08 & 1.08$\pm$0.08 & 1.21$\pm$0.12 & 25.68 & 26.72 & 25.98 & 25.07  \\ 
Field7  & 1.04$\pm$0.05 & 0.83$\pm$0.10 & 0.95$\pm$0.09 & 1.42$\pm$0.11 & 25.55 & 26.81 & 26.06 & 24.87  \\ 
Field9  & 1.38$\pm$0.07 & 1.20$\pm$0.08 & 0.97$\pm$0.09 & 0.84$\pm$0.08 & 25.30 & 26.83 & 26.15 & 25.37  \\ 
Field10 & 1.34$\pm$0.05 & 1.15$\pm$0.04 & 1.02$\pm$0.12 & 1.09$\pm$0.07 & 25.66 & 26.77 & 26.16 & 25.24  \\ 
Field11 & 1.27$\pm$0.05 & 1.06$\pm$0.12 & 1.09$\pm$0.11 & 1.15$\pm$0.06 & 25.25 & 26.51 & 26.02 & 25.04  \\ 
Field12 & 1.15$\pm$0.06 & 0.83$\pm$0.10 & 1.04$\pm$0.10 & 1.17$\pm$0.10 & 25.69 & 26.74 & 26.09 & 25.04  \\ 
Field13 & 1.10$\pm$0.05 & 0.91$\pm$0.06 & 1.03$\pm$0.06 & 1.16$\pm$0.07 & 25.39 & 26.83 & 26.14 & 25.44  \\ 
Field14 & 1.34$\pm$0.06 & 1.18$\pm$0.08 & 0.96$\pm$0.09 & 0.86$\pm$0.07 & 25.28 & 26.70 & 26.00 & 25.29  \\ 
Field15 & 1.30$\pm$0.04 & 1.13$\pm$0.05 & 0.90$\pm$0.07 & 0.97$\pm$0.06 & 25.37 & 26.60 & 26.14 & 25.12  \\ 
Field16 & 1.31$\pm$0.04 & 1.26$\pm$0.07 & 0.94$\pm$0.08 & 1.08$\pm$0.09 & 25.52 & 26.68 & 26.09 & 25.21  \\ 
Field17 & 1.27$\pm$0.04 & 1.11$\pm$0.12 & 0.87$\pm$0.08 & 1.01$\pm$0.08 & 25.35 & 26.54 & 26.21 & 25.17  \\ 
Field18 & 1.11$\pm$0.06 & 0.95$\pm$0.08 & 1.03$\pm$0.09 & 1.12$\pm$0.11 & 25.33 & 26.79 & 26.17 & 25.43  \\ 
Field19 & 1.26$\pm$0.04 & 1.14$\pm$0.13 & 0.89$\pm$0.07 & 0.87$\pm$0.08 & 25.25 & 26.70 & 26.14 & 25.23  \\ 
Field20 & 1.30$\pm$0.06 & 1.22$\pm$0.07 & 0.95$\pm$0.09 & 1.08$\pm$0.07 & 25.29 & 26.46 & 26.06 & 25.04  \\ 
Field21 & 1.22$\pm$0.05 & 1.12$\pm$0.06 & 0.78$\pm$0.05 & 0.88$\pm$0.07 & 25.13 & 26.51 & 25.84 & 25.28  \\ 
Field22 & -$\pm$-       & 1.04$\pm$0.07 & 0.81$\pm$0.05 & 0.85$\pm$0.07 & - & 26.52 & 25.90 & 25.16  \\ 
Field25 & -$\pm$-       & 1.11$\pm$0.10 & 0.77$\pm$0.06 & 0.85$\pm$0.07 & - & 26.63 & 25.84 & 25.11  \\ 
Field26 & -$\pm$-       & 0.93$\pm$0.07 & 0.81$\pm$0.05 & 0.91$\pm$0.07 & - & 25.89 & 25.96 & 25.06  \\ 
Field27 & -$\pm$-       & 1.07$\pm$0.10 & 0.78$\pm$0.06 & 0.89$\pm$0.10 & - & 26.39 & 25.63 & 24.88  \\ 
Field28 & -$\pm$-       & 1.08$\pm$0.14 & 0.79$\pm$0.09 & 0.92$\pm$0.09 & - & 26.31 & 25.57 & 24.89  \\ 
Field31 & 1.33$\pm$0.05 & 1.22$\pm$0.13 & 1.00$\pm$0.08 & 0.86$\pm$0.08 & 25.11 & 26.58 & 25.86 & 24.98  \\ 
Field33 & -$\pm$-       & 1.09$\pm$0.07 & 0.84$\pm$0.07 & 0.83$\pm$0.13        & - & 26.40 & 25.74 & 24.80  \\

\hline 
\end{tabular} 
\end{table*} 

\begin{table*} 
\caption{ Parameters of the inner {\it PSF} fits for each field. The first column gives the number of the field, the next five columns give the parameters of the g'-band fit, and the next five columns for the r'-band fit. The fitted parameters are defined in Eq. 5. $\alpha$ and $\sigma$ are in arcseconds. }
\label{table:psf_fits} 
\centering\begin{tabular}{l | c c c c c | c c c c c} 
\hline\hline 
       & g'   &  &  &  &  & r' & & & & \\
Field  & $I_{0,Gauss}$ & $\sigma$ & $I_{0,Mof}$ & $\alpha$ & $\beta$ & $I_{0,Gauss}$ & $\sigma$ & $I_{0,Mof}$ & $\alpha$ & $\beta$ \\ 
\hline 

Field1 & 0.181 & 0.765 & 0.819 & 0.763 & 1.69 & 0.093 & 0.638 & 0.907 & 0.681 & 1.67 \\ 
Field2 & 0.152 & 0.574 & 0.848 & 0.651 & 1.72 & 0.000 & 0.016 & 1.000 & 0.524 & 1.69 \\ 
Field4 & 0.211 & 0.768 & 0.789 & 0.768 & 1.66 & 0.103 & 0.705 & 0.897 & 0.692 & 1.66 \\ 
Field5 & 0.139 & 0.677 & 0.861 & 0.692 & 1.70 & 0.163 & 0.903 & 0.837 & 0.708 & 1.50 \\ 
Field6 & 0.000 & 0.000 & 1.000 & 0.608 & 1.90 & 0.081 & 0.623 & 0.919 & 0.636 & 1.63 \\ 
Field7 & 0.000 & 0.076 & 1.000 & 0.544 & 1.74 & 0.070 & 0.523 & 0.930 & 0.558 & 1.60 \\ 
Field9 & 0.165 & 0.657 & 0.835 & 0.718 & 1.73 & 0.079 & 0.521 & 0.921 & 0.595 & 1.67 \\ 
Field10 & 0.129 & 0.651 & 0.871 & 0.687 & 1.72 & 0.042 & 0.724 & 0.957 & 0.561 & 1.52 \\ 
Field11 & 0.085 & 0.613 & 0.915 & 0.648 & 1.73 & 0.061 & 0.757 & 0.938 & 0.635 & 1.60 \\ 
Field12 & 0.000 & 0.000 & 1.000 & 0.532 & 1.71 & 0.061 & 0.597 & 0.938 & 0.603 & 1.61 \\ 
Field13 & 0.000 & 0.000 & 1.000 & 0.590 & 1.76 & 0.050 & 0.591 & 0.950 & 0.635 & 1.67 \\ 
Field14 & 0.141 & 0.633 & 0.859 & 0.704 & 1.75 & 0.088 & 0.480 & 0.912 & 0.576 & 1.66 \\ 
Field15 & 0.128 & 0.621 & 0.872 & 0.664 & 1.71 & 0.000 & 0.118 & 1.000 & 0.535 & 1.67 \\ 
Field16 & 0.173 & 0.693 & 0.827 & 0.720 & 1.70 & 0.086 & 0.500 & 0.914 & 0.543 & 1.59 \\ 
Field17 & 0.135 & 0.711 & 0.865 & 0.704 & 1.69 & 0.000 & 0.000 & 1.000 & 0.565 & 1.80 \\ 
Field18 & 0.065 & 0.546 & 0.935 & 0.593 & 1.71 & 0.053 & 0.560 & 0.947 & 0.633 & 1.68 \\ 
Field19 & 0.177 & 0.573 & 0.823 & 0.662 & 1.72 & 0.000 & 0.000 & 1.000 & 0.519 & 1.68 \\ 
Field20 & 0.175 & 0.681 & 0.825 & 0.687 & 1.65 & 0.053 & 0.606 & 0.946 & 0.544 & 1.55 \\ 
Field21 & 0.114 & 0.647 & 0.886 & 0.683 & 1.72 & 0.000 & 0.000 & 1.000 & 0.455 & 1.66 \\ 
Field22 & 0.080 & 0.592 & 0.920 & 0.646 & 1.75 & 0.000 & 0.000 & 1.000 & 0.487 & 1.68 \\ 
Field25 & 0.102 & 0.629 & 0.898 & 0.691 & 1.76 & 0.000 & 0.000 & 1.000 & 0.487 & 1.76 \\ 
Field26 & 0.000 & 0.000 & 1.000 & 0.655 & 1.85 & 0.000 & 0.000 & 1.000 & 0.479 & 1.68 \\ 
Field27 & 0.088 & 0.607 & 0.913 & 0.662 & 1.75 & 0.000 & 0.000 & 1.000 & 0.461 & 1.66 \\ 
Field28 & 0.091 & 0.615 & 0.909 & 0.674 & 1.76 & 0.000 & 0.000 & 1.000 & 0.459 & 1.66 \\ 
Field31 & 0.185 & 0.647 & 0.815 & 0.729 & 1.74 & 0.100 & 0.491 & 0.900 & 0.593 & 1.68 \\

\hline 
\end{tabular} 
\end{table*}

\section{Quantitative test for the effects of the redshift on the morphological and structural parameters}

To test how the increasing redshift changes the measured parameters of the galaxies, we selected a group of spectroscopically confirmed Fornax cluster galaxies with different morphological classes and artificially put them to different distances. We first rebinned the images by a factor of $z/0.005$ (Fornax cluster is located at the redshift of 0.005) and then convolved the data with the OmegaCAM's {\it PSF}. Since the convolution reduces the pixel noise in the images, we empirically tested how much the noise is reduced, and added the required amount of noise to match the image quality with the one of the original data.

\begin{figure*}
        \includegraphics[width=17cm]{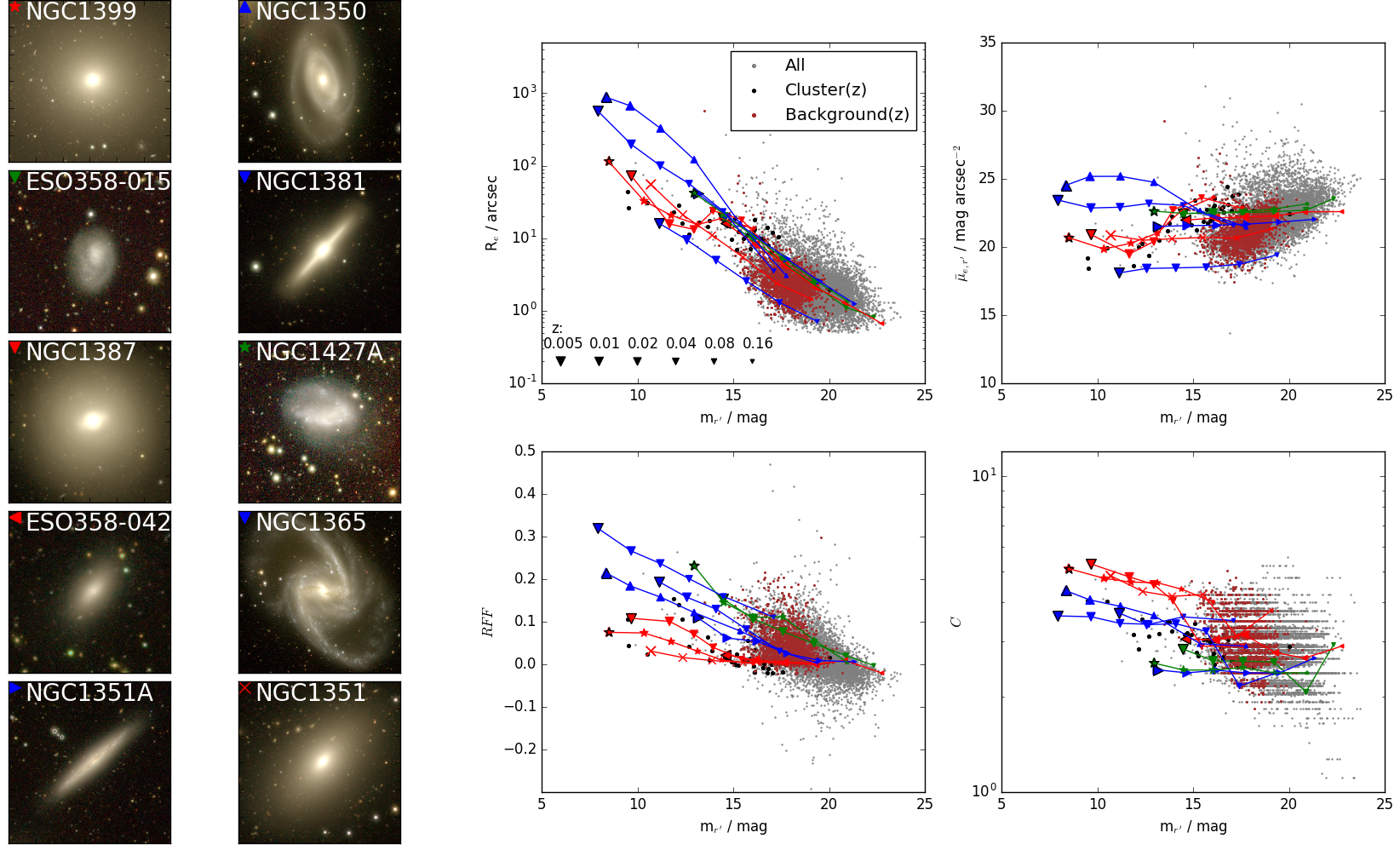}
        \caption{Effect of redshift on the photometric parameters of the galaxies. As a function of galaxy magnitude (m$_{r'}$), shown are the effective radius (R$_e$), the mean effective surface brightness ($\bar{\mu}_{e,r'}$), the Residual Flux Fraction ($RFF$), and the  concentration parameter (C). The galaxies in the images on the left are presented in the right-side panels with the symbols shown in the upper left corner of the images. The different symbol sizes in the panels correspond to the different redshifts, as indicated in the upper left panel. The gray points in the panels correspond to all the galaxies in our sample, with the spectroscopically confirmed cluster and background galaxies indicated with the black and red filled circles, respectively. This figure is best viewed on-screen.}

                \label{fig:effect_of_redshift}
\end{figure*}

\indent In Fig. \ref{fig:effect_of_redshift} we use the photometric parameters (R$_e$, $\mu_e$, RFF and C) to show how the galaxies move in the magnitude - photometric parameter space, as a function of redshift. As expected, the r'-band apparent magnitude (m$_r$) and the effective radius (R$_e$) decrease, and the surface brightness ($\mu_e$) stays almost constant with increasing redshift. The upper right panel in Fig. \ref{fig:effect_of_redshift} shows that there is not much contamination expected from the background galaxies fainter than $\bar{\mu}_{e,r'}$ > 23 mag / arcsec$^{-2}$.  For the  galaxies brighter than that, contamination is expected since the parameters of the redshifted galaxies overlap with the cluster galaxies. The two lower panels in the right side show that the $RFF$ and $C$ parameters are also affected by redshift. This can be explained by {\it PSF} effects, as the relative size of the {\it PSF} compared to the angular size of the galaxies increases and thus blurs the structures in the galaxies. Regardless of the redshift dependence, the different morphological types can still be clearly separated at the different redshifts using the $RFF$ and $C$.

\section{Filter transformations}
To transform the MegaCAM filters into the SDSS filters we used the following formula provided at the MegaCAM web pages (\url{http://www1.cadc-ccda.hia-iha.nrc-cnrc.gc.ca/community/CFHTLS-SG/docs/extra/filters.html}):

\begin{equation}
\begin{array}{l}
u_{Mega} = u' - 0.241 (u' - g')\\
g_{Mega} = g' - 0.153 (g' - r')\\
r_{Mega} = r' - 0.024 (g' - r')\\
i_{Mega} = i' - 0.085 (r' - i'),
\end{array}
\end{equation}
where $x_{Mega}$ correspond to MegaCAM magnitudes and x' correspond to SDSS magnitudes.

\indent To transform the V-I colors into g'-i' colors, we used the transformations of \citet{Jordi2006}
\begin{equation}
V-I   =     (0.671 \pm 0.002)*(g'-i')  + (0.359 \pm 0.002)
\end{equation}

\end{appendix}

\end{document}